\documentclass[screen, manuscript]{acmart}
\usepackage{graphicx}
\usepackage{array}
\usepackage{float}
\usepackage{tabularx}
\usepackage{natbib}
\usepackage{subcaption}

\usepackage[table]{xcolor}
\usepackage{booktabs}
\usepackage{color}
\definecolor{main}{HTML}{CFCFCF}
\definecolor{sub}{HTML}{CFCFCF}

\PassOptionsToPackage{hyphens}{url}
\usepackage{xurl}
\PassOptionsToPackage{colorlinks=true,breaklinks=true}{hyperref}
\usepackage{float}
\usepackage{booktabs}
\usepackage{multirow}
\usepackage{tablefootnote}
\usepackage{flushend}
\usepackage{ragged2e}
\newcolumntype{L}[1]{>{\RaggedRight\arraybackslash}p{#1}}
\usepackage{xstring}

\usepackage{caption}
\usepackage[export]{adjustbox}

\usepackage{listings}

\usepackage{todonotes}

\usepackage{tikz,pgfplots,multirow}
\usetikzlibrary{positioning}
\pgfplotsset{compat=1.18}
\definecolor{tabcat1}{rgb}{0.122,0.467,0.706}
\definecolor{piecat1}{rgb}{0.122,0.467,0.706}
\definecolor{tabcat2}{rgb}{1.000,0.498,0.055}
\definecolor{piecat2}{rgb}{0.682,0.780,0.910}
\definecolor{tabcat3}{rgb}{0.173,0.627,0.173}
\definecolor{piecat3}{rgb}{1.000,0.498,0.055}
\definecolor{tabcat4}{rgb}{0.839,0.153,0.157}
\definecolor{piecat4}{rgb}{1.000,0.733,0.471}
\definecolor{tabcat5}{rgb}{0.580,0.404,0.741}
\definecolor{piecat5}{rgb}{0.173,0.627,0.173}
\definecolor{tabcat6}{rgb}{0.549,0.337,0.294}
\definecolor{piecat6}{rgb}{0.596,0.875,0.541}
\definecolor{tabcat7}{rgb}{0.890,0.467,0.761}
\definecolor{piecat7}{rgb}{0.839,0.153,0.157}
\definecolor{tabcat8}{rgb}{0.498,0.498,0.498}
\definecolor{piecat8}{rgb}{1.000,0.596,0.588}
\definecolor{tabcat9}{rgb}{0.737,0.741,0.133}
\definecolor{piecat9}{rgb}{0.580,0.404,0.741}
\definecolor{tabcat10}{rgb}{0.090,0.745,0.812}
\definecolor{piecat10}{rgb}{0.773,0.690,0.835}
\definecolor{tabcat11}{rgb}{0.122,0.467,0.706}
\definecolor{piecat11}{rgb}{0.549,0.337,0.294}
\definecolor{tabcat12}{rgb}{1.000,0.498,0.055}
\definecolor{piecat12}{rgb}{0.769,0.612,0.580}
\definecolor{tabcat13}{rgb}{0.173,0.627,0.173}
\definecolor{piecat13}{rgb}{0.890,0.467,0.761}
\definecolor{tabcat14}{rgb}{0.839,0.153,0.157}
\definecolor{piecat14}{rgb}{0.969,0.714,0.824}
\definecolor{tabcat15}{rgb}{0.580,0.404,0.741}
\definecolor{piecat15}{rgb}{0.498,0.498,0.498}
\definecolor{tabcat16}{rgb}{0.549,0.337,0.294}
\definecolor{piecat16}{rgb}{0.780,0.780,0.780}
\definecolor{tabcat17}{rgb}{0.890,0.467,0.761}
\definecolor{piecat17}{rgb}{0.737,0.741,0.133}
\definecolor{tabcat18}{rgb}{0.498,0.498,0.498}
\definecolor{piecat18}{rgb}{0.859,0.859,0.553}
\definecolor{tabcat19}{rgb}{0.737,0.741,0.133}
\definecolor{piecat19}{rgb}{0.090,0.745,0.812}
\definecolor{tabcat20}{rgb}{0.090,0.745,0.812}
\definecolor{piecat20}{rgb}{0.620,0.855,0.898}
\definecolor{tabcat21}{rgb}{0.122,0.467,0.706}
\definecolor{piecat21}{rgb}{0.122,0.467,0.706}
\definecolor{tabcat22}{rgb}{1.000,0.498,0.055}
\definecolor{piecat22}{rgb}{0.682,0.780,0.910}
\definecolor{tabcat23}{rgb}{0.173,0.627,0.173}
\definecolor{piecat23}{rgb}{1.000,0.498,0.055}
\definecolor{tabcat24}{rgb}{0.839,0.153,0.157}
\definecolor{piecat24}{rgb}{1.000,0.733,0.471}
\definecolor{tabcat25}{rgb}{0.580,0.404,0.741}
\definecolor{piecat25}{rgb}{0.173,0.627,0.173}
\definecolor{tabcat26}{rgb}{0.549,0.337,0.294}
\definecolor{piecat26}{rgb}{0.596,0.875,0.541}
\definecolor{tabcat27}{rgb}{0.890,0.467,0.761}
\definecolor{piecat27}{rgb}{0.839,0.153,0.157}
\definecolor{tabcat28}{rgb}{0.498,0.498,0.498}
\definecolor{piecat28}{rgb}{1.000,0.596,0.588}
\usepackage{fontawesome}

\usepackage[framemethod=tikz]{mdframed}

\usepackage{etoolbox} 

\definecolor{boxbg}{gray}{0.92}      % light gray background
\definecolor{boxborder}{gray}{0.55}  % medium gray border

\newmdenv[
  skipabove=0.5\baselineskip,
  skipbelow=0.7\baselineskip,
  linewidth=0.5pt,
  linecolor=boxborder,
  backgroundcolor=boxbg,
  roundcorner=6pt,
  innerleftmargin=10pt,
  innerrightmargin=10pt,
  innertopmargin=8pt,
  innerbottommargin=8pt
]{boxC}

\newcommand{\boxCTitle}[1]{%
  \noindent\textbf{#1}.\par\vspace{4pt}%
}

\newcounter{keyTakeAwaysCounter}
\newenvironment{keyTakeAways}[1][Key Take Away]{%
  \refstepcounter{keyTakeAwaysCounter}%
  \begin{boxC}%
  \boxCTitle{\faLightbulbO~\thekeyTakeAwaysCounter. \textbf{#1}}%
\noindent}{%
  \end{boxC}%
}

\newcounter{keyRQAnswerCounter}
\newenvironment{keyRQAnswer}[1][RQAnswer]{%
  \refstepcounter{keyRQAnswerCounter}%
  \begin{boxC}%
  \boxCTitle{\faKey~\thekeyRQAnswerCounter.\ \textbf{#1}}%
\noindent}{%
  \end{boxC}%
}

\newcounter{RQCounter}

\newcounter{keyLimitationsCounter}

\renewcommand{\arraystretch}{1.4}

\usepackage{booktabs,longtable,geometry,array}
\newcolumntype{L}[1]{>{\raggedright\arraybackslash}p{#1}}
\begin{document}
% Short title and authors for the header (optional)

\title {Architectural Degradation: How to Measure and to Remediate}
\author{Noman Ahmad}
\orcid{0009-0005-4228-2493}
\affiliation{%
  \institution{University of Oulu}
  \city{Oulu}
  \country{Finland}
}
\email{noman.ahmad@oulu.fi}

\author{Ruoyu Su}
\orcid{0009-0008-6206-8787}
\affiliation{%
  \institution{University of Oulu}
  \city{Oulu}
  \country{Finland}
}
\email{ruoyu.su@oulu.fi}

\author{Matteo Esposito}
\orcid{0000-0002-8451-3668}
\affiliation{%
  \institution{University Southern Denmark}
  \city{Vejle}
  \country{Denmark}
}
\email{esposito@imada.sdu.dk}

\author{Andrea Janes}
\orcid{0000-0002-1423-6773}
\affiliation{%
  \institution{Free University of Bozen-Bolzano}
  \country{Italy}
}
\email{andrea.janes@unibz.it}

\author{Valentina Lenarduzzi}
\orcid{0000-0003-0511-5133}
\affiliation{%
  \institution{University Southern Denmark}
  \city{Vejle}
  \country{Denmark}
}
\affiliation{%
  \institution{University of Oulu}
  \city{Oulu}
  \country{Finland}
}
\email{lenarduzzi@imada.sdu.dk}

\author{Davide Taibi}
\orcid{0000-0002-3210-3990}
\affiliation{%
  \institution{University Southern Denmark}
  \city{Vejle}
  \country{Denmark}
}
\affiliation{%
  \institution{University of Oulu}
  \city{Oulu}
  \country{Finland}
}
\email{taibi@imada.sdu.dk}


\begin{abstract}
\textbf{Context.} Architectural degradation undermines software maintainability, evolvability, and quality. However, existing research remains fragmented across measurement approaches, metrics, tools, and remediation strategies, limiting our understanding of how these elements relate across the degradation lifecycle.

\noindent\textbf{Aim.} We consolidate the state of the art on architectural degradation by examining how researchers measure it, which metrics and tools support its assessment, and how existing approaches address remediation.

\noindent\textbf{Method.} We conducted a Multivocal Literature Review of 284 peer-reviewed and grey-literature studies. We supported screening, data extraction, and classification with a locally executed LLM-assisted pipeline combining Retrieval-Augmented Generation, multi-model validation, and human adjudication. We then analyzed the resulting taxonomies and their cross-dimensional relationships.

\noindent\textbf{Results and Conclusions.} We identified 277 measurement approaches, 357 metrics, 238 tools, and 395 remediation approaches. Research strongly concentrates on static and structural analysis, structural metrics, and detection-oriented tools. In contrast, remediation spans heterogeneous code-level, architectural, and organizational interventions and shows substantially less consolidation. Overall, the field has developed a mature diagnostic apparatus but has made less progress in connecting degradation detection with effective remediation. Our results provide a structured view of the available techniques and identify the diagnosis--remediation gap as a key direction for future research.\end{abstract}


\maketitle

\section{Introduction}
Software architecture underpins the long-term maintainability, modifiability, scalability, and evolvability of a system~\citep{herold2016}.
Yet as systems evolve to meet new requirements, technologies, and business constraints, their implementation inevitably drifts away from the intended design.
This phenomenon, \emph{architectural degradation} or \emph{erosion}, gradually erodes architectural integrity and threatens sustainability~\citep{herold2014,perry1992foundations}.

Degradation is rarely the result of a single poor decision.
It accumulates through continuous evolution, undocumented changes, architectural violations, growing dependency complexity, and routine maintenance.
Over time, systems become harder to understand, maintain, and extend, while organizations face rising costs and mounting architectural technical debt~\citep{lenarduzzi2021systematic}.
Understanding how to identify, measure, and remediate architectural degradation has therefore become a pressing research and industrial challenge.

Over the past two decades, researchers have proposed a broad spectrum of techniques to tackle this challenge, including architecture conformance checking, architectural smell detection, dependency analysis, architecture recovery, software metrics, static and dynamic analysis, and, more recently, AI-assisted approaches~\citep{DESILVA2012132,li2021}.
Numerous tools support degradation assessment, and remediation strategies have been proposed at the code, architectural, organizational, and runtime levels.
Despite this substantial body of work, the knowledge remains fragmented.
Existing secondary studies typically examine only isolated aspects, metrics, recovery, or erosion detection, making it difficult to see how measurement approaches, metrics, tools, and remediation strategies relate to one another.
Consequently, researchers and practitioners still lack a unified view of the architectural degradation landscape and the connections among its fundamental building blocks.

To fill this gap, we conducted a Multivocal Literature Review (MLR)~\citep{MLRguidelines} that synthesizes evidence from both peer-reviewed and grey literature.
To handle the large-scale analysis, we employed an LLM-assisted pipeline based on Retrieval-Augmented Generation (RAG), multi-model validation, and human adjudication.
Beyond independently classifying the existing literature, we connect the resulting taxonomies through cross-dimensional and maturity analyses, providing an integrated perspective of the architectural degradation ecosystem.
This study makes the following contributions:
\begin{itemize}
    \item a comprehensive multivocal literature review on architectural degradation;
    \item four evidence-based taxonomies;
    \item an integrated and maturity analysis of the research landscape.
\end{itemize}

Our synthesis reveals a striking convergence: the community has built a mature, structurally focused diagnostic apparatus, static analysis, structural design metrics, and detection-oriented tooling dominate the field.
In contrast, remediation remains fragmented across code-level, architectural, and organizational interventions, and the transition from detecting degradation to acting on it is still comparatively immature.
These findings offer researchers a structured agenda for closing the diagnosis--remediation gap and give practitioners an evidence-based map of the techniques available for assessing and managing architectural degradation.

\textbf{Paper structure:} Section~\ref{sec:RW} describes the background and the related works, while  Section~\ref{sec:methodology} the design of this study. Section~\ref{sec:results} presents the obtained results and Section~\ref{sec:discussion} discusses them. Section~\ref{sec:T2V} identifies some threats to validity while Section~\ref{sec:Conclusion} draws the conclusion. 
\section{Background and Related Works}
\label{sec:RW}

%\todo[inline]{need to change some text}

In this section, we review existing studies and present a detailed comparison with our own research.

The evolution of software architecture topic over time has been extensively investigated in the software engineering literature. This section summarizes the most relevant contributions in this area, covering a variety of approaches and findings. The existing work addresses both conceptual frameworks and practical solutions, with a particular focus on improving recovery techniques and mitigating associated risks. However, a comprehensive understanding of architectural degradation requires examining not only how it is detected and measured, but also the tools that support its assessment and the strategies proposed for its remediation. In this regard, there remains a need for a comprehensive investigation integrating these complementary dimensions of architectural degradation.

To show how research in this area has evolved, we present the selected studies in chronological order. The earliest review was conducted by De Silva and Balasubramaniam~\cite{DESILVA2012132}, while the most recent, to the best of our knowledge, is by Baabad et al.~\cite{Baabad202222915}. The primary objective of each study is summarized in Table~\ref{tab:SLR}.
\begin{table}[tb]
\centering
\caption{Related Reviews} 
\label{tab:SLR} 
\resizebox{\linewidth}{!}{%
\begin{tabular}{m{0.5cm} m{0.5cm} m{2.5cm} m{1.5cm} m{4.5cm} m{4cm} m{4cm}}
\hline
\textbf{Ref.} & \textbf{Year} & \textbf{Focus} & \textbf{Method} & \textbf{Findings} & \textbf{Contributions} & \textbf{Limitations} \\ \hline
\cite{DESILVA2012132} & 2012 & Erosion strategies & Survey & A single strategy is insufficient; needs a holistic approach & Categorization framework & Lacks empirical validation \\ \hline
\cite{herold2016} & 2016 & Empirical evidence on degradation & Literature Review & Outlines research roadmap & Critique of methods & Limited scope and depth \\ \hline
\cite{zahid2017} & 2017 & Recovery technique evolution & Survey & Shift to automation due to industry needs & Links industry needs to recovery evolution & Lacks critical analysis of real-world use \\ \hline
\cite{zipani2018} & 2018 & Recovery in SPLs & Mapping study & Techniques don’t fit SPL needs & Highlights SPL-specific issues & No practical solutions proposed \\ \hline
\cite{baabad2020} & 2020 & Degradation in OSS & Literature Review & Frequent changes and smells are key causes & Spotlights erosion in OSS & More focused studies needed \\ \hline
\cite{li2021} & 2021 & Practitioner views on erosion & Survey & Structural + quality concerns, tech + non-tech causes & Reports detection and mitigation practices & Needs stronger empirical basis \\ \hline
\cite{Baabad202222915} & 2022 & Erosion metrics & Mapping study & Classifies metrics and their impact & Characterizes metric types & More study on metric mechanisms needed \\ \hline
\end{tabular}%
}
\end{table}
% 3. Controlling Software Architecture Erosion

The study by De Silva and Balasubramaniam~\cite{DESILVA2012132} provides structured overviews of architectural erosion, classifying existing techniques into prevention, minimization, and repair. The authors analyze the methodologies, tools, and processes associated with each category, emphasizing that architectural erosion cannot be effectively addressed through a single strategy. Instead, they advocate for a combined approach that integrates multiple techniques to manage the problem more comprehensively.

Herold et al.~\cite{herold2016} conducted a literature review focusing on empirical studies related to architectural degradation and consistency checking. The objective was to assess the current state of research and identify areas that have been underexplored. Their findings indicate that many proposed solutions to remediate architectural degradation rely on case studies with limited generalizability. As a result, the authors highlight the need for more rigorous empirical validation, including controlled experiments and large-scale studies, to better understand the effectiveness of these approaches. These insights were later reinforced in their subsequent work, where they outlined a roadmap for future research \citep{herold2016}.

The work by \cite{zahid2017} examines the evolution of architecture recovery techniques, particularly the transition from manual to automated approaches. The study reflects the growing influence of industry demands for faster development cycles, which has driven the need for more efficient recovery methods. The authors emphasize the importance of maintaining architectural integrity in rapidly evolving systems and suggest that future research should focus on automated and semi-automated solutions.

Sinkala et al.~\cite{zipani2018} carried out a mapping study investigating the application of architecture recovery techniques in software product lines (SPLs). The study highlights the challenges of adapting approaches designed for single systems to the more complex SPL context. In particular, it identifies limitations in the handling of shared assets and the management of variability. The findings suggest that existing techniques do not adequately address architectural degradation in SPLs, especially in systems that evolve from cloned or ad hoc codebases and accumulate technical debt over time.

Baabad et al.~\cite{baabad2020} reviewed the architectural degradation in open-source software, identifying key contributing factors such as frequent modifications, rapid system evolution, and limited architectural awareness among developers. The study also highlights common indicators of degradation, including code smells and architectural smells, and emphasizes the need for more research to better understand these causes and improve remediation strategies.

Li et al.~\cite{li2021} conducted a survey to explore architectural erosion from a practitioner’s perspective. The study aims to understand how developers perceive, identify, and manage erosion in practice. The results indicate that architectural erosion is viewed as a structural issue affecting software quality, driven by both technical and organizational factors. Practitioners typically recognize erosion through its symptoms and attempt to mitigate its impact, although the study highlights the need for stronger empirical support for existing tools and techniques.

Finally, Baabad et al.~\cite{Baabad202222915} presented a mapping study focusing on metrics for architectural degradation. The study categorizes existing metrics and demonstrates their role in identifying architectural decay. However, it also highlights the need for further investigation into how these metrics function and how effectively they capture different aspects of degradation.

Although previous secondary studies have examined architectural erosion, recovery techniques, degradation metrics, and practitioner perceptions, they generally address these dimensions separately. None provides an integrated synthesis connecting measurement approaches, metrics, tool support, and remediation strategies across the architectural degradation lifecycle. Our work addresses this gap through a multivocal literature review that combines academic and practitioner evidence and analyzes both the individual dimensions and their cross-dimensional relationships.

In summary, existing secondary studies provide valuable but largely dimension-specific views of architectural degradation. What remains missing is a synthesis that jointly examines how degradation is measured, which metrics operationalize it, which tools support its assessment, and which strategies are available for remediation. Our study addresses this gap by integrating these four dimensions and examining their relationships across the architectural degradation lifecycle.

\section{Methodology}
\label{sec:methodology}
This section addresses methodology. we defined the research goal and the search process such as inclusion and exclusion criteria.

\subsection{Goal and Research Questions}
The purpose of this study is to investigate the phenomenon of architectural degradation and identify the approaches to measure and remediate architectural degradation. Therefore, we conducted a multivocal literature review supported by Large Language Models.
Based on the objectives of our study, we defined the following Research Questions (RQs). 

\begin{boxC}{\textbf{RQ$_1$}}
Which approaches have been proposed for measuring architectural degradation?
\end{boxC}

The purpose of  RQ$_1$ is to explore the different methods that have been proposed or developed to measure software architectural degradation. 
Architectural degradation has been investigated through a broad range of approaches, including metric-based analyses, architecture conformance checking, smell detection, dependency analysis, graph-based techniques, and machine learning methods. However, these approaches have been proposed independently, making it difficult to obtain a comprehensive view of the available solutions and their characteristics. This analysis provides researchers and practitioners with a consolidated overview of the current state of the art and reveals areas requiring further methodological advances.

\begin{boxC}{\textbf{RQ$_2$}}
Which metrics have been considered for measuring architectural degradation?
\end{boxC}

RQ$_2$ addresses the metrics used to assess software architectural degradation. Architectural degradation cannot be directly observed or quantified. Instead, existing studies rely on a variety of proxy metrics that capture structural properties, architectural symptoms, software evolution, and quality-related consequences. Despite the extensive body of research, there is currently no consolidated understanding of which metrics are adopted, which architectural dimensions they capture, or whether a common measurement framework has emerged. RQ$_2$ therefore aims to systematically identify and classify the metrics used to assess architectural degradation, providing a comprehensive overview of the architectural properties they measure and highlighting opportunities for more standardized measurement frameworks. RQ$_2$ aims to systematically identify and classify the metrics used to assess architectural degradation, providing a comprehensive overview of the architectural properties they measure. This analysis highlights inconsistencies, underexplored dimensions, and opportunities for developing more standardized and comprehensive measurement frameworks.

\begin{boxC}{\textbf{RQ$_3$}}
Which tools have been proposed to support architectural degradation analysis?
\end{boxC}

The research question RQ$_3$ aims to find several tools and approaches  by concentrating on these solutions, which can assist in measuring software architectural degradation and understanding how they address different aspects of architectural degradation.
Although numerous techniques for detecting architectural degradation have been proposed, their practical adoption largely depends on the availability of supporting tools. Existing tools differ substantially in terms of supported analyses, automation capabilities, integration with development environments, scalability, and maintenance support. However, there is currently no comprehensive overview of the available tool ecosystem. RQ3 aims to identify and classify the tools proposed for assessing architectural degradation, analyzing their supported functionalities, underlying approaches, and level of automation. This provides practitioners with an overview of existing tool support while highlighting research opportunities for more integrated and continuously operating solutions.

\begin{boxC}{\textbf{RQ$_4$}}
Which approaches and techniques have been proposed for remediating architectural degradation?
\end{boxC}

We need to systematically understand the approaches and techniques proposed in the literature that can help to remediate architectural degradation. Identifying and classifying approaches in RQ$_4$ can provide insights to solve the concern of architectural degradation. Moreover, by uncovering the gaps in current techniques and approaches, this research RQ$_4$ aims to direct future architectural sustainability research and development. Detecting architectural degradation represents only the first step toward preserving software architecture. Long-term software sustainability requires effective remediation strategies capable of preventing, mitigating, or repairing architectural degradation throughout the software lifecycle. Although numerous studies focus on degradation detection, considerably less attention has been devoted to remediation. RQ4 aims to systematically identify and classify the approaches proposed to remediate architectural degradation, including preventive, corrective, and continuous remediation techniques. By synthesizing existing solutions, this research question highlights current limitations and identifies research gaps toward proactive and automated architectural sustainability.

%\todo[inline]{@Noman: Please update the methodology. The following text is copy/pasted from the Motivation's paper}

\subsection{Search Strategy}
\begin{figure}[tb]
    \centering
    \includegraphics[width=0.6\linewidth]{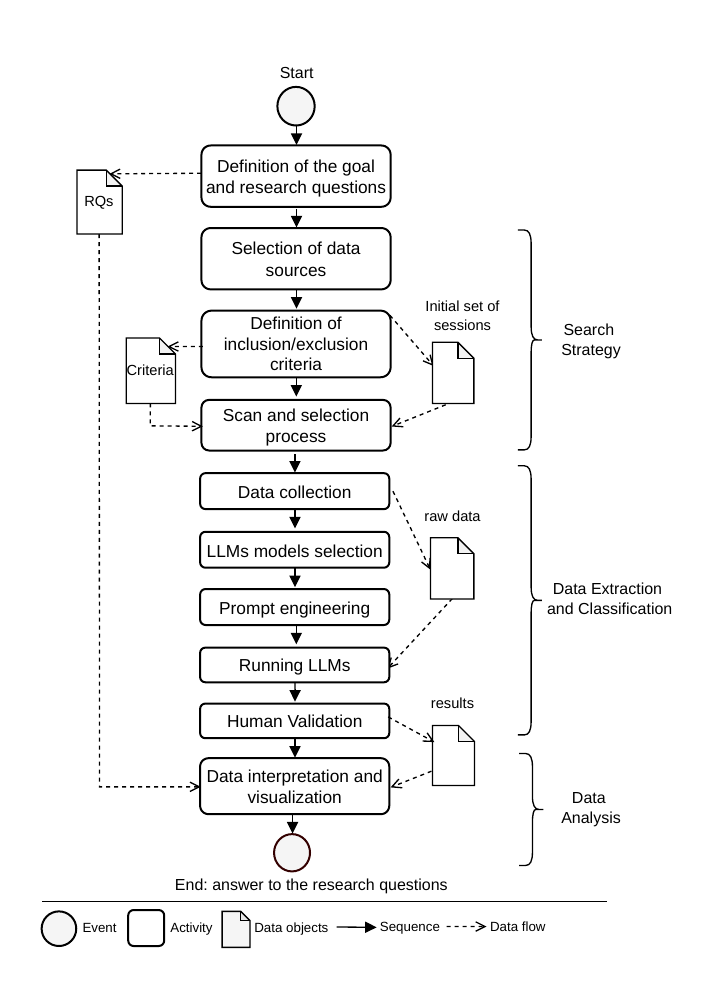}
    \caption{Research Process Overview~\citep{su2026emerging}}
    \label{fig:sasprocess}
\end{figure}

Our search methodology included the following steps: selection of bibliographic sources, definition of inclusion and exclusion criteria, selection of search terms, data extraction, and data classification (Figure~\ref{fig:sasprocess}).

We used the LRM/LLM configuration (rules and steps) described in Section~\ref{sec:LLMSetup} for the search and selection, data extraction, and data classification processes (Figure~\ref{fig:sasprocess}) and carried out two execution rounds: one for the comprehensive reading and another for the title and abstract selection. Additionally, we provided a full description of the outcomes produced by LRM/LLM for each stage in Table~\ref{tab:LLM}.

All the steps were performed within a secure environment of the university's supercomputer locally. We didn’t use any cloud services or external API’s. PDFs documents were utilized only for inference and were not used to fine-tune or train any model. In addition, all downloaded PDFs were removed after processing, and each job was operated in an isolated environment without persistent storage, guaranteeing that no data remained on the system after completion.

\subsubsection{Search String} 

To provide complete coverage of the literature, multiple terms associated with software architecture degradation were utilized in the search strategy, including degradation, aging, erosion, decay, drift, deterioration, degeneration, consistency, adherence, and conformance. Since several of these terms are used interchangeably in prior studies, they were included as alternative keywords. Furthermore, wildcard characters (*) were applied (e.g., software architec and degrad`) to capture different lexical forms of the terms, such as singular and plural nouns, verb forms, and other word variations. Accordingly, the following search string was defined:

\begin{boxC}{\textbf{Search String}}\newline
\textit{``software architec*'' 
\textbf{AND} 
(degrad* OR aging* OR erosion* OR decay* OR drift* OR deteriorat* OR degenerat* OR consisten* OR adher* OR conform*)}
\end{boxC}

\subsubsection{Bibliographic Sources} 
To identify peer-reviewed studies, we selected a set of well-established bibliographic databases based on the guidelines proposed by Kitchenham and Charters~\citep{Kitchenham2013}. These sources are widely regarded as representative of the software engineering research community and have been extensively utilized in previous secondary studies. Accordingly, the selected databases included \textit{Scopus}, \textit{Springer Nature}, \textit{Web of Science (WoS)}, \textit{IEEE Xplore}, and the \textit{ACM Digital Library}.
To complement the academic literature with gray literature, data were collected from several web-based sources, including \textit{Yahoo}, \textit{Bing}, \textit{Google}, \textit{Google Scholar} (for non-peer-reviewed publications), and \textit{Stack Overflow}. These platforms were chosen because they provide access to practitioner knowledge, industrial experiences, technical reports, blogs, and other forms of gray literature that may not be indexed in traditional academic databases. Their use is consistent with established guidelines for conducting Multivocal Literature Reviews (MLRs)~\citep{MLRguidelines}. For example, Janes et al.~\citep{Janes2023} employed general-purpose search engines and practitioner-oriented platforms such as \textit{Stack Overflow} to supplement academic evidence in their systematic gray literature review on microservices.

\subsubsection{Inclusion and Exclusion Criteria}
We explained the criteria for inclusion and exclusion to be applied to the title and abstract (T/A), the full text (F), or both cases (All), as reported in Table~\ref{tab:peer_criteria} and Table~\ref{tab:grey_criteria}.

% \begin{table}[tb]
% \centering
% \small
% \caption{Inclusion and Exclusion Criteria} 
% \label{tab:Criteria} 
% \begin{tabular}
% {@{}p{0.3cm}|p{11cm}|p{0.8cm}@{}}
% \hline
% \textbf{ID} & \textbf{Criteria} & \textbf{Step} \\ \hline
% \textbf{I$_1$} & The paper discusses about software architectural degradation or decay or aging or erosion or drift or deteriorate or degeneration or consistency, adherence, conformance & T/A \\ 
% \hline
% \textbf{E$_1$} & Paper not written in English & T/A \\ 
% \textbf{E$_2$} & Duplicated / Extension has been included & T/A \\ 
% \textbf{E$_3$} & Out of topic & All \\ 
% \textbf{E$_4$} & Not accessible by institution & T/A \\ \hline
% \end{tabular}
% \end{table}

\begin{table}[t]
\centering
\small
\caption{Inclusion and Exclusion Criteria for Peer-reviewed Literature}
\label{tab:peer_criteria}
\begin{tabular}{p{0.4cm}|p{14cm}}
\hline
\textbf{ID} & \textbf{Criteria} \\ \hline
I$_1$ & The study discusses software architectural degradation, erosion, decay, drift, deterioration, degeneration, consistency, adherence, or architecture conformance. \\ \hline \hline 
E$_1$ & The study is not written in English. \\ \hline
E$_2$ & The study is a duplicate or a previous version of another included publication. In such cases, only the most complete version is retained. \\ \hline
E$_3$ & The study is out of scope and does not address software architectural degradation or related architectural evolution issues. \\ \hline
E$_4$ & The study is a secondary study (e.g., systematic literature review, mapping study, survey, tutorial, editorial, or position paper) without providing original evidence. \\ \hline
E$_5$ & The full text is not accessible through institutional subscriptions or publicly available repositories. \\ \hline
E$_6$ & After full-text reading, the study does not report any approach, metric, tool, or remediation strategy related to architectural degradation. \\ \hline
\end{tabular}
\end{table}

\begin{table}[t]
\centering
\small
\caption{Inclusion and Exclusion Criteria for Grey Literature}
\label{tab:grey_criteria}
\begin{tabular}{p{0.4cm}|p{14cm}}
\hline
\textbf{ID} & \textbf{Criteria} \\ \hline
I$_2$ & The source discusses software architectural degradation, erosion, decay, drift, deterioration, degeneration, consistency, adherence, or architecture conformance. \\ \hline \hline 
E$_7$ & The source is not written in English. \\ \hline
E$_8$ & The source is duplicated. \\ \hline
E$_9$ & The source is out of scope and does not address software architectural degradation or related architectural evolution issues. \\ \hline
E$_{10}$ & The source does not provide sufficient technical detail to extract evidence (e.g., advertisements, news articles, or superficial blog posts). \\ \hline
E$_{11}$  & The source is anonymous or lacks an identifiable and trustworthy author or organization. \\ \hline
E$_{12}$  & The source is purely promotional or commercial, without technical or empirical content. \\ \hline
E$_{13}$  & After full-text reading, the source does not report any approach, metric, tool, or remediation strategy related to architectural degradation. \\ \hline
\end{tabular}
\end{table}

\subsubsection{Search and Selection Process} 
The search and study selection process was conducted in January 2026 and covered all publications accessible at that time (Table~\ref{tab:search-process}). The execution of the search strategy resulted in 8672 retrieved peer-reviewed and grey articles. After removing duplicates, we obtained 7263 unique articles, comprising 6413 peer-reviewed and 850 gray.

To facilitate and accelerate the screening process, we used a local reasoning model (LRM) and incorporated the inclusion criteria directly into the prompting strategy. Studies were considered relevant when their titles or abstracts contained terms corresponding to the defined search string. The outputs generated by the LRM were subsequently verified using three independent large language models (LLMs) and further reviewed by the researchers to ensure consistency and reliability.

Following the initial screening phase, the full-text PDFs of the selected studies were provided to the LRM for systematic data extraction. To further improve the quality of the extracted information, our implementation integrated a Retrieval-Augmented Generation (RAG) mechanism, enabling the model to retrieve and utilize relevant contextual evidence during the extraction process. Similarly to the title and abstract screening stage, the extracted data were first cross-validated using multiple LLMs and then examined by the humans.
The complete set of prompts, scripts, and results is also available in the replication package\footref{Package}.

\begin{enumerate}
    \item \textit{Testing the applicability of the inclusion and exclusion criteria.} Before applying the inclusion and exclusion criteria, we conducted a pilot assessment on 20 randomly selected studies from the retrieved set, which were independently evaluated by three authors, following the recommendations of \citet{Kitchenham2013}.
    \item \textit{Applying the inclusion and exclusion criteria to titles and abstracts.}  We applied the refined inclusion and exclusion criteria to their titles and abstracts using our automated script. This step filtered out clearly irrelevant studies while retaining potentially relevant ones for further assessment. As a result, 421 studies were selected for full-text review.
    \item \textit{Full-text reading.} We performed a full-text assessment of the 421 studies that passed the title and abstract screening. The full text of each study was analyzed using prompts designed to identify approaches related to software architecture degradation. Studies that did not report any relevant approach were excluded. Following this review, 284 studies were retained as Primary Studies (PS). 
    \item \textit{Quality and Assessment Criteria:} Before proceeding with the review, we checked whether the quality of the selected articles was sufficient to support our goal and whether the quality of each article reached a certain quality level. To evaluate the selected peer-reviewed articles, we prepared a checklist according to the protocol proposed by Dyb{\aa} and Dings{\o}yr~\cite{Dyba2008} (Table~\ref{tab:QualityCriteriaWhite}) with a set of specific questions. We rank each answer, assigning a score on a five-point Likert scale (0=poor, 4=excellent). A paper satisfied the quality assessment criteria if it achieved a rating higher than (or equal to) 2. For the grey literature, we rely on the checklist proposed by the MLR guideline~\cite{MLRguidelines}. 
\end{enumerate}

\begin{table}
\centering
\small
\caption{Quality Assessment Criteria - Peer-Reviewed Papers} 
\label{tab:QualityCriteriaWhite} 
\begin{tabular}{p{2.2cm}|p{11cm}}
\hline
\textbf{Criteria ID (QA$_s$)} & \textbf{Questions} \\ \hline
QA$_1$&Is the paper based on research (or is it merely a ``lessons learned'' report based on expert opinion)? \\ \hline 
QA$_2$&Is there a clear statement of the aims of the research? \\\hline 
QA$_3$&Is there an adequate description of the context in which the research was carried out?  \\\hline 
QA$_4$&Was the research design appropriate to address the aims of the research? \\\hline 
QA$_5$ & Was the recruitment strategy appropriate for the aims of the research?  \\\hline 
QA$_6$&Was there a control group with which to compare treatments?\\ \hline 
QA$_7$&Was the data collected in a way that addressed the research issue? \\\hline 
QA$_8$&Was the data analysis sufficiently rigorous? \\\cline{1-2}
QA$_9$&Has the relationship between researcher and participants been considered to an adequate degree?\\\cline{1-2}
QA$_{10}$&Is there a clear statement of findings?\\\cline{1-2}
QA$_{11}$&Is the study of value for research or practice?\\\hline
\multicolumn{2}{l}{\small{Response scale: 4 (Excellent), 3 (Very Good), 2 (Good), 1 (Fair), 0 (Poor)}}
\end{tabular}%
\end{table}

\begin{table}
    \centering
\footnotesize
    \caption{Quality Assessment Criteria - Grey literature}
    \label{tab:QualityCriteria}
        \begin{tabular}{p{1.9cm}|p{6cm}|p{6cm}}
\hline
\textbf{Criteria}    & \textbf{Questions} & \textbf{Possible Answers}\\ \hline 
Producer authority & Is the publishing organization reputable?	& 1: reputable and well known organization 	\\ \cline{3-3} 
& & 0.5: existing organization but not well known, 0: unknown or low-reputation organization	\\ \cline{2-3}
& Is an individual author associated with a reputable organization?	& 1: true	\\ \cline{3-3}
& & 0: false	\\ \cline{2-3}

& Has the author published other work in the field?	 & 1: Published more than three other work\\  \cline{3-3}
&&0.5: published 1-2 other works, 0: no other works published.	\\  \cline{2-3}

& Does the author have expertise in the area? (e.g., job title principal software engineer)	& 1: author job title is principal software engineer, cloud engineer, front-end developer or similar\\  \cline{3-3} 
&& 0: author job not related to any of the previously mentioned groups. ) \\ \hline 

Methodology & Does the source have a clearly stated aim? & 1: yes\\  \cline{3-3}
&&  0: no	\\ \cline{2-3}

& Is the source supported by authoritative, documented references?	& 1: references pointing to reputable sources \\\cline{3-3}
&&0.5: references to non-highly reputable sources\\  \cline{3-3}
&& 0: no references\\  \cline{2-3}

& Does the work cover a specific question?	& 1: yes\\ \cline{3-3}
&& 0.5: not explicitly\\ \cline{3-3}
&& 0: no\\ \hline 

Objectivity & Does the work seem to be balanced in presentation	& 1: yes\\ \cline{3-3}
&& 0.5: partially\\ \cline{3-3}
&& 0: no\\  \cline{2-3}

& Is the statement in the sources as objective as possible? Or, is the statement a subjective opinion?	 & 1: objective\\ \cline{3-3}
&& 0.5 partially objective\\ \cline{3-3}
&& 0: subjective\\  \cline{2-3}

& Are the conclusions free of bias or is there vested interest? & 1=no interest\\ \cline{3-3}
&& 0.5: partial or small interest\\ \cline{3-3}
&& 0: strong interest\\  \cline{2-3}

& Are the conclusions supported by the data? & 1: yes \\ \cline{3-3}
&& 0.5: partially\\ \cline{3-3}
&& 0: no\\ \hline 

Date & Does the item have a clearly stated date? & 1: yes\\ \cline{3-3}
&& 0: no \\  \hline 

Position & Have key related GL or formal sources been linked to/discussed? & 1: yes\\ \cline{3-3}
&& 0: no \\\hline 

Novelty & Does it enrich or add something unique to the research?	& 1: yes\\ \cline{3-3}
&& 0.5: partially\\ \cline{3-3}
&& 0: no \\\hline 

Outlet type & Outlet Control & 1:  high outlet control/ high credibility: books, magazines, theses, government reports, white papers \\ \cline{3-3}
& & moderate outlet control/ moderate credibility (e.g. annual reports, videos, Q/A sites, wiki articles) \\ \cline{3-3}
& &  0: low outlet control/low credibility (e.g. tweet, presentations, emails) \\ \hline 

\end{tabular}%
\end{table} 

\begin{table}[t]
\centering
\small
\caption{Primary Study (PS) Search and Selection Process}
\label{tab:search-process}

\begin{subtable}[t]{0.47\linewidth}
\centering
\begin{tabular}{l|r} \hline
\textbf{Step} & \textbf{\#} \\ \hline
-Scopus & +1382 \\
-Springer & +566 \\
-Wos & +898 \\
-IEEE & +2731 \\
-ACM & +1916 \\ \hline
\rowcolor{gray!20} Retrieved & 7493 \\ \hline
\rowcolor{gray!20} Duplicates & -1080 \\ \hline
\rowcolor{gray!20} Final retrieved & 6413 \\ \hline
\rowcolor{gray!20} Title \& abstract & -6026 \\ \hline
\rowcolor{gray!20} Full reading & -120 \\ \hline
\rowcolor{gray!20} Quality assessment & -1 \\ \hline
\textbf{PS} & \textbf{266} \\ \hline
\end{tabular}
\caption{Peer-reviewed Literature}
\label{tab:peer}
\end{subtable}
\hfill
\begin{subtable}[t]{0.47\linewidth}
\centering
\begin{tabular}{l|r} \hline
\textbf{Step} & \textbf{\#} \\ \hline
-Yahoo & +421 \\
-Bing & +308 \\
-Google & +16 \\
-Google Scholar & +390 \\
-Stackoverflow & +44 \\ \hline
\rowcolor{gray!20} Retrieved & 1179\\ \hline
\rowcolor{gray!20} Duplicates & -329 \\ \hline
\rowcolor{gray!20} Final retrieved & 850 \\ \hline
\rowcolor{gray!20} Title\& abstract & -816 \\ \hline
\rowcolor{gray!20} Full reading & -15 \\ \hline
\rowcolor{gray!20} Quality assessment & -1 \\ \hline
\textbf{PS} & \textbf{18} \\ \hline
\end{tabular}
\caption{Grey Literature}
\label{tab:grey}
\end{subtable}
\end{table}

\begin{boxC} {\textbf{Search and Selection Process.}}
Starting from the initial \textbf{7263 unique papers}, following the process, we finally included \textbf{284} papers, called \textbf{Primary Study (PS)}. 
\end{boxC}

\subsection{LRM/LLM Setup}
\label{sec:LLMSetup}

In empirical research, categorization and rating tasks were some of the most laborious and error-prone activities, frequently prone to inconsistencies, bias, and learning effects in human judgment, especially if conducted through large-scale manual coding~\citep{esposito_large_2024,esposito_call_2025}.

LRM/LLM are becoming more common and better at what they do, which has created new possibilities for people in Software Engineering (SE). They can now help with tasks that humans used to do accurately and efficiently. So, as an alternative to a manual, expert-driven coding and labeling procedure, we used LRM/LLM as judges to improve objectivity, scalability, and reproducibility. 

Traditional NLP methods can also do this task, but they needed feature engineering that was specific to the job, training data, and a lot of tweaking for each classification dimension. However, LLMs enabled us to: 
\begin{itemize}
    \item Extract multi-label, semantically rich data from brief sentences without the need for custom pipeline engineering;
    \item Use similar prompt logic throughout dimensions;
    \item Evaluate results at scale using model agreement and sampling.
\end{itemize}

The recent studies performed by \citep{esposito_generative_2025}, \citep{robredo2025were}, \citep{su2026emerging} show that LLMs can help practitioners and researchers with encouraging levels of dependability across a wide range of software engineering tasks, such as classification, summarization, and decision-making.

Previous studies synthesizing evidence~\cite{silva2012controlling} on architectural degradation have predominantly relied on manual coding, keyword-based techniques, or predefined taxonomies, approaches that can become increasingly resource-intensive when applied to large and heterogeneous bodies of literature~\cite{li2021,li2022understanding,}.

In contrast, our approach, successfully applied in similar tasks~\cite{robredo2025were,su2026emerging,su2026evaluating} integrates LLM-based semantic extraction with multi-model cross-validation and human adjudication, enabling the systematic analysis of a large and heterogeneous corpus of academic and gray literature. This pipeline supports the extraction and classification of measurement approaches, metrics, tools, and remediation strategies at scale while preserving human oversight throughout the process. Importantly, our aim is not to replace established empirical review methods, but to support their application at a larger scale while maintaining methodological rigor and traceability.

In our study, we analyzed peer-reviewed and grey literature on software architectural degradation and extracted information on measurement approaches, metrics, supporting tools, and remediation strategies. Our methodology followed five main steps:

\begin{enumerate}
    \item Using a large reasoning model to make a decision about papers inclusion and data extraction;
    \item Validation of the output with three LLMs and human validation;
    \item Clean and unify the extracted data;
    \item Categorize the extracted data from the papers by using LLM;
    \item Analyze co-occurrence patterns and map relationships
\end{enumerate}

\subsection{LRM/LLMIntegration, Experimental Design and Validation}
\paragraph{LRM/LLM Model Roles:}
We assigned a unique role to each model based on its responsibility in the data analysis pipeline as follows (Table~\ref{tab:model_overview}):

\begin{itemize}
    \item \textbf{Large Reasoning Model (LRM):} Marco-o1 is responsible to extract the required information. For each prompt, it receives input data (e.g., title/abstract, full paper),  it makes decision first of all about title and abstract selection and then full read the selected papers and later on classifies the data into categories, and provides information. Its output formed the basis for further verification and was referred to as LRM responses throughout the study.
    \item \textbf{Validation Models (V1/V2/V3):} Llama, Mistral-NeMo, and Qwen  are used for validation task. They receive the similar input as the LRM and used to evaluate against the LRM's reasoning and output.
\end{itemize}

\paragraph{Prompting Techniques:} 
We applied prompt engineering strategies to steer LRM/LLM in classifying approaches to measure and remediate software architectural degradation. Prior research indicates that in-context learning via chat-based prompting can achieve performance comparable to or exceeding that of fine-tuning, while avoiding its higher computational cost \citep{esposito_beyond_2024}. In our setup, each prompt comprised two elements: a \textit{system message}, which defined the assistant’s role and specified the required output structure, and a \textit{user message}, which supplied the contextual input  (e.g title/abstract).
The model was instructed to produce a structured JSON output including: 
\begin{itemize}
    \item The predicted category;
    \item A brief \texttt{description};
    \item The corresponding \texttt{reasoning}.
\end{itemize}

We employed a few-shot learning approach combined with \texttt{Chain-of-Thought} (CoT) prompting \citep{wei2022chain} to encourage more transparent and accurate classification.

\paragraph{Retrieval-Augmented Generation (RAG):} 
We implemented a Retrieval-Augmented Generation (RAG) mechanism into our pipeline for both full-document processing. RAG enhances the model's input with relevant external knowledge by retrieving semantically similar, previously labeled instances from the dataset using embedding-based similarity search. In our case, scientific PDFs exceed typical model context limits, each document was first converted into text, segmented into overlapping chunks, and indexed using embedding-based similarity search (FAISS).  The model was then instructed to generate a structured JSON output containing one concise answer per category. This approach enables efficient processing of long documents while ensuring that classification decisions are evidence-based, context-aware, and less prone to hallucination.

\paragraph{Running LRM/LLM:} For executing this big scale analysis, we CSC’s Mahti supercomputer, using vLLM to serve the LRM/LLM for large-scale inference\footnote{\url{http://csc.fi}}. Mahti is a high-performance computing platform intended for computationally and data-intensive research, providing more than 180,000 CPU cores connected through a fast interconnect network. Our jobs leveraged up to four NVIDIA A100 GPUs, ensuring high-speed and memory-efficient inference to process large volumes of data effectively. We employed deterministic decoding (temperature = 0, max\_tokens = 1900) with a strict no-hallucination JSON prompt to ensure reproducible, evidence-grounded extraction and to minimize stochastic variability.

\paragraph{Human Validation:} Since no previous study reported findings on the accuracy of LLMs for the task at hand, we designed a validation involving three human experts mimicking the LLM validator's roles. 
The goal was to assess the quality and correctness of the model-generated outputs and to determine whether the validation models provided consistent judgments.
Our validation followed a three-step protocol. 

\begin{itemize}
    \item One expert independently reviewed the same input provided to the LRM and the three validation models (V1–V3), and manually evaluated the correctness of each model's motivation. For each case, the expert indicated whether they \texttt{agreed} or \texttt{disagreed} with the LRM’s motivation, noted the majority decision among the validation models, and identified the models they considered correct.
    \item  A second expert repeated the same evaluation independently and documented their level of agreement with the first expert’s judgments. 
    \item  In cases of disagreement between the first two reviewers, a third expert was brought in to independently assess the same outputs. Final decisions were made through majority voting among the three validators.   
\end{itemize}

\begin{table*}[hbtp]
  \centering
  \footnotesize
  \caption{Overview of Selected LLMs}
  \begin{tabular}{p{0.5cm} p{2cm} | l p{4cm} p{4cm}}
    \hline
   \multicolumn{2}{l|}{\textbf{Model}} 
     & \textbf{Parameters/Quantization} & \textbf{Details} & \textbf{Highlights} \\
    \hline
\textbf{(LRM) } & \textbf{Marco o1\textsuperscript{a}} & 7.6B, not quantized & Inspired by OpenAI’s o-1 & Fine-tuned on CoT datasets, uses MCTS + softmax scoring, excels at math, coding, and logic tasks\\ \hline 
\textbf{(LLM)} & \textbf{Mistral NeMo Instruct 2407\textsuperscript{b}} & 12.2B, not quantized & Fine-tuned version of Mistral-Nemo-Base-2407 & Alignment fine-tuned, supports 128K tokens, outperforms similarly-sized models \\ 
\textbf{(LLM)}& \textbf{Qwen2.5 14B\textsuperscript{c}} & 14.8B, not quantized & Decoding Transformer-based & Fine-tuned with enhanced instruction-following, excels in math, programming, and dialogue \\
\textbf{(LLM)}&  \textbf{Llama 3.1 8B\textsuperscript{d} }& 8B, not quantized & Meta-developed decoding architecture & High performance, strong language understanding and generation, lightweight \\
\hline
\multicolumn{5}{l}{\textsuperscript{a} \url{https://huggingface.co/AIDC-AI/Marco-o1}, \textsuperscript{b} \url{https://huggingface.co/mistralai/Mistral-Nemo-Instruct-2407}}\\
\multicolumn{5}{l}{\textsuperscript{c} \url{https://huggingface.co/Qwen/Qwen2.5-14B-Instruct}, \textsuperscript{d} \url{https://huggingface.co/meta-llama/Llama-3.1-8B-Instruct}}\\ 
  \end{tabular}
  \label{tab:model_overview}
\end{table*}
\subsubsection{Data Extraction and Classification}
\label{sec:DataExtraction}
% For initial step, we preprocessed the data removing duplicate records and, then, we used the LRM/LLM process to extract the relevant data.
% Our scripts processed each paper (based on the ID), extracted the required information, and stored the results in a CSV file corresponding to the assigned paper IDs. 
% Then, we obtained the data corresponding to the approaches related to software architectural degradation such as data about measuring approaches, remediation approaches, tools and metrics. 

% % \subsubsection{Data Classification}
% % \label{sec:DataClassification}

% We classified the extracted data using thematic and axial coding~\citep{CorbinStrauss2008} through multiple iterative rounds to refine the categories and ensure consistency. The validation with three LLMs and human reviewers was performed only in the final round for each RQ, as earlier rounds focused on improving and consolidating the categories. This allowed us to evaluate the quality of the finalized classification while avoiding repeated validation of intermediate versions.

% The data extraction form, together with the mapping of the information needed to answer each RQ, is summarized in Table~\ref{tab:DataExtraction}.

After identifying the primary studies, we designed a structured data extraction form to ensure that all information required to answer the research questions was collected consistently across studies. The extracted fields were defined before the extraction process and were directly mapped to the research questions. Table~\ref{tab:data-extraction} summarizes the extracted data items, their purpose, and their relationship with each research question.

Each primary study was processed using the LRM/LLM-based extraction pipeline (Section~\ref{sec:LLMSetup}). The full text of each paper was provided to the LRM together with prompts specifically designed to extract the information required for the review. The LRM generated structured JSON outputs containing the extracted information, which were automatically converted into CSV files to facilitate subsequent analysis.
To improve the reliability of the extracted data, the generated outputs were validated using the multi-model validation process (Section~\ref{sec:LLMSetup}). Whenever inconsistencies or low-confidence extractions were detected, the corresponding studies were manually inspected and the extracted information was corrected accordingly.
As summarized in Table~\ref{tab:data-extraction}, we extracted information regarding measurement approaches, architectural degradation metrics, supporting tools, and remediation strategies. This structured extraction allowed a consistent comparison across the primary studies and provided the evidence base for the thematic coding and synthesis.
Finally, all extracted data were stored in a structured repository, maintaining traceability between each extracted element and its corresponding primary study, thus facilitating the reproducibility and auditability of the review process.

%\todo[inline]{@Noman: please report Inter rater agreement }

%\todo[inline]{@Noman: adjust the methodology to actually reflect what has been done in this paper removing the references to 5W1H paper}

% \begin{table*}[t]
% \centering
% \caption{Data extraction form used to answer the research questions.}
% \label{tab:data_extraction}
% \small
% \begin{tabularx}{\textwidth}{p{0.4cm}|XX}
% \hline
% \textbf{RQ} & \textbf{Extracted Data} & \textbf{Purpose} \\
% \hline
% RQ$_1$ &
% Approach name, underlying technique, input artifacts, degradation aspect addressed, output produced. &
% To identify and classify the existing approaches proposed to assess architectural degradation. \\
% \hline
% RQ$_2$ &
% Metric name, metric category, architectural property measured, analysis granularity (e.g., system, component, class), measurement purpose. &
% To identify the metrics used to assess architectural degradation and understand which architectural properties are considered indicators of degradation. \\
% \hline
% RQ$_3$ &
% Evaluation context, subject systems, datasets, application domain, industrial/open-source setting, system size. &
% To understand where and under which conditions the proposed approaches have been evaluated. \\
% \hline
% RQ$_4$ &
% Validation strategy, comparison baseline, reported benefits, identified limitations, threats to validity. &
% To assess the maturity, effectiveness, and limitations of the existing approaches. \\
% \hline
% \end{tabularx}
% \end{table*}

\begin{table}[t]
\centering
\small
\caption{Data extraction form used to answer the research questions.}
\label{tab:data-extraction}
\renewcommand{\arraystretch}{1.1}
\begin{tabularx}{\linewidth}{c|p{4.2cm}|X}
\hline
\textbf{RQ} &
\textbf{Extracted Data} &
\textbf{Purpose} \\
\hline

RQ$_1$ &
Measurement approach; approach description; supported analysis &
Identify and classify the approaches proposed to assess architectural degradation and construct the taxonomy of measurement approaches. \\
\hline

RQ$_2$ &
Metric name; metric definition; measured architectural property &
Identify the metrics used to assess architectural degradation, normalize synonymous metrics, and classify them into coherent metric categories. \\
\hline

RQ$_3$ &
Tool name; tool category; supported functionality; automation level; availability &
Identify the software tools supporting architectural degradation assessment, classify them according to their primary functionality, and characterize the current tool ecosystem. \\
\hline

RQ$_4$ &
Remediation approach; remediation level; remediation objective; automation level &
Identify and classify the approaches proposed to remediate architectural degradation and characterize the available remediation strategies. \\
\hline

\end{tabularx}
\end{table}

\begin{table*}[tb]
\centering
\small
\caption{Manual validation results for LRM/LLM outputs}
\label{tab:LLM}
\begin{tabular}{l|l|rrlrrr}
\hline
\textbf{Step} & \textbf{RQ} & \textbf{Round (\#)} & \textbf{Dataset (\#)} &
\textbf{Model} & \textbf{Sample (\#)} & \textbf{Disagreement (\#)} & \textbf{Accuracy (\%)} \\
\hline

\multirow{3}{*}{Search \& Selection} & \multirow{3}{*}{All} & \multirow{3}{*}{1} & \multirow{3}{*}{7263}
& Llama   & 369 & 84 & 77.24 \\
& & & & Mistral & 369 & 54 & 85.37 \\
& & & & Qwen & 369 & 73 & 80.22 \\

\hline

\multirow{12}{*}{Data Extraction}
& \multirow{3}{*}{RQ$_1$} & \multirow{3}{*}{1} & \multirow{3}{*}{284} % ((173-Disagreements)/Disagreements) X 100
& Llama   & 164 & 22 & 86.59 \\
& & & & Mistral & 164 & 19 & 88.41 \\
& & & & Qwen    & 164 & 22 & 86.59 \\ \cline{2-8}

& \multirow{3}{*}{RQ$_2$} & \multirow{3}{*}{1} & \multirow{3}{*}{284}
& Llama   & 164 & 37 & 77.44 \\
& & & & Mistral & 164 & 39 & 76.22 \\
& & & & Qwen    & 164 & 44 & 73.17 \\ \cline{2-8}

& \multirow{3}{*}{RQ$_3$} & \multirow{3}{*}{1} & \multirow{3}{*}{284}
& Llama   & 164 & 18 & 89.02 \\
& & & & Mistral & 164 & 18 & 89.02 \\
& & & & Qwen    & 164 & 27 & 83.54 \\ \cline{2-8}

& \multirow{3}{*}{RQ$_4$} & \multirow{3}{*}{1} & \multirow{3}{*}{284}
& Llama   & 164 & 26 & 84.15 \\
& & & & Mistral & 164 & 41 & 75.00 \\
& & & & Qwen    & 164 & 45 & 72.56 \\ \cline{2-8}

\hline

\multirow{12}{*}{Data Classification}
& \multirow{3}{*}{RQ$_1$} & \multirow{3}{*}{3} & \multirow{3}{*}{16}
& Llama   & 16 & 2  & 87.50 \\
& & & & Mistral & 16 & 2 & 87.50 \\
& & & & Qwen    & 16 & 4 & 75.00 \\ \cline{2-8}

& \multirow{3}{*}{RQ$_2$} & \multirow{3}{*}{3} & \multirow{3}{*}{14}
& Llama   & 14 & 2  & 85.71 \\
& & & & Mistral & 14 & 2  & 85.71 \\
& & & & Qwen    & 14 & 5 & 64.29 \\ \cline{2-8}

& \multirow{3}{*}{RQ$_3$} & \multirow{3}{*}{3} & \multirow{3}{*}{16}
& LLaMA   & 16 & 7  & 56.25 \\
& & & & Mistral & 16 & 5 & 68.75 \\
& & & & Qwen    & 16 & 13 & 18.75 \\ \cline{2-8}

& \multirow{3}{*}{RQ$_4$} & \multirow{3}{*}{3} & \multirow{3}{*}{15}
& Llama   & 15 & 2  & 86.67 \\
& & & & Mistral & 15 & 4  & 73.33 \\
& & & & Qwen    & 15 & 6  & 60.00 \\ \cline{2-8}

\hline
\end{tabular}
\end{table*}

\subsection{Verifiability and Replicability}
\label{sec:Replicability}
To support the verifiability and replication of our study, we provide a replication package \footnote{\label{Package}} containing the complete search protocol, inclusion and exclusion criteria, prompts, LLM configurations, processing scripts, extracted data, classification outputs, and analysis results. The package preserves traceability between the extracted evidence and the corresponding primary studies, enabling independent inspection of the review process and reproduction of the analyses reported in this paper. 

\section{Results}\label{sec:results}

For each table (Tables~\ref{tab:rq1_combined}--\ref{tab:rq4_combined}), the first column (\textit{Category}) reports the high-level category together with the number and percentage of primary studies mentioning that category. The second column (\textit{Subcategory}) reports the corresponding subcategories derived through the LLM-supported thematic/axial coding process (Section~\ref{sec:LLMSetup}). The third (\textit{\#PS}) and fourth (\textit{\%PS}) columns indicate, respectively, the number and percentage of primary studies mentioning each subcategory. Since a single primary study may contribute to multiple categories or subcategories, the \#PS values are non-exclusive and therefore do not necessarily sum to the total number of primary studies. Finally, the last column reports the number of distinct approaches, metrics, tools, or remediation techniques identified within each subcategory.
For each research question, we first present the overall distribution of the identified categories and then discuss the three most frequently reported categories together with the least represented one. This organization highlights both the dominant research directions and the areas that have received comparatively limited attention. 

\subsection{Approaches to Measure Architectural Degradation (RQ$_1$)}
% \todo[inline]{@Davide: decidiamo se top 3+1 last o diversamente.}   

\begin{figure}[tb]
    \centering
    \includegraphics[width=\linewidth]{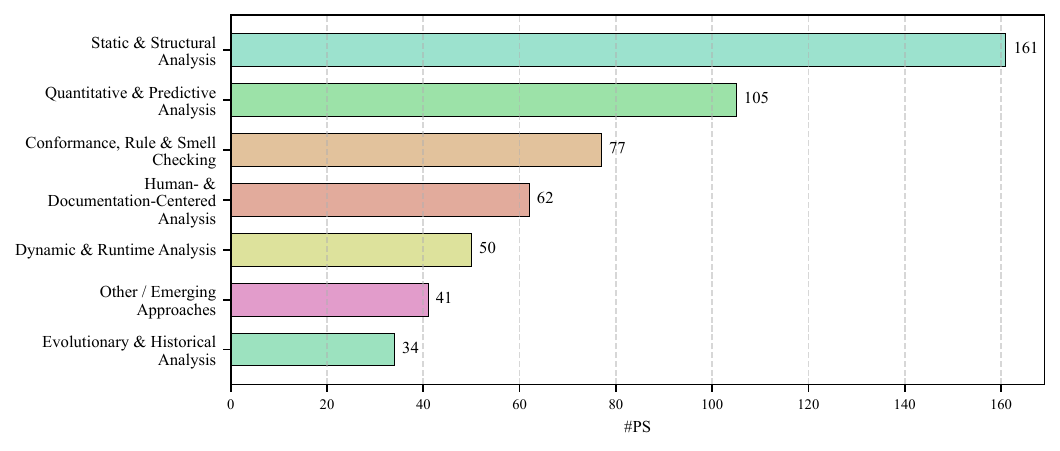}
    \caption{Distribution of measurement approaches across categories (RQ$_1$).}
    \label{fig:RQ1}
\end{figure}

We identified 277 techniques for measuring architectural degradation and grouped them into seven categories according to their underlying analysis strategy. Figure~\ref{fig:RQ1} and Table~\ref{tab:rq1_combined} show a clear concentration around static and code-centric assessment. Researchers rely primarily on approaches that inspect source code, structural dependencies, architectural rules, and measurable software properties without observing how the system evolves or behaves during execution.

\textbf{Static \& Structural Analysis} dominates the literature and appears in \textbf{56.7\%} of the primary studies. Static Analysis alone appears in 40.5\% of the the studies \ref{WL3}, \ref{WL5}, \ref{WL9}, \ref{WL10}, \ref{WL13}, \ref{WL16}, \ref{WL17}, \ref{WL19}, \ref{WL21}, \ref{WL32}, \ref{WL33}, \ref{WL38}, \ref{WL39}, \ref{WL44}, upto \ref{GL280} represents the most frequently reported subcategory in the entire taxonomy. Dependency Analysis accounts for 8.8\% \ref{WL1}, \ref{WL38}, \ref{WL39},\ref{WL185}, \ref{WL207}, \ref{WL266}, while Graph-Based Analysis accounts for 7.4\% \ref{WL1}, \ref{WL38}, \ref{WL43}, \ref{WL45}, \ref{WL51}, \ref{WL52}, \ref{WL56}, \ref{WL241}. Together, these approaches inspect source code, analyze relationships among components, and represent the architecture through structural or dependency graphs to identify anomalies, undesirable configurations, and deviations from the intended design. Researchers frequently adopt these techniques because they scale to large systems, support repeatable analyses, and integrate naturally into continuous integration and quality assurance workflows.

\textbf{Quantitative \& Predictive Analysis} forms the second-largest category and appears in \textbf{37.0\%} of the studies. Metric-Based Assessment drives this category and accounts for 27.8\%  \ref{WL1}, \ref{WL7}, \ref{WL21}, \ref{WL26}, \ref{WL27}, \ref{WL29},\ref{GL279}, while Machine Learning and Prediction reaches 8.1\% \ref{WL2}, \ref{WL5}, \ref{WL20}, \ref{WL38}, \ref{WL133},\ref{GL276}. Technical Debt Assessment remains uncommon and appears in only 1.1\% of the studies \ref{WL147}, \ref{WL150}, \ref{GL272}. This distribution shows that researchers frequently quantify degradation through predefined measurements but still make limited use of predictive techniques that could identify emerging degradation patterns before they become severe.

At the opposite end, \textbf{Evolutionary \& Historical Analysis} represents the smallest category and appears in only \textbf{12.0\%} of the studies ref{WL4}, \ref{WL37}, \ref{WL38}, \ref{WL49},\ref{WL259}, \ref{WL260}, \ref{GL268}. These approaches analyze version histories, architectural changes, and degradation patterns across releases to understand how architecture evolves over time. The difference between Static \& Structural Analysis and Evolutionary \& Historical Analysis approaches a factor of five. Although architectural degradation develops through continuous change, researchers still assess it mainly through structural properties observable in individual system snapshots.

\begin{figure}
    \centering
    \includegraphics[width=\linewidth]{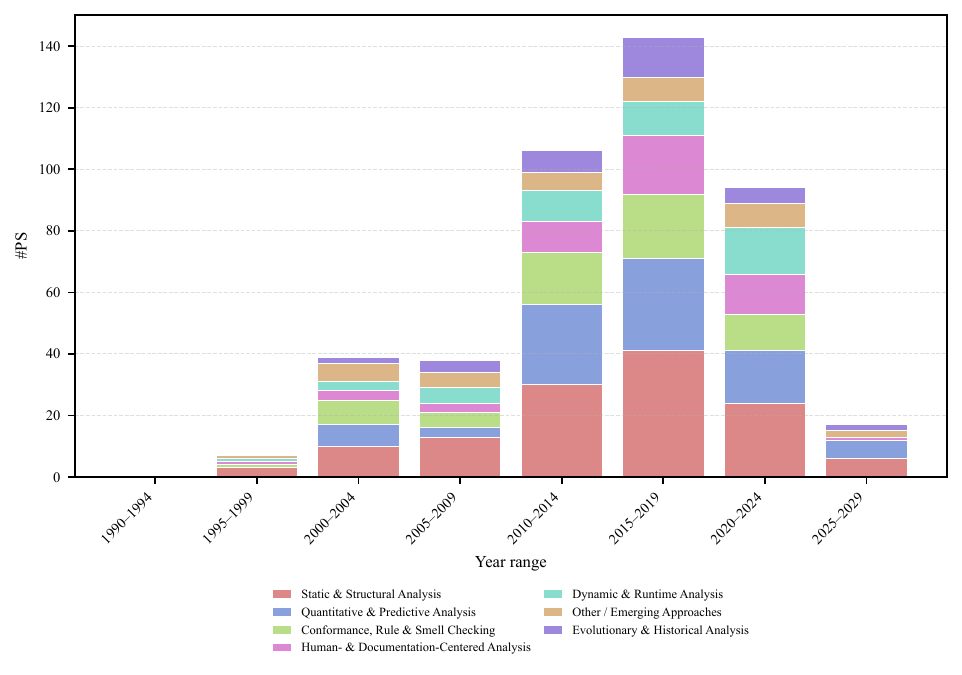}
    \caption{Evolution of architectural degradation measurement approaches over time}
    \label{fig:placeholder}
\end{figure}

\begin{keyRQAnswer}[Measuring Approaches]
Researchers measure architectural degradation primarily through static and structural analysis (56.7\%), followed by quantitative and predictive analysis (37.0\%) and conformance, rule, and smell checking (27.1\%). Evolutionary and historical analysis represents the least explored category (12.0\%). The literature therefore characterizes degradation mainly through structural properties observable in individual software snapshots rather than through the process by which degradation develops over time.
\end{keyRQAnswer}

\begin{table}[tb]
\centering
\footnotesize
\caption{Categories and Subcategories with PS citation counts (RQ$_1$).}
\label{tab:rq1_combined}
\begin{tabular}{L{4.8cm}L{4.0cm}rrrr}
\hline
Category (\#, \%) & Subcategory & \#PS & PS (\%) & Category (\%) & \#Approaches \\
\hline
\multirow{3}{=}{\parbox{4.4cm}{Static \& Structural Analysis \\ (161 - 56.7\%)}} & Static Analysis & 115 & 40.5 & 71.4 & 48 \\
 & Dependency Analysis & 25 & 8.8 & 15.5 & 12 \\
 & Graph-Based Analysis & 21 & 7.4 & 13.0 & 18 \\
\hline
\multirow{3}{=}{\parbox{4.6cm}{Quantitative \& Predictive Analysis \\ (105 - 37.0\%)}} & Metric-Based Assessment & 79 & 27.8 & 75.2 & 49 \\
 & Machine Learning and Prediction & 23 & 8.1 & 21.9 & 13 \\
 & Technical Debt Assessment & 3 & 1.1 & 2.9 & 1 \\
\hline
\multirow{3}{=}{\parbox{4.4cm}{Conformance{,} Rule \& Smell Checking \\ (77 - 27.1\%)}} & Rule and Constraint Checking & 49 & 17.3 & 63.6 & 30 \\
 & Architectural Smell Detection & 17 & 6.0 & 22.1 & 8 \\
 & Pattern and Anti-Pattern Analysis & 11 & 3.9 & 14.3 & 11 \\
\hline
\multirow{4}{=}{\parbox{4.4cm}{Human- \& Documentation-Centered Analysis \\ (62 - 21.8\%)}} & Manual Expert Evaluation & 25 & 8.8 & 40.3 & 9 \\
 & Visualization-Based Analysis & 23 & 8.1 & 37.1 & 8 \\
 & Quality Attribute Assessment & 9 & 3.2 & 14.5 & 9 \\
 & Documentation and Traceability Analysis & 5 & 1.8 & 8.1 & 6 \\
\hline
\multirow{2}{=}{\parbox{4.4cm}{Dynamic \& Runtime Analysis \\ (50 - 17.6\%)}} & Runtime Monitoring & 50 & 17.6 & -- & 21 \\
 & & & & & \\
\hline
\multirow{2}{=}{\parbox{4.4cm}{Other / Emerging Approaches \\ (41 - 14.4\%)}} & Other & 41 & 14.4 & -- & 27 \\
 & & & & & \\
\hline
\multirow{2}{=}{\parbox{4.4cm}{Evolutionary \& Historical Analysis \\ (34 - 12.0\%)}} & Historical and Evolutionary Analysis & 34 & 12.0 & -- & 7 \\
 & & & & & \\
\hline
\end{tabular}
\end{table}

\subsection{Software Metrics for Architectural Degradation (RQ$_2$)}

\begin{figure}[tb]
    \centering
    \includegraphics[width=\linewidth]{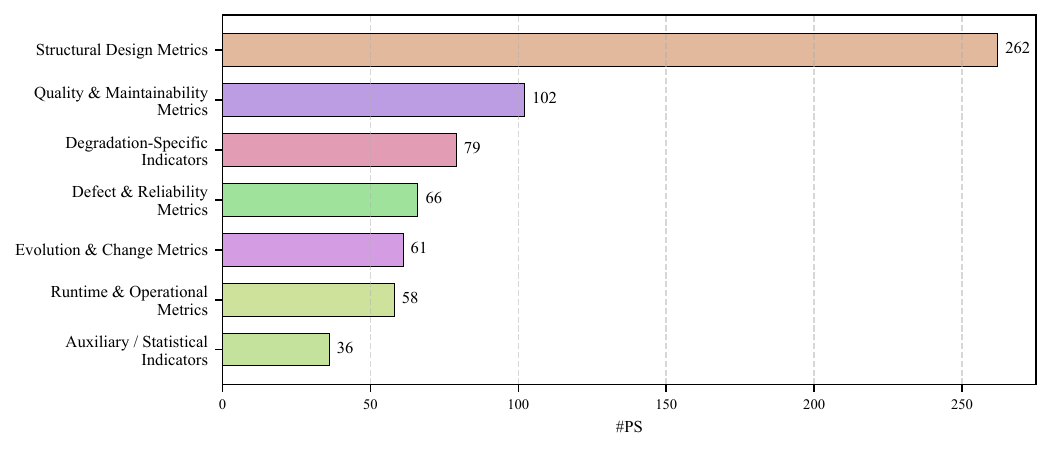}
    \caption{Distribution of metrics across categories (RQ$_2$).}
    \label{fig:RQ2}
\end{figure}

We identified 357 metrics used to evaluate architectural degradation and grouped them into seven categories according to the architectural properties they capture. Figure~\ref{fig:RQ2} and Table~\ref{tab:rq2_combined} show a highly concentrated measurement landscape in which structural metrics dominate both the number of studies and the variety of identified items.

\textbf{Structural Design Metrics} appears in \textbf{92.3\%} of the primary studies. Nearly every study that quantifies architectural degradation uses at least one metric from this category, which makes structural measurement the common language of the field. Dependency and Modularity Metrics lead at 28.9\% \ref{WL1}, \ref{WL3}, \ref{WL4}, \ref{WL5},\ref{GL273}, \ref{GL278} and \ref{GL280}, followed by Coupling Metrics at 19.7\% \ref{WL3}, \ref{WL4}, \ref{WL5}, \ref{WL262}, and \ref{WL266}, Size Metrics at 16.2\% \ref{WL2}, \ref{WL4}, \ref{WL5}, \ref{WL9}, \ref{WL15},\ref{GL278}, and \ref{GL280}, Complexity Metrics at 14.8\% \ref{WL4}, \ref{WL5}, \ref{WL6}, \ref{WL7},\ref{WL265}, \ref{GL271}, and Cohesion Metrics at 12.7\%\ref{WL3}, \ref{WL19}, \ref{WL23}, \ref{WL32}, \ref{WL39}, \ref{WL45}, \ref{WL256}, \ref{WL258}, and \ref{WL262}. Together, these metrics characterize how software components interact, how responsibilities spread across modules, how complex individual elements become, and how closely the implementation respects modular boundaries.

\textbf{Quality \& Maintainability Metrics} forms the second-largest category and appears in \textbf{35.9\%} of the studies. Maintainability and Quality Metrics account for most of this category and appear in 32.4\% of the studies \ref{WL1}, \ref{WL2}, \ref{WL4}, \ref{WL5}, \ref{WL7}, \ref{WL11},, \ref{GL280}, and \ref{GL281}. Documentation and Traceability Metrics reach only 2.5\% \ref{WL12}, \ref{WL26}, \ref{WL80}, \ref{WL81}, \ref{WL98}, \ref{WL103}, \ref{WL118}, while Testability Metrics appear in 1.1\% \ref{WL23}, \ref{WL215}, \ref{WL244}. Researchers therefore assess the quality-related consequences of degradation mainly through general maintainability indicators, while they devote considerably less attention to documentation quality, architectural traceability, and testability.

\textbf{Degradation-Specific Indicators} represent the third-largest category and appear in \textbf{27.8\%} of the studies. Architectural Smell Metrics account for 14.4\% \ref{WL1}, \ref{WL4}, \ref{WL5}, \ref{WL7}, \ref{WL8}, \ref{WL19},\ref{WL233}, \ref{WL235}, \ref{GL268}, \ref{GL280}, while Technical Debt Metrics account for 13.4\%  \ref{WL1}, \ref{WL2}, \ref{WL5}, \ref{WL6}, \ref{WL12}, \ref{WL23}, \ref{WL31},\ref{GL273}, \ref{GL278}, and \ref{GL280}. These metrics directly operationalize concepts associated with architectural degradation, including recurring smells, accumulated debt, structural inconsistencies, and violations. Their presence shows that part of the literature has moved beyond general-purpose structural metrics and developed indicators that target degradation more explicitly.

At the opposite end, \textbf{Auxiliary / Statistical Indicators} represents the smallest category and appears in \textbf{12.7\%} of the studies. Researchers use these indicators to complement architectural measurements through statistical summaries, derived indices, and context-specific variables. Their limited presence further confirms that the literature prioritizes direct structural measurements over supporting statistical evidence and less conventional indicators.

\begin{keyRQAnswer}[Software Metrics]
Structural design metrics dominate the measurement landscape and appear in 92.3\% of the primary studies, followed by quality and maintainability metrics (35.9\%) and degradation-specific indicators (27.8\%). Auxiliary and statistical indicators represent the smallest category (12.7\%). Researchers therefore quantify degradation mainly through structural properties and their maintainability consequences.
\end{keyRQAnswer}

\begin{table}[tb]
\centering
\footnotesize
\caption{Categories and Subcategories with PS citation counts (RQ$_2$).}
\label{tab:rq2_combined}
\begin{tabular}{L{4.6cm}L{4.0cm}rrrr}
\hline
Category (\#, \%) & Subcategory & \#PS & PS (\%) & Category (\%) & \#Metrics \\
\hline
\multirow{5}{=}{\parbox{4.4cm}{Structural Design Metrics \\ (262 - 92.3\%)}} & Dependency and Modularity Metrics & 82 & 28.9 & 31.3 & 48 \\
 & Coupling Metrics & 56 & 19.7 & 21.4 & 34 \\
 & Size Metrics & 46 & 16.2 & 17.6 & 16 \\
 & Complexity Metrics & 42 & 14.8 & 16.0 & 31 \\
 & Cohesion Metrics & 36 & 12.7 & 13.7 & 13 \\
\hline
\multirow{3}{=}{\parbox{4.4cm}{Quality \& Maintainability Metrics \\ (102 - 35.9\%)}} & Maintainability and Quality Metrics & 92 & 32.4 & 90.2 & 43 \\
 & Documentation and Traceability Metrics & 7 & 2.5 & 6.9 & 4 \\
 & Testability Metrics & 3 & 1.1 & 2.9 & 2 \\
\hline
\multirow{2}{=}{\parbox{4.4cm}{Degradation-Specific Indicators \\ (79 - 27.8\%)}} & Architectural Smell Metrics & 41 & 14.4 & 51.9 & 23 \\
 & Technical Debt Metrics & 38 & 13.4 & 48.1 & 5 \\
\hline
\multirow{2}{=}{\parbox{4.4cm}{Defect \& Reliability Metrics \\ (66 - 23.2\%)}} & Defect and Fault Metrics & 66 & 23.2 & -- & 17 \\
 & & & & & \\
\hline
\multirow{2}{=}{\parbox{4.4cm}{Evolution \& Change Metrics \\ (61 - 21.5\%)}} & Change and Evolution Metrics & 61 & 21.5 & -- & 45 \\
 & & & & & \\
\hline
\multirow{2}{=}{\parbox{4.4cm}{Runtime \& Operational Metrics \\ (58 - 20.4\%)}} & Runtime and Performance Metrics & 58 & 20.4 & -- & 51 \\
 & & & & & \\
\hline
\multirow{2}{=}{\parbox{4.4cm}{Auxiliary / Statistical Indicators \\ (36 - 12.7\%)}} & Other & 36 & 12.7 & -- & 25 \\
 & & & & & \\
\hline
\end{tabular}
\end{table}

\subsection{Tools to Measure Architectural Degradation (RQ$_3$)}

\begin{figure}[tb]
    \centering
    \includegraphics[width=\linewidth]{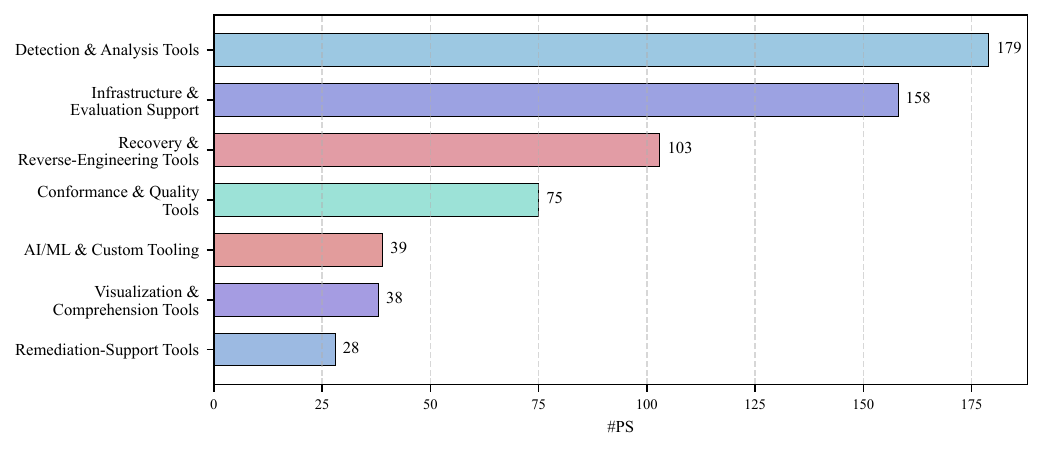}
    \caption{Distribution of tools across categories (RQ$_3$).}
    \label{fig:RQ3}
\end{figure}

We classified the tools reported in the literature into seven categories according to their primary function in the degradation assessment process. Figure~\ref{fig:RQ3} and Table~\ref{tab:rq3_combined} show a mature ecosystem for detecting and analyzing degradation, together with extensive supporting infrastructure, but a considerably less developed ecosystem for assisting remediation.

\textbf{Detection \& Analysis Tools} dominates the literature and appears in \textbf{63.0\%} of the primary studies. Static Analysis Tools account for 20.4\% \ref{WL3}, \ref{WL5}, \ref{WL7}, \ref{WL10}, \ref{GL275}, \ref{GL280}, \ref{GL283}, while Dependency Analysis Tools account for 20.1\% \ref{WL1}, \ref{WL5}, \ref{WL7}, \ref{WL9}, \ref{WL16}, \ref{GL273}, \ref{GL279}, \ref{GL280}. Graph-Based Tools follow at 10.2\% \ref{WL1}, \ref{WL9}, \ref{WL33}, \ref{WL59}, \ref{WL68}, \ref{WL79}, \ref{WL103},\ref{WL256}, and \ref{WL257}, Smell Detection Tools at 7.0\% \ref{WL6}, \ref{WL21}, \ref{WL27}, \ref{WL33}, \ref{WL36}, \ref{WL39}, \ref{WL241}, \ref{GL269}, \ref{GL273}, and Clustering Tools at 5.3\% \ref{WL3}, \ref{WL5}, \ref{WL9}, \ref{WL45}, \ref{WL54},\ref{WL200}, \ref{WL235}, \ref{WL266}. Together, these tools inspect source code, identify structural dependencies, model architectural relationships, detect recurring anomalies, and group software elements according to their structural characteristics. The prevalence of these tools closely reflects the dominance of static and structural measurement approaches reported for RQ$_1$.

\textbf{Infrastructure \& Evaluation Support} forms the second-largest category and appears in \textbf{55.6\%} of the studies. Other Tools account for 32.4\% \ref{WL3}, \ref{WL5}, \ref{WL10}, \ref{WL13}, \ref{WL15}, \ref{WL18}, \ref{GL280}, \ref{GL281}, \ref{GL282}, while Benchmark and Evaluation Infrastructure accounts for 23.2\% \ref{WL14}, \ref{WL15}, \ref{WL19}, \ref{WL22},\ref{GL269}, \ref{GL274}, \ref{GL280}, and \ref{GL281}. These tools support experimentation, data preparation, repository processing, benchmarking, and comparative evaluation rather than degradation assessment alone. Their prominence indicates that researchers frequently combine specialized architectural tools with general-purpose infrastructures, custom scripts, and experimental environments.

\textbf{Recovery \& Reverse-Engineering Tools} represents the third-largest category and appears in \textbf{36.3\%} of the studies. Architecture Recovery Tools dominate this category and appear in 28.9\% \ref{WL1}, \ref{WL3}, \ref{WL4}, \ref{WL7}, \ref{WL9}, \ref{WL13}, \ref{GL268}, \ref{GL276}, \ref{GL280}, followed by Reverse Engineering Tools at 6.0\% \ref{WL5}, \ref{WL69}, \ref{WL97}, \ref{WL113}, \ref{WL125}, \ref{WL126}, \ref{WL247}, \ref{GL268}, \ref{GL269} and Mining and Repository Analysis Tools at 1.4\% \ref{WL10}, \ref{WL241}, \ref{WL245}, \ref{WL266}. These tools reconstruct architectural views, recover component relationships, and extract architectural knowledge from source code or repositories when explicit documentation no longer provides an accurate representation of the system. Their strong presence indicates that researchers often need to recover the architecture before they can identify its degradation.

At the opposite end, \textbf{Remediation-Support Tools} represents the smallest category and appears in only \textbf{9.9\%} of the studies. Refactoring Tools constitute the entire category \ref{WL6}, \ref{WL20}, \ref{WL27}, \ref{WL35}, \ref{WL45}, \ref{WL57}, \ref{WL87}, \ref{WL88}, \ref{WL90}, \ref{WL95}, \ref{WL248}, \ref{GL273}, \ref{GL276}. The difference between Detection \& Analysis Tools and Remediation-Support Tools exceeds a factor of six. The tool ecosystem therefore provides extensive support for identifying degradation but substantially less support for selecting, validating, or executing corrective actions.

\begin{keyRQAnswer}[Tools]
Detection and analysis tools dominate the ecosystem (63.0\%), followed by infrastructure and evaluation support (55.6\%) and recovery and reverse-engineering tools (36.3\%). Remediation-support tools represent the smallest category (9.9\%). The available tooling therefore supports diagnosis and architectural reconstruction much more extensively than corrective action.
\end{keyRQAnswer}

\begin{table}[tb]
\centering
\footnotesize
\caption{Categories and Subcategories with PS citation counts (RQ$_3$).}
\label{tab:rq3_combined}
\begin{tabular}{L{4.6cm}L{4.0cm}rrrr}
\hline
Category (\#, \%) & Subcategory & \#PS & PS (\%) & Category (\%) & \#Tools \\
\hline
\multirow{5}{=}{\parbox{4.4cm}{Detection \& Analysis Tools \\ (179 - 63.0\%)}} & Static Analysis Tools & 58 & 20.4 & 32.4 & 31 \\
 & Dependency Analysis Tools & 57 & 20.1 & 31.8 & 29 \\
 & Graph-Based Tools & 29 & 10.2 & 16.2 & 17 \\
 & Smell Detection Tools & 20 & 7.0 & 11.2 & 14 \\
 & Clustering Tools & 15 & 5.3 & 8.4 & 9 \\
\hline
\multirow{2}{=}{\parbox{4.4cm}{Infrastructure \& Evaluation Support \\ (158 - 55.6\%)}} & Other Tools & 92 & 32.4 & 58.2 & 98 \\
 & Benchmark and Evaluation Infrastructure & 66 & 23.2 & 41.8 & 45 \\
\hline
\multirow{3}{=}{\parbox{4.4cm}{Recovery \& Reverse-Engineering Tools \\ (103 - 36.3\%)}} & Architecture Recovery Tools & 82 & 28.9 & 79.6 & 45 \\
 & Reverse Engineering Tools & 17 & 6.0 & 16.5 & 7 \\
 & Mining and Repository Analysis Tools & 4 & 1.4 & 3.9 & 4 \\
\hline
\multirow{2}{=}{\parbox{4.4cm}{Conformance \& Quality Tools \\ (75 - 26.4\%)}} & Architecture Conformance Checking Tools & 63 & 22.2 & 84.0 & 27 \\
 & Code Quality and Maintainability Tools & 12 & 4.2 & 16.0 & 8 \\
\hline
\multirow{2}{=}{\parbox{4.4cm}{AI/ML \& Custom Tooling \\ (39 - 13.7\%)}} & Custom Scripts and Prototypes & 21 & 7.4 & 53.8 & 14 \\
 & Machine Learning and AI-Based Tools & 18 & 6.3 & 46.2 & 26 \\
\hline
\multirow{2}{=}{\parbox{4.4cm}{Visualization \& Comprehension Tools \\ (38 - 13.4\%)}} & Visualization Tools & 38 & 13.4 & -- & 37 \\
 & & & & & \\
\hline
\multirow{2}{=}{\parbox{4.4cm}{Remediation-Support Tools \\ (28 - 9.9\%)}} & Refactoring Tools & 28 & 9.9 & -- & 17 \\
 & & & & & \\
\hline
\end{tabular}
\end{table}

%\todo[inline]{Vale @Matteo: verificare 0 nella Tab~\ref{tab:rq3_combined}}

\subsection{Remediation Approaches (RQ$_4$)}

\begin{figure}[tb]
    \centering
    \includegraphics[width=\linewidth]{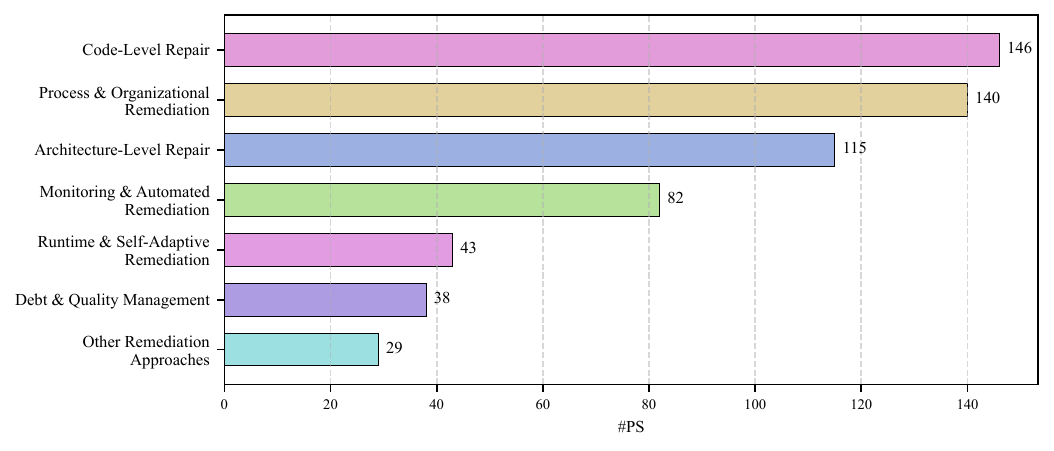}
    \caption{Distribution of remediation approaches across categories (RQ$_4$).}
    \label{fig:RQ4}
\end{figure}

We classified the remediation approaches into seven categories according to the level and timing of their intervention. Figure~\ref{fig:RQ4} and Table~\ref{tab:rq4_combined} show that the literature emphasizes corrective actions at the code, organizational, and architectural levels, while it devotes less attention to less conventional forms of remediation.

\textbf{Code-Level Repair} represents the most common remediation category and appears in \textbf{51.4\%} of the primary studies. Refactoring and Restructuring drives this category and appears in 39.4\% of the studies \ref{WL7}, \ref{WL11}, \ref{WL17}, \ref{WL20}, \ref{WL24}, \ref{WL25},\ref{GL275}, \ref{GL279}, \ref{GL280}, \ref{GL282}. Smell and Violation Resolution accounts for 6.7\% \ref{WL6}, \ref{WL27}, \ref{WL33}, \ref{WL35}, \ref{WL36}, \ref{WL38}, \ref{WL39}, \ref{WL44},\ref{GL269}, \ref{GL273}, while Dependency and Modularity Improvement accounts for 5.3\% \ref{WL9}, \ref{WL15}, \ref{WL78}, \ref{WL86}, \ref{WL249}, \ref{WL251}. These approaches modify the implementation to improve its internal structure while preserving its expected external behavior. They include operations such as extracting methods or classes, moving elements across modules, reducing dependencies, and reorganizing package or module boundaries.

\textbf{Process \& Organizational Remediation} follows closely and appears in \textbf{49.3\%} of the studies. Process, Governance, and Decision Support accounts for 33.1\% \ref{WL2}, \ref{WL3}, \ref{WL6}, \ref{WL11}, \ref{WL12}, \ref{WL25}, \ref{GL273}, \ref{GL274}, \ref{GL275}, \ref{GL276}, \ref{GL280}, while Documentation and Knowledge Management accounts for 15.5\% \ref{WL3}, \ref{WL6}, \ref{WL12}, \ref{WL27}, \ref{WL36}, \ref{WL46}, \ref{WL48}, \ref{GL269}, \ref{GL271}, \ref{GL280}. Human-Centered Collaboration and Expert Involvement appears in only 0.7\% \ref{WL66}, \ref{WL120}. These approaches address degradation by improving architectural decision-making practices, governance mechanisms, review activities, documentation, knowledge sharing, and development processes rather than by changing source code directly.

\textbf{Architecture-Level Repair} represents the third-largest category and appears in \textbf{40.5\% }of the studies. Architecture Recovery and Reconstruction accounts for 20.8\% \ref{WL1}, \ref{WL7}, \ref{WL11}, \ref{WL17}, \ref{WL19}, \ref{GL276}, \ref{GL280}, \ref{GL281}, followed by Architecture Conformance and Compliance Improvement at 14.1\% \ref{WL13}, \ref{WL31}, \ref{WL32}, \ref{WL42},, \ref{WL252}, \ref{WL256}, \ref{GL274} and Migration and Modernization at 5.6\% \ref{WL16}, \ref{WL34}, \ref{WL94}, \ref{WL95}, \ref{WL105}, \ref{WL140}, \ref{WL222}, \ref{WL266}, \ref{GL273}. These approaches intervene above the level of individual code elements and aim to restore architectural structures, realign the implementation with architectural decisions, or replace outdated architectural solutions.

At the opposite end, \textbf{Other Remediation Approaches} represents the smallest category and appears in \textbf{10.2\%} of the studies \ref{WL20}, \ref{WL26}, \ref{WL54}, \ref{WL81}, \ref{WL104}, \ref{GL272}, \ref{GL280}, \ref{GL283}. This category collects remediation techniques that do not fit the established code-level, architecture-level, organizational, monitoring, runtime, or quality-management categories. Its limited presence indicates that the literature largely converges on a recognizable set of remediation levels, even though individual studies implement them through different techniques.

\begin{keyRQAnswer}[Remediation Approaches]
Code-level repair represents the most frequently reported remediation category (51.4\%), followed closely by process and organizational remediation (49.3\%) and architecture-level repair (40.5\%). Other remediation approaches form the smallest category (10.2\%). The literature therefore concentrates remediation efforts on technical repair, organizational change, and architectural restructuring.
\end{keyRQAnswer}

\begin{table}[tb]
\centering
\footnotesize
\caption{Categories and Subcategories with PS citation counts (RQ$_4$)}
\label{tab:rq4_combined}
\begin{tabular}{L{4.6cm}L{4.0cm}rrrr}
\hline
Category (\#, \%) & Subcategory & \#PS & PS (\%) & Category (\%) & \#Approaches \\
\hline
\multirow{3}{=}{\parbox{4.4cm}{Code-Level Repair \\ (146 - 51.4\%)}} & Refactoring and Restructuring & 112 & 39.4 & 76.7 & 42 \\
 & Smell and Violation Resolution & 19 & 6.7 & 13.0 & 13 \\
 & Dependency and Modularity Improvement & 15 & 5.3 & 10.3 & 11 \\
\hline
\multirow{3}{=}{\parbox{4.4cm}{Process \& Organizational Remediation \\ (140 - 49.3\%)}} & Process{,} Governance{,} and Decision Support & 94 & 33.1 & 67.1 & 55 \\
 & Documentation and Knowledge Management & 44 & 15.5 & 31.4 & 7 \\
 & Human-Centered Collaboration and Expert Involvement & 2 & 0.7 & 1.4 & 3 \\
\hline
\multirow{3}{=}{\parbox{4.4cm}{Architecture-Level Repair \\ (115 - 40.5\%)}} & Architecture Recovery and Reconstruction & 59 & 20.8 & 51.3 & 29 \\
 & Architecture Conformance and Compliance Improvement & 40 & 14.1 & 34.8 & 23 \\
 & Migration and Modernization & 16 & 5.6 & 13.9 & 12 \\
\hline
\multirow{2}{=}{\parbox{4.4cm}{Monitoring \& Automated Remediation \\ (82 - 28.9\%)}} & Monitoring{,} Feedback{,} and Continuous Improvement & 67 & 23.6 & 81.7 & 48 \\
 & Tool-Supported Remediation and Automation & 15 & 5.3 & 18.3 & 9 \\
\hline
\multirow{2}{=}{\parbox{4.4cm}{Runtime \& Self-Adaptive Remediation \\ (43 - 15.1\%)}} & Runtime Adaptation and Self-Adaptive Remediation & 43 & 15.1 & -- & 30 \\
 & & & & & \\
\hline
\multirow{2}{=}{\parbox{4.4cm}{Debt \& Quality Management \\ (38 - 13.4\%)}} & Quality Attribute Improvement & 21 & 7.4 & 55.3 & 12 \\
 & Technical Debt Management and Repayment & 17 & 6.0 & 44.7 & 5 \\
\hline
\multirow{2}{=}{\parbox{4.4cm}{Other Remediation Approaches \\ (29 - 10.2\%)}} & Other Remediation Approaches & 29 & 10.2 & -- & 19 \\
 & & & & & \\
\hline
\end{tabular}
\end{table}

\begin{table}[t]
\centering
\footnotesize
\caption{Most frequent end-to-end architectural degradation workflows identified across the primary studies.}
\label{tab:top10_paths}
\begin{tabularx}{\textwidth}{r|X|c|c}
\hline
\textbf{Rank} &
\textbf{Workflow (RQ$_1$ $\rightarrow$ RQ$_2$ $\rightarrow$ RQ$_3$ $\rightarrow$ RQ$_4$)} &
\textbf{\#PS} &
\textbf{\%PS} \\
\hline 

1 &
Static \& Structural Analysis $\rightarrow$
Structural Design Metrics $\rightarrow$
Detection \& Analysis Tools $\rightarrow$
Code-Level Repair &
18 & 10.2 \\

2 &
Static \& Structural Analysis $\rightarrow$
Structural Design Metrics $\rightarrow$
Detection \& Analysis Tools $\rightarrow$
Process \& Organizational Remediation &
12 & 6.8 \\

3 &
Static \& Structural Analysis $\rightarrow$
Structural Design Metrics $\rightarrow$
Infrastructure \& Evaluation Support $\rightarrow$
Code-Level Repair &
7 & 4.0 \\

4 &
Quantitative \& Predictive Analysis $\rightarrow$
Structural Design Metrics $\rightarrow$
Detection \& Analysis Tools $\rightarrow$
Code-Level Repair &
6 & 3.4 \\

5 &
Static \& Structural Analysis $\rightarrow$
Structural Design Metrics $\rightarrow$
Detection \& Analysis Tools $\rightarrow$
Architecture-Level Repair &
5 & 2.8 \\

6 &
Static \& Structural Analysis $\rightarrow$
Structural Design Metrics $\rightarrow$
Detection \& Analysis Tools $\rightarrow$
Monitoring \& Automated Remediation &
4 & 2.3 \\

7 &
Static \& Structural Analysis $\rightarrow$
Quality \& Maintainability Metrics $\rightarrow$
Detection \& Analysis Tools $\rightarrow$
Code-Level Repair &
4 & 2.3 \\

8 &
Static \& Structural Analysis $\rightarrow$
Structural Design Metrics $\rightarrow$
Recovery \& Reverse-Engineering Tools $\rightarrow$
Code-Level Repair &
4 & 2.3 \\

9 &
Static \& Structural Analysis $\rightarrow$
Structural Design Metrics $\rightarrow$
Conformance \& Quality Tools $\rightarrow$
Code-Level Repair &
3 & 1.7 \\

10 &
Static \& Structural Analysis $\rightarrow$
Quality \& Maintainability Metrics $\rightarrow$
Detection \& Analysis Tools $\rightarrow$
Architecture-Level Repair &
3 & 1.7 \\

\hline 
\end{tabularx}
\end{table}

\section{Discussion}\label{sec:discussion}

Architectural degradation has attracted sustained research attention for more than two decades, resulting in a rich body of techniques for measuring, detecting, and mitigating architectural erosion. Previous secondary studies have shown that degradation emerges from a combination of technical and organizational factors and have documented a broad spectrum of approaches for detecting and addressing it \cite{DESILVA2012132,Li2022ArchitectureErosionSMS}. The evidence synthesized in this review suggests that progress has not been uniform across the architectural degradation lifecycle. The literature has advanced considerably in understanding how degradation can be identified, whereas substantially less agreement exists on how it should be prevented, remediated, and continuously managed.

This imbalance appears consistently across all four research questions. Measurement approaches, metric families, and tooling exhibit a high degree of convergence, indicating that the community largely shares a common understanding of how architectural degradation should be assessed. In contrast, remediation strategies remain fragmented across technical, architectural, and organizational interventions, with no comparable consensus emerging. More than a decade ago, De Silva and Balasubramaniam argued that effectively controlling architectural erosion requires combining prevention, detection, and restoration strategies rather than relying on isolated technical solutions \cite{DESILVA2012132}. The evidence collected in this review suggests that this objective has not yet been fully achieved. While the literature provides a mature ecosystem of techniques for recognizing degradation, considerably less attention has been devoted to understanding how different remediation strategies should be selected, coordinated, and sustained over time.

Perhaps the most consistent finding is the strong \textbf{structural bias} that characterizes the literature. Static \& Structural Analysis dominates the measurement approaches, while Structural Design Metrics appear in more than 90\% of the primary studies. This observation agrees with previous work showing that architectural degradation is primarily identified through structural symptoms, architectural anomalies, and conformance violations rather than runtime evidence \cite{Macia2012Symptoms,Baabad202222915}. Our results indicate that this predominance extends beyond the choice of metrics. Structural reasoning shapes the entire assessment process, influencing how degradation is measured, how evidence is collected, and how supporting tools are designed.

By comparison, relatively few studies investigate degradation through system execution. Dynamic analysis, runtime metrics, and runtime-aware remediation remain markedly underrepresented. This imbalance is understandable because structural artifacts are easier to collect, deterministic to analyze, and supported by mature tooling. However, architectural degradation is inherently an evolutionary phenomenon. It emerges through continuous software evolution, changing requirements, operational pressures, and accumulated design decisions rather than from static structure alone. Consequently, assessment approaches centred primarily on structural evidence are likely to identify degradation only after its effects have already materialized within the implemented architecture. This predominance of structural assessment contrasts with the broader view that architectural quality should be continuously preserved throughout software evolution rather than periodically evaluated \cite{Koziolek2011Sustainability}.

This interpretation also complements the observations of Li et al. \cite{Li2022ArchitectureErosionSMS}, who argue that architectural erosion should not be viewed solely as a structural phenomenon but also as the result of organizational and evolutionary processes. While the literature increasingly acknowledges these broader causes, the methods used to assess degradation remain overwhelmingly centred on structural properties. The evidence therefore reveals a disconnect between how architectural degradation is conceptualized and how it is empirically investigated.

The end-to-end paths reported in Figure~\ref{fig:placeholder} reinforce this interpretation. The most informative observation is not that the dominant path represents only 18 studies, but that nearly all dominant paths follow the same sequence of static analysis, structural metrics, and detection tools before diverging during remediation. This remarkable stability indicates that methodological diversity is concentrated almost entirely after degradation has been identified. In other words, the community has largely converged on how degradation should be observed, while substantially different views remain on how corrective actions should be designed, prioritized, and combined.

This pattern also clarifies the conclusions of previous secondary studies. De Silva and Balasubramaniam organized the literature into complementary prevention, detection, and restoration strategies \cite{DESILVA2012132}, whereas Li et al. highlighted the broad spectrum of techniques proposed to detect and address architectural erosion \cite{Li2022CodeReview}. The evidence synthesized here suggests that this diversity is not uniformly distributed across the degradation lifecycle. Instead, it emerges primarily after degradation has been diagnosed. Before that point, the literature exhibits a remarkably stable methodological pipeline centred on structural assessment. The principal challenge therefore appears to have shifted from improving degradation detection toward understanding how different remediation strategies should interact to preserve architectural integrity throughout software evolution.

\begin{keyTakeAways}
\textbf{Architectural degradation research has reached methodological maturity in diagnosis but not in remediation.} The literature consistently converges on static analysis, structural metrics, and detection tools to identify degradation. In contrast, remediation remains fragmented across code-level, architectural, and organizational interventions, suggesting that future advances are more likely to arise from integrating corrective strategies than from proposing additional detection techniques.
\end{keyTakeAways}
\subsection{Detection Without Action}
A second concern emerging from the findings is the clear \textbf{imbalance between diagnosis and intervention}. The literature provides a rich ecosystem of approaches for identifying architectural degradation, including structural metrics, rule checking, smell detection, and dedicated analysis tools. By contrast, considerably fewer studies investigate how degradation should be addressed once it has been detected. This imbalance is particularly evident in the tooling landscape, where Detection \& Analysis Tools largely outnumber Remediation-Support Tools. The evidence therefore suggests that the community has invested substantially more effort in answering the question \emph{``Is the architecture degrading?''} than \emph{``What should be done next?''}

The remediation landscape reveals an additional level of fragmentation. Code-Level Repair, Architecture-Level Repair, and Process \& Organizational Remediation all receive substantial attention, yet they are rarely investigated as complementary components of the same intervention strategy. Most studies focus on one level of action while treating the others as external concerns. However, architectural degradation rarely originates from a single cause. It accumulates through design decisions, implementation shortcuts, evolving requirements, and organizational constraints acting simultaneously. This observation closely resembles the challenges discussed in the Architectural Technical Debt literature, where long-term architectural quality depends on continuously managing the consequences of local design decisions rather than periodically correcting accumulated problems \cite{Kruchten2012ATD,Nord2012ATDMetric}.

The unexpectedly high presence of Recovery \& Reverse-Engineering tools reinforces this interpretation. Rather than preventing degradation during software evolution, a considerable portion of the literature concentrates on reconstructing architectural knowledge after it has already been lost. Similar observations motivated previous work on architecture-preserving refactorings, where architectural recovery is viewed as a prerequisite for restoring architectural integrity rather than preventing degradation in the first place \cite{Terra2012Refactoring}. Together, these findings suggest that a significant part of the literature remains inherently reactive, addressing degradation after architectural knowledge has already deteriorated.

Another notable observation concerns predictive approaches. Despite the increasing availability of software repositories, architectural histories, and development analytics, Machine Learning and Prediction remain among the least represented measurement approaches, while AI/ML-based tooling also occupies only a limited portion of the landscape. Rather than indicating a lack of interest in artificial intelligence, these results suggest that predictive architectural degradation remains an emerging research direction. Existing studies continue to rely predominantly on structural indicators and threshold-based reasoning, whereas comparatively little attention has been devoted to forecasting degradation trajectories before architectural symptoms become observable. This gap appears particularly significant considering that architectural degradation is inherently progressive and cumulative, making early intervention substantially more valuable than late diagnosis.

\begin{keyTakeAways}
\textbf{The literature excels at diagnosing architectural degradation but provides comparatively limited support for deciding how to respond.} Detection techniques and structural assessment have reached a high degree of maturity, whereas remediation remains fragmented across technical, architectural, and organizational perspectives. At the same time, predictive approaches remain relatively unexplored, despite offering the opportunity to shift architectural degradation management from reactive recovery toward proactive prevention.
\end{keyTakeAways}

\begin{figure}[tb]
    \centering
    \includegraphics[width=\linewidth]{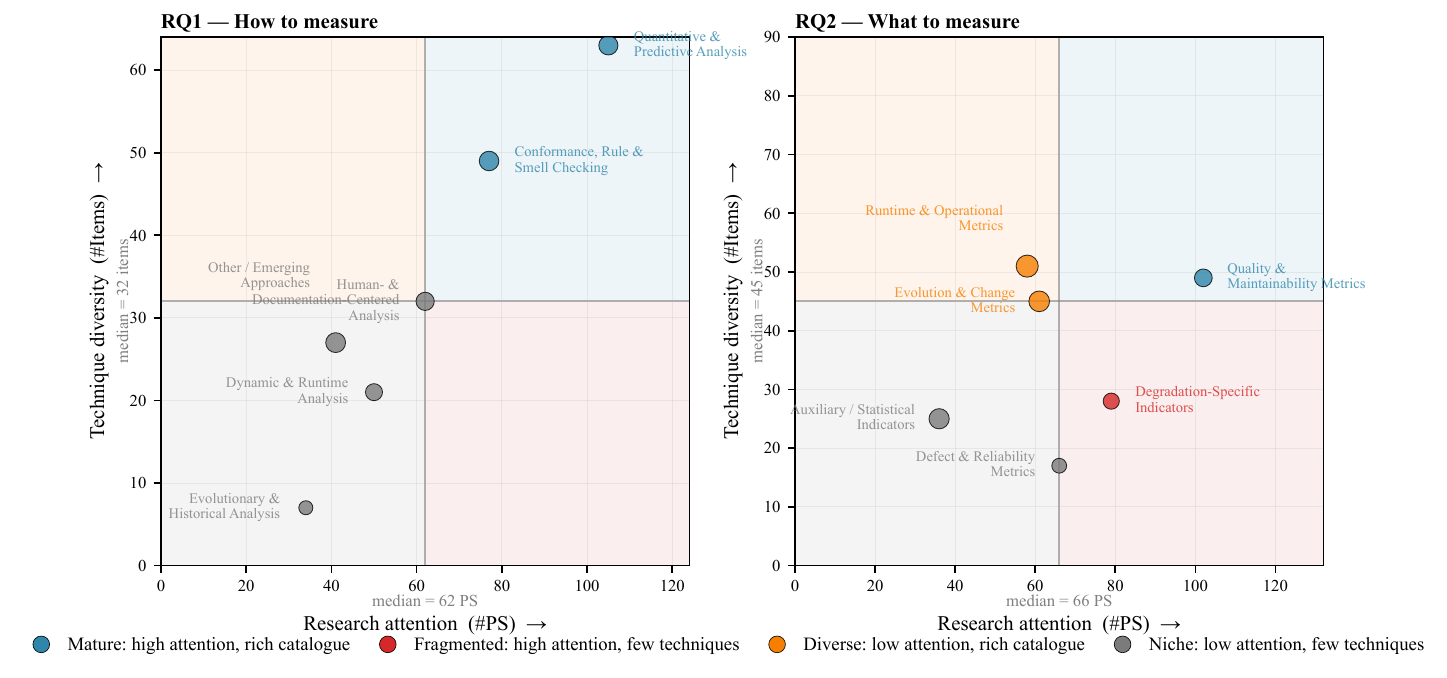}
    \caption{Maturity analysis of RQ$_1$ and RQ$_2$ categories based on research attention (\#PS) and technique diversity (\#Items).}
    \label{fig:maturity_rq1_rq2}
\end{figure}

\begin{figure}[tb]
    \centering
    \includegraphics[width=\linewidth]{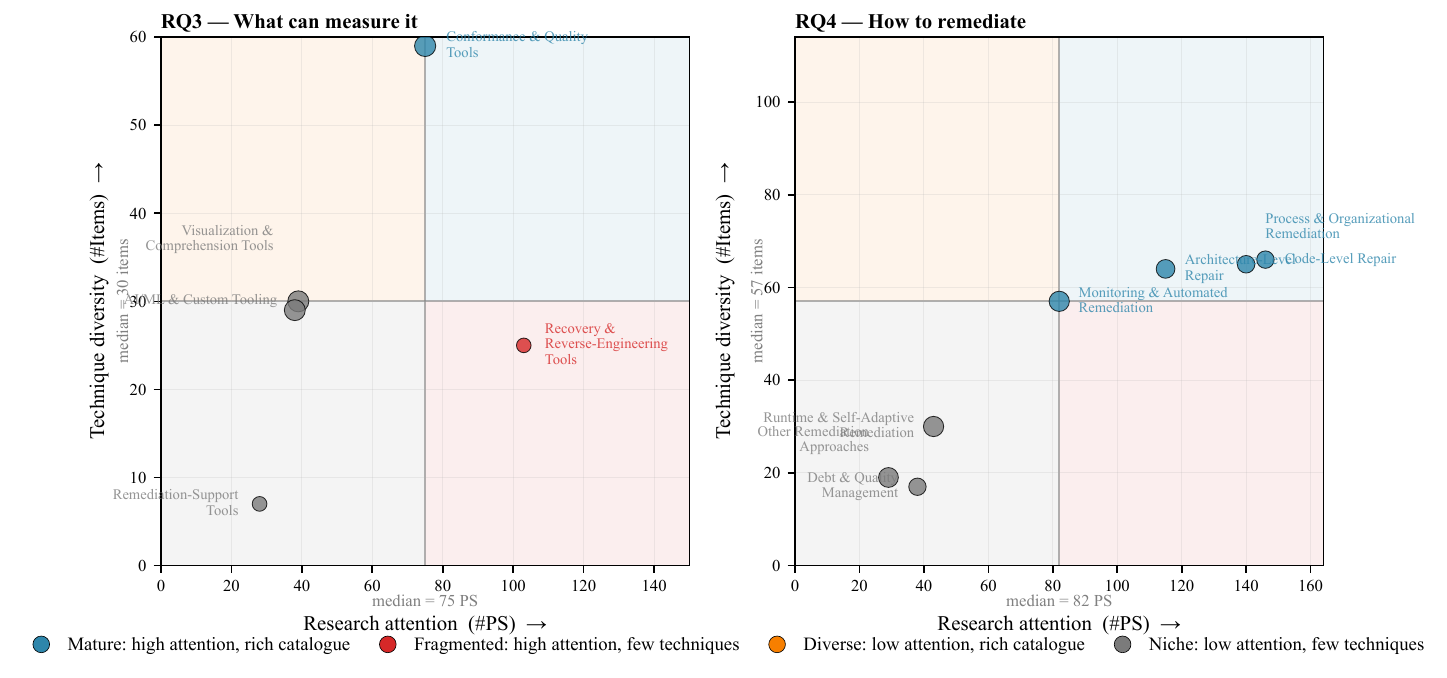}
    \caption{Maturity analysis of RQ$_3$ and RQ$_4$ categories based on research attention (\#PS) and technique diversity (\#Items).}
    \label{fig:maturity_rq3_rq4}
\end{figure}

\begin{table}[tb]
\footnotesize
\centering
\caption{Maturity classification of categories by research question. Thresholds: median \#PS = 70, median \#Items = 38. Mature = above both medians; Fragmented = above \#PS median, below \#Items median; Diverse = below \#PS median, above \#Items median; Niche = below both.}
\label{tab:maturity_rq1_rq2}
\begin{tabular}{L{4.8cm}rrrc}
\hline
Category & \#PS & \#Items & Items/PS & Maturity \\
\hline
\multicolumn{5}{l}{\textbf{RQ1 — How to measure}}\\
\cmidrule(lr){1-5}
Static \& Structural Analysis & 161 & 78 & 0.48 & Mature \\
Quantitative \& Predictive Analysis & 105 & 63 & 0.60 & Mature \\
Conformance, Rule \& Smell Checking & 77 & 49 & 0.64 & Mature \\
Human- \& Documentation-Centered Analysis & 62 & 32 & 0.52 & Niche \\
Dynamic \& Runtime Analysis & 50 & 21 & 0.42 & Niche \\
Other / Emerging Approaches & 41 & 27 & 0.66 & Niche \\
Evolutionary \& Historical Analysis & 34 & 7 & 0.21 & Niche \\
\multicolumn{5}{l}{\textbf{RQ2 — What to measure}}\\
\cmidrule(lr){1-5}
Structural Design Metrics & 262 & 142 & 0.54 & Mature \\
Quality \& Maintainability Metrics & 102 & 49 & 0.48 & Mature \\
Degradation-Specific Indicators & 79 & 28 & 0.35 & Fragmented \\
Defect \& Reliability Metrics & 66 & 17 & 0.26 & Niche \\
Evolution \& Change Metrics & 61 & 45 & 0.74 & Diverse \\
Runtime \& Operational Metrics & 58 & 51 & 0.88 & Diverse \\
Auxiliary / Statistical Indicators & 36 & 25 & 0.69 & Niche \\
\hline
\end{tabular}
\end{table}

\begin{table}[tb]
\centering
\footnotesize
\caption{Maturity classification of categories by research question. Thresholds: median \#PS = 70, median \#Items = 38. Mature = above both medians; Fragmented = above \#PS median, below \#Items median; Diverse = below \#PS median, above \#Items median; Niche = below both.}
\label{tab:maturity_rq3_rq4}
\begin{tabular}{L{4.8cm}rrrc}
\hline
Category & \#PS & \#Items & Items/PS & Maturity \\
\hline
\multicolumn{5}{l}{\textbf{RQ3 — What can measure it}}\\
\cmidrule(lr){1-5}
Detection \& Analysis Tools & 179 & 121 & 0.68 & Mature \\
Infrastructure \& Evaluation Support & 158 & 59 & 0.37 & Mature \\
Recovery \& Reverse-Engineering Tools & 103 & 25 & 0.24 & Fragmented \\
Conformance \& Quality Tools & 75 & 59 & 0.79 & Mature \\
AI/ML \& Custom Tooling & 39 & 30 & 0.77 & Niche \\
Visualization \& Comprehension Tools & 38 & 29 & 0.76 & Niche \\
Remediation-Support Tools & 28 & 7 & 0.25 & Niche \\
\multicolumn{5}{l}{\textbf{RQ4 — How to remediate}}\\
\cmidrule(lr){1-5}
Code-Level Repair & 146 & 66 & 0.45 & Mature \\
Process \& Organizational Remediation & 140 & 65 & 0.46 & Mature \\
Architecture-Level Repair & 115 & 64 & 0.56 & Mature \\
Monitoring \& Automated Remediation & 82 & 57 & 0.69 & Mature \\
Runtime \& Self-Adaptive Remediation & 43 & 30 & 0.70 & Niche \\
Debt \& Quality Management & 38 & 17 & 0.45 & Niche \\
Other Remediation Approaches & 29 & 19 & 0.66 & Niche \\
\hline
\end{tabular}
\end{table}
\subsection{Research Attention and Technique Diversity}
We examined how the architectural degradation literature combines \textbf{research attention}, measured by the number of unique primary studies associated with each category, with \textbf{technique diversity}, measured by the number of distinct normalized approaches, metrics, tools, or remediation strategies. Tables~\ref{tab:maturity_rq1_rq2} and~\ref{tab:maturity_rq3_rq4} report the classification, while Figures~\ref{fig:maturity_rq1_rq2} and~\ref{fig:maturity_rq3_rq4} illustrate the distributions. Together, these measures describe the coverage and breadth of the literature, without directly establishing effectiveness, adoption, or methodological maturity.

The concentration of approaches, metrics, and tools around \textbf{structural assessment} gives researchers several ways to examine dependencies, modular boundaries, and architectural violations. Far fewer approaches explicitly investigate how these properties change: the catalogue contains 78 static and structural approaches, compared with seven evolutionary and historical approaches. This difference highlights an opportunity to connect structural observations across successive versions and examine when architectural changes indicate deterioration. Static techniques can contribute to this investigation, provided that their application and interpretation account for the system's evolution.

Choosing among the available structural techniques requires understanding \textbf{what additional evidence each contributes}. Techniques may examine similar properties while relying on different assumptions, inputs, or thresholds. Combining them may broaden the assessment, but it may also reproduce the same information through several indicators. Comparative evaluations would help distinguish complementary measurements from redundant ones and establish whether their conclusions remain consistent across systems and stages of evolution.

The breadth of remediation extends across implementation, architecture, and development practices, with 66 code-level, 64 architectural, and 65 organizational strategies. This range challenges a general characterization of remediation as immature and makes \textbf{intervention selection} a more precise concern. An architectural violation, for example, may require both a technical correction and changes to the practices that allowed it to recur. The catalogue identifies possible responses, but does not establish which combinations are appropriate under particular conditions. Studies connecting the diagnosed problem to the intervention rationale, implementation costs, and outcomes over subsequent releases would provide stronger guidance for these decisions.

Repeated use of a smaller set of recovery tools or degradation-specific indicators could indicate \textbf{consolidation}, particularly when independent studies evaluate the same techniques under different conditions. The same concentration could also reflect limited alternatives or broad grouping during classification. Aggregate counts cannot distinguish these explanations. Establishing whether knowledge is accumulating requires examining how studies reuse earlier techniques, test their assumptions, and compare their findings. A small catalogue supported by repeated validation may offer more dependable guidance than a larger catalogue whose techniques remain difficult to compare.

The 51 runtime and operational metrics and 45 evolutionary and change metrics provide varied ways to examine dimensions that receive less attention than structural assessment. Their contribution to degradation management depends on whether they inform \textbf{actionable architectural decisions}. Evaluations could trace whether changes in these indicators reveal a problem, influence the choice of an intervention, and capture its subsequent effects. This requires distinguishing execution behaviour from architectural change across versions: each provides different evidence, and neither alone establishes how degradation develops.

Comparative studies would make extensive catalogues easier to use by clarifying technique selection and applicability. For smaller catalogues, evaluations should establish whether existing techniques adequately cover the relevant degradation phenomena and produce reliable findings. Following \textbf{the connection from measurement to intervention and sustained outcomes} would address both needs, showing how the available contributions support degradation management in practice. This would provide empirical support for the combined use of prevention, detection, and repair advocated by previous reviews \cite{DESILVA2012132,Li2022CodeReview}.

\begin{keyTakeAways}
\textbf{The breadth of available techniques makes their selection and integration a central research challenge.} Structural assessment and remediation offer extensive catalogues, but their size alone does not establish maturity or effectiveness. Comparative evaluations are needed to clarify which techniques provide complementary evidence, when particular interventions are appropriate, and whether their benefits persist over time. Shared benchmarks and longitudinal datasets could support these evaluations, strengthening \textbf{the connection between diagnosis, corrective action, and sustained architectural improvement}.
\end{keyTakeAways}

\subsection{Implications for Practice and Research}

Our review brings together techniques for observing architectural degradation and strategies for addressing it, while showing a stronger emphasis on structural assessment within measurement approaches and tooling. We therefore identify a practical need to distinguish \textbf{architectural anomalies from evidence of architectural degradation}. Although coupling, complexity, or a detected smell can identify a concern, establishing degradation requires examining its relationship with architectural intent and its consequences for system evolution. Conformance checks and recovered architectural views can support this interpretation; however, recovered views describe the implemented structure and may reproduce the dependencies under investigation. We consequently recommend distinguishing recovered structure from intended constraints before using it to guide corrective action.

Through RQ$_4$, we identified strategies addressing both software structure and development practices, which motivates connecting \textbf{architectural repair with measures that prevent recurrence}. While refactoring may remove an unwanted dependency, architectural rules, decision documentation, and review responsibilities may help prevent its reintroduction. We therefore recommend comparing technical repair alone with repair accompanied by changes to architectural guidance or development practices. Such comparisons would establish whether organizational measures reduce recurring violations sufficiently to justify their cost, thereby giving empirical substance to combined erosion-control strategies \cite{DESILVA2012132}.

Evaluating those interventions also requires distinguishing \textbf{measurements of architectural deterioration from descriptions of structural change}. Although RQ$_2$ identifies a broad catalogue of structural metrics, changes in coupling or complexity do not necessarily imply worsening maintainability or increasing change effort. We thus propose examining when metric changes coincide with architectural violations, changes spreading across component boundaries, or more costly modifications, while comparing cases in which similar structural changes support legitimate evolution. The resulting evidence would help determine both when a measurement warrants intervention and whether an observed improvement indicates successful remediation.

Across RQ$_3$ and RQ$_4$, we observed an emphasis on detection-oriented tooling alongside a broad range of remediation strategies; we consequently identify \textbf{explicit links between degradation symptoms and corrective actions} as a research priority. While diagnostic evidence identifies the concern, selecting a repair also requires considering architectural constraints and expected evolution costs. We therefore propose investigating how those factors justify particular interventions, so that supporting tools can explain the rationale and trade-offs behind their recommendations. Evaluations could then establish whether the recommended actions address the diagnosed degradation without transferring dependencies or maintenance burdens elsewhere \cite{Kruchten2012ATD}.

\begin{keyTakeAways}[Architectural degradation should increasingly be managed as a continuous architectural governance activity rather than a periodic assessment task]  Future research should prioritize runtime-aware assessment, decision-support for remediation, and predictive architectural governance to complement the mature ecosystem of structural analysis and detection techniques.
\end{keyTakeAways}

% \begin{figure}
%     \centering
%     \includegraphics[width=\linewidth]{Figures/Summary.png}
%     \caption{Caption}
%     \label{fig:placeholder}
% \end{figure}
\section{Threats to Validity}
\label{sec:T2V}

This section discusses the potential threats to validity following the categories proposed by \citet{wohlin_experimentation_2024}.

\textbf{Construct Validity.} Construct validity concerns whether the collected data accurately represents the phenomenon under investigation \citep{wohlin_experimentation_2024}. We investigated approaches, metrics, tools, and remediation techniques related to architectural degradation across both peer-reviewed and grey literature. However, architectural degradation lacks a universally accepted terminology, and closely related concepts such as erosion, decay, drift, aging, technical debt, and architecture conformance frequently overlap. In particular, several primary studies treat architecture conformance as evidence or a proxy of architectural degradation, whereas others consider it a distinct research topic. Consequently, some included studies may operationalize degradation differently, potentially affecting the resulting taxonomy. Moreover, our LLM-supported extraction and classification pipeline may have interpreted ambiguous descriptions differently from human experts, particularly in heterogeneous grey literature. Our manual validation also showed that one validation model (Qwen) consistently achieved lower agreement than the other models for several extraction and classification tasks. Rather than relying on a single model, we mitigated these threats by defining explicit inclusion criteria, grounding model responses through Retrieval-Augmented Generation (RAG), validating every extraction using three independent LLMs, identifying disagreements through model consensus, and resolving them through human adjudication and majority voting.

\textbf{Internal Validity.} Internal validity concerns whether the reported findings originate from the analyzed evidence rather than methodological artifacts \citep{wohlin_experimentation_2024}. Our study relied on LLMs to support study selection, information extraction, and thematic classification. Different LLMs may encode different prior knowledge, reasoning strategies, or alignment policies, potentially producing different classifications for ambiguous cases. Likewise, prompt formulation and category consolidation may influence the final taxonomy. We reduced them by assigning specialized roles to the models, validating every output through a multi-model consensus process, performing several rounds of iterative refinement before final validation, and involving multiple independent human validators to resolve disagreements through majority voting.

\textbf{External Validity.} External validity concerns the generalizability of the findings beyond the analyzed dataset \citep{wohlin_experimentation_2024}. We analyzed 284 primary studies retrieved from multiple scientific databases together with grey literature, providing broad coverage of both academic and industrial perspectives. Nevertheless, our dataset only includes English-language publications indexed by the selected databases and search engines that were available at the time of the search. Furthermore, grey literature varies substantially in reporting quality and methodological rigor. Although we applied identical inclusion criteria, extraction procedures, and validation processes to both peer-reviewed and grey literature, some industrial practices, non-English publications, or emerging trends may remain underrepresented. Future updates of the multivocal literature review may therefore identify additional approaches, metrics, tools, or remediation techniques.

\textbf{Conclusion Validity.} Conclusion validity concerns whether the conclusions are adequately supported by the available evidence \citep{wohlin_experimentation_2024}. Our study synthesizes recurring patterns across 284 primary studies rather than establishing statistical or causal relationships. Therefore, frequently reported approaches, metrics, tools, or remediation strategies should not be interpreted as evidence of superior effectiveness, but as the most consistently reported evidence in the literature.
The use of LLMs introduces additional threats due to the inherent variability and limited transparency of their outputs. We mitigated these threats through local inference with fixed model versions, deterministic decoding (temperature = 0), fixed parameters, RAG, and validation using three independent LLMs followed by human majority voting. We also documented the complete pipeline and provide a replication package. Nevertheless, different models or execution environments may produce minor variations; reproducibility should therefore be considered at the level of taxonomies, categories, and aggregate findings rather than exact outputs.
\section{Conclusion}
\label{sec:Conclusion}

This paper presented a Multivocal Literature Review of 284 primary studies, delivering the first integrated view of architectural degradation across four dimensions: measurement approaches, metrics, tools, and remediation strategies.
The resulting taxonomies and cross-dimensional analysis reveal a field that has matured impressively at \emph{detecting} degradation, static analysis, structural metrics, and detection tools dominate the landscape, but remains fragmented and comparatively immature when it comes to \emph{acting} on what is detected.
Remediation strategies are scattered across code-level, architectural, and organizational interventions with no shared vision, and the tooling ecosystem offers extensive support for diagnosis yet very little for deciding or executing corrective actions.

\textbf{The central implication is clear: the next generation of research must shift from building yet more detection techniques toward closing the diagnosis--remediation gap.}
Priorities include decision-support frameworks that help practitioners select and prioritize remediation strategies, runtime-aware and predictive approaches that catch degradation before it solidifies, and continuous architectural governance that treats degradation management as an ongoing process rather than a periodic assessment.
For practitioners, the evidence synthesized here provides an actionable map of what is available today and where the blind spots remain.

Architectural degradation is not a solved problem, but the path forward is now visible.
We hope the taxonomies, maturity analysis, and conceptual framework offered in this work serve both as a reference for future research and as a practical foundation for building the next generation of intelligent, continuous, and automated degradation-management solutions.

\section*{Data Availability Statement}
We have published the complete raw data in the replication package to allow our study to be replicated \footnote{\url{https://doi.org/10.5281/zenodo.23031891}}.

\section*{Acknowledgment}
This work has been funded by the Research Council of Finland (grants n. 359861 and 349488 - MuFAno), by Business Finland (grant 6GSoft~\citep{akbar_6gsoft_2024}), and FAST, the Finnish Software Engineering Doctoral Research Network, funded by the Ministry of Education and Culture, Finland. 

\section*{Declaration on the use of generative AI}
The authors used ChatGPT for suggestions on improving textual clarity. All research design, data analysis, interpretations, and manuscripts were created by the authors themselves.

\bibliographystyle{ACM-Reference-Format}
\bibliography{main.bib}
\newpage
\appendix
\section*{Primary Studies}
% ------------------------------------------------------------------
% Automatically generated reference list
% ------------------------------------------------------------------
%
% Required package:
%   \usepackage{hyperref}
%
% The macro below makes:
%
%   \ref{BibTeXKey}
%
% return the BibTeX key itself.
%
% Example:
%
%   See reference~\ref{Esposito2025}.
%
% ------------------------------------------------------------------

\makeatletter
\newcommand{\keyedreference}[1]{%
    \item[\texttt{\detokenize{#1}}]%
    \def\@currentlabel{#1}%
    \phantomsection%
    \label{#1}%
}
\makeatother

\begin{description}

\keyedreference{WL1}
Zamfirov, Filip, Radulescu, Andrei, Krüger, Jacob, Chaudron, Michel R. V., Lessons from Visualizing Software Architecture Structure Conformance at Thermo Fisher Scientific, \textit{Lecture Notes in Computer Science} 16083 LNCS (2026) 372 – 389, \url{https://doi.org/10.1007/978-3-032-04207-1_25}.

\keyedreference{WL2}
Johansson, Nils, Caporuscio, Mauro, Application of AI in Software Architecture Development at Volvo Construction Equipment, \textit{Lecture Notes in Computer Science} 15982 LNCS (2026) 119 – 126, \url{https://doi.org/10.1007/978-3-032-04403-7_12}.

\keyedreference{WL3}
Thielen, Sven, Salgert, Björn, Franz, Thomas, From Lab to Market: Architectural Evolution in Open Source Transition, \textit{Lecture Notes in Computer Science} 15929 LNCS (2026) 306 – 322, \url{https://doi.org/10.1007/978-3-032-02138-0_20}.

\keyedreference{WL4}
Baabad, Ahmed, Zulzalil, Hazura Binti, Hassan, Sa'adah, Baharom, Salmi Binti, An Empirical Analysis of Approach-Based Metrics Model for Architectural Erosion Detection, \textit{Software - Practice and Experience} 55 (9) (2025) 1464 – 1488, \url{https://doi.org/10.1002/spe.3431}.

\keyedreference{WL5}
Zhan, Wenjing, Mo, Ran, Jiang, Yingjie, Wang, Dongyu, Just-in-Time Prediction of Software Architectural Changes Through Commit-Level Analyses, \textit{IEEE Transactions on Software Engineering} 51 (8) (2025) 2285 – 2304, \url{https://doi.org/10.1109/TSE.2025.3587849}.

\keyedreference{WL6}
Gnoyke, Philipp, Schulze, Sandro, Krüger, Jacob, Evolution patterns of software-architecture smells: An empirical study of intra- and inter-version smells, \textit{Journal of Systems and Software} 217 (2024), \url{https://doi.org/10.1016/j.jss.2024.112170}.

\keyedreference{WL7}
Wang, Tong, Li, BiXin, EsArCost: Estimating repair costs of software architecture erosion using slice technology, \textit{Journal of Systems and Software} 208 (2024), \url{https://doi.org/10.1016/j.jss.2023.111875}.

\keyedreference{WL8}
Kumar, Neeraj, Singh, Randeep, Rathee, Amit, \textit{Need for architecture recovery in OSS: A decade study}, 2024, \url{https://doi.org/10.1201/9781003433958-1}.

\keyedreference{WL9}
Al-Refai, Mohammed, Hammad, Mahmoud M., An Evaluation of ACDC in Recovering JPMS-Compliant Components for Java Applications, 2024, \url{https://doi.org/10.1109/ICICS63486.2024.10638314}.

\keyedreference{WL10}
Genfer, Patric, Zdun, Uwe, Exploring Architectural Evolution in Microservice Systems Using Repository Mining Techniques and Static Code Analysis, \textit{Lecture Notes in Computer Science (including subseries Lecture Notes in Artificial Intelligence and Lecture Notes in Bioinformatics)} 14889 LNCS (2024) 157 – 173, \url{https://doi.org/10.1007/978-3-031-70797-1_10}.

\keyedreference{WL11}
Rao, A. Eashaan, Chimalakonda, Sridhar, Towards Connecting Bugs and Architecture in Software Systems: A Perspective, 2024, pp. 100 – 104, \url{https://doi.org/10.1109/ICSA-C63560.2024.00025}.

\keyedreference{WL12}
Anthony, Emilie, Berntsson, Astrid, Santilli, Tiziano, Wohlrab, Rebekka, We're Drifting Apart: Architectural Drift from the Developers' Perspective, 2024, pp. 101 – 111, \url{https://doi.org/10.1109/ICSA59870.2024.00018}.

\keyedreference{WL13}
Bucaioni, Alessio, Di Salle, Amleto, Iovino, Ludovico, Mariani, Leonardo, Pelliccione, Patrizio, Continuous Conformance of Software Architectures, 2024, pp. 112 – 122, \url{https://doi.org/10.1109/ICSA59870.2024.00019}.

\keyedreference{WL14}
Garcia, Joshua, Mirakhorli, Mehdi, Xiao, Lu, Malek, Sam, Kazman, Rick, Cai, Yuanfang, Medvidovic, Nenad, SAIN: A Community-Wide Software Architecture INfrastructure, 2023, pp. 336 – 337, \url{https://doi.org/10.1109/ICSE-Companion58688.2023.00095}.

\keyedreference{WL15}
Michalkowski, Carissa, Janhsen, Jan, Springer, Patrick, Concept for a generic modular software architecture for the integration of quality relevant data and sample implementation for a laser sintering system, \textit{Progress in Additive Manufacturing} 8 (1) (2023) 67 – 73, \url{https://doi.org/10.1007/s40964-022-00390-8}.

\keyedreference{WL16}
Júnior, Jackson, Coutinho, Pedro, Restructuring the Software Architecture: A Case Study of the CoolBiz Core Banking Platform, 2023, pp. 348 – 363, \url{https://doi.org/10.18803/capsi.v23.348-363}.

\keyedreference{WL17}
Agouf, Nour Jihene, Labsari, Soufyane, Ducasse, Stephane, Etien, Anne, Anquetil, Nicolas, A Visualization for Client-Server Architecture Assessement, 2023, pp. 1 – 11, \url{https://doi.org/10.1109/VISSOFT60811.2023.00010}.

\keyedreference{WL18}
Prajapati, Amarjeet, A particle swarm optimization approach for large-scale many-objective software architecture recovery, \textit{Journal of King Saud University - Computer and Information Sciences} 34 (10) (2022) 8501 – 8513, \url{https://doi.org/10.1016/j.jksuci.2021.08.027}.

\keyedreference{WL19}
Cerny, Tomas, Abdelfattah, Amr S., Bushong, Vincent, Al Maruf, Abdullah, Taibi, Davide, Microservice Architecture Reconstruction and Visualization Techniques: A Review, 2022, pp. 39 – 48, \url{https://doi.org/10.1109/SOSE55356.2022.00011}.

\keyedreference{WL20}
Ge, Ning, Wang, Ze, Zhang, Li, Zhao, Jiuang, Zhou, Yufei, Liu, Zewei, ArchTacRV: Detecting and Runtime Verifying Architectural Tactics in Code, 2022, pp. 566 – 576, \url{https://doi.org/10.1109/SANER53432.2022.00074}.

\keyedreference{WL21}
Baabad, Ahmed, Zulzalil, Hazura Binti, Hassan, Sa'Adah, Baharom, Salmi Binti, Characterizing the Architectural Erosion Metrics: A Systematic Mapping Study, \textit{IEEE Access} 10 (2022) 22915 – 22940, \url{https://doi.org/10.1109/ACCESS.2022.3150847}.

\keyedreference{WL22}
Murwantara, I Made, Yugopuspito, Pujianto, An adaptive IoT architecture using combination of concept-drift and dynamic software product line engineering, \textit{Telkomnika (Telecommunication Computing Electronics and Control)} 19 (4) (2021) 1226 – 1233, \url{https://doi.org/10.12928/TELKOMNIKA.v19i4.19012}.

\keyedreference{WL23}
Semlali, Badr-Eddine Boudriki, Amrani, Chaker El, Ortiz, Guadalupe, Boubeta-Puig, Juan, Garcia-de-Prado, Alfonso, SAT-CEP-monitor: An air quality monitoring software architecture combining complex event processing with satellite remote sensing, \textit{Computers and Electrical Engineering} 93 (2021), \url{https://doi.org/10.1016/j.compeleceng.2021.107257}.

\keyedreference{WL24}
Knieke, Christoph, Rausch, Andreas, Schindler, Mirco, Tackling Software Architecture Erosion: Joint Architecture and Implementation Repairing by a Knowledge-based Approach, 2021, pp. 19 – 20, \url{https://doi.org/10.1109/APR52552.2021.00011}.

\keyedreference{WL25}
Verdecchia, Roberto, Kruchten, Philippe, Lago, Patricia, Malavolta, Ivano, Building and evaluating a theory of architectural technical debt in software-intensive systems, \textit{Journal of Systems and Software} 176 (2021), \url{https://doi.org/10.1016/j.jss.2021.110925}.

\keyedreference{WL26}
Rahmati, Zahed, Tanhaei, Mohammad, Ensuring software maintainability at software architecture level using architectural patterns, \textit{AUT Journal of Mathematics and Computing} 2 (1) (2021) 81 – 102, \url{https://doi.org/10.22060/ajmc.2021.19232.1044}.

\keyedreference{WL27}
Gnoyke, Philipp, Schulze, Sandro, Kruger, Jacob, An Evolutionary Analysis of Software-Architecture Smells, 2021, pp. 413 – 424, \url{https://doi.org/10.1109/ICSME52107.2021.00043}.

\keyedreference{WL28}
Lewis, Grace A., Ozkaya, Ipek, Xu, Xiwei, Software Architecture Challenges for ML Systems, 2021, pp. 634 – 638, \url{https://doi.org/10.1109/ICSME52107.2021.00071}.

\keyedreference{WL29}
Portugal, Telma, Barata, João, Enterprise architecture erosion: A definition and research framework, 2021, \url{https://www.scopus.com/inward/record.uri?eid=2-s2.0-85118633266&partnerID=40&md5=355954e244e0b25b356856fb13336263}.

\keyedreference{WL30}
Konersmann, Marco, Goedicke, Michael, \textit{Same but different: Consistently developing and evolving software architecture models and their implementation}, 2020, \url{https://doi.org/10.1007/978-3-030-58617-1_6}.

\keyedreference{WL31}
Whiting, Erik, Andrews, Sharon, Drift and Erosion in Software Architecture: Summary and Prevention Strategies, 2020, pp. 132 – 138, \url{https://doi.org/10.1145/3404663.3404665}.

\keyedreference{WL32}
Martin, Daniel San, Siqueira, Bento, De Camargo, Valter Vieira, Ferrari, Fabiano, Characterizing Architectural Drifts of Adaptive Systems, 2020, pp. 389 – 399, \url{https://doi.org/10.1109/SANER48275.2020.9054856}.

\keyedreference{WL33}
Herold, Sebastian, An initial study on the association between architectural smells and degradation, \textit{Lecture Notes in Computer Science (including subseries Lecture Notes in Artificial Intelligence and Lecture Notes in Bioinformatics)} 12292 LNCS (2020) 193 – 201, \url{https://doi.org/10.1007/978-3-030-58923-3_13}.

\keyedreference{WL34}
Bohé, Ilse, Willocx, Michiel, Naessens, Vincent, SMIOT: A software architecture for maintainable internet-of-things applications, \textit{International Journal of Cloud Computing} 9 (1) (2020) 75 – 94, \url{https://doi.org/10.1504/IJCC.2020.105875}.

\keyedreference{WL35}
Kadri, Salim, Aouag, Sofiane, Hedjazi, Djalal, Multi-level approach for controlling architecture quality with Alloy, 2019, \url{https://doi.org/10.1109/ICTAACS48474.2019.8988128}.

\keyedreference{WL36}
Fontana, Francesca Arcelli, Pigazzini, Ilaria, Raibulet, Claudia, Basciano, Stefano, Roveda, Riccardo, PageRank and criticality of architectural smells, 2019, pp. 197 – 204, \url{https://doi.org/10.1145/3344948.3344982}.

\keyedreference{WL37}
Wang, Tong, Li, Bixin, Analyzing Software Architecture Evolvability Based on Multiple Architectural Attributes Measurements, 2019, pp. 204 – 215, \url{https://doi.org/10.1109/QRS.2019.00037}.

\keyedreference{WL38}
Tommasel, Antonela, Applying Social Network Analysis Techniques to Architectural Smell Prediction, 2019, pp. 254 – 261, \url{https://doi.org/10.1109/ICSA-C.2019.00053}.

\keyedreference{WL39}
Lenhard, Jörg, Blom, Martin, Herold, Sebastian, Exploring the suitability of source code metrics for indicating architectural inconsistencies, \textit{Software Quality Journal} 27 (1) (2019) 241 – 274, \url{https://doi.org/10.1007/s11219-018-9404-z}.

\keyedreference{WL40}
Xu, Jia, A software architecture for handling complex critical section constraints on multiprocessors in a fault-tolerant real-time embedded system, 2019, pp. 251 – 261, \url{https://doi.org/10.29007/brkj}.

\keyedreference{WL41}
Dorado, Santiago Hyun, Hurtado, Julio Ariel, Documenting architectural rationale using source code annotations: An exploratory study, 2019, pp. 204 – 214, \url{https://doi.org/10.29007/f5md}.

\keyedreference{WL42}
Shaikh, Mohsin, Ibarhimov, Dilshod, Zardari, Baqir, Assessing architectural sustainability during software evolution using package-modularization metrics, \textit{International Journal of Advanced Computer Science and Applications} 10 (12) (2019) 592 – 608, \url{https://doi.org/10.14569/ijacsa.2019.0101277}.

\keyedreference{WL43}
Murty, Rohan, Padhye, Jitendra, Wolman, Alec, Welsh, Matt, Dyson: An architecture for extensible wireless LANs, 2019, pp. 201 – 214, \url{https://doi.org/}.

\keyedreference{WL44}
Jasser, Stefanie, Constraining the Implementation Through Architectural Security Rules: An Expert Study, \textit{Lecture Notes in Computer Science (including subseries Lecture Notes in Artificial Intelligence and Lecture Notes in Bioinformatics)} 11915 LNCS (2019) 203 – 219, \url{https://doi.org/10.1007/978-3-030-35333-9_15}.

\keyedreference{WL45}
Sinkala, Zipani Tom, Blom, Martin, Herold, Sebastian, A mapping study of software architecture recovery for software product lines, 2018, \url{https://doi.org/10.1145/3241403.3241454}.

\keyedreference{WL46}
Zapalowski, Vanius, Nunes, Daltro José, Nunes, Ingrid, Understanding architecture non-conformance: Why is there a gap between conceptual architectural rules and source code dependencies?, 2018, pp. 22 – 31, \url{https://doi.org/10.1145/3266237.3266261}.

\keyedreference{WL47}
Verdecchia, Roberto, Architectural Technical Debt Identification: Moving Forward, 2018, pp. 43 – 44, \url{https://doi.org/10.1109/ICSA-C.2018.00018}.

\keyedreference{WL48}
Bandara, Vidudaya, Perera, Indika, Identifying software architecture erosion through code comments, 2018, pp. 62 – 69, \url{https://doi.org/10.1109/ICTER.8615560}.

\keyedreference{WL49}
Venters, Colin C., Capilla, Rafael, Betz, Stefanie, Penzenstadler, Birgit, Crick, Tom, Crouch, Steve, Nakagawa, Elisa Yumi, Becker, Christoph, Carrillo, Carlos, Software sustainability: Research and practice from a software architecture viewpoint, \textit{Journal of Systems and Software} 138 (2018) 174 – 188, \url{https://doi.org/10.1016/j.jss.2017.12.026}.

\keyedreference{WL50}
Phillips, Dewanne M., Mazzuchi, Thomas A., Sarkani, Shahram, An architecture, system engineering, and acquisition approach for space system software resiliency, \textit{Information and Software Technology} 94 (2018) 150 – 164, \url{https://doi.org/10.1016/j.infsof.2017.10.006}.

\keyedreference{WL51}
Altinisik, Metin, Ersoy, Ersin, Sözer, Hasan, Evaluating software architecture erosion for PL/SQL programs, 2017, pp. 159 – 165, \url{https://doi.org/10.1145/3129790.3129811}.

\keyedreference{WL52}
Naim, Sheikh Motahar, Damevski, Kostadin, Hossain, M. Shahriar, Reconstructing and evolving software architectures using a coordinated clustering framework, \textit{Automated Software Engineering} 24 (3) (2017) 543 – 572, \url{https://doi.org/10.1007/s10515-017-0211-8}.

\keyedreference{WL53}
Behnamghader, Pooyan, Le, Duc Minh, Garcia, Joshua, Link, Daniel, Shahbazian, Arman, Medvidovic, Nenad, A large-scale study of architectural evolution in open-source software systems, \textit{Empirical Software Engineering} 22 (3) (2017) 1146 – 1193, \url{https://doi.org/10.1007/s10664-016-9466-0}.

\keyedreference{WL54}
Usach, Hèctor, Torens, Christoph, Adolf, Florian-M., Vila, Juan, Architectural considerations towards automated contingency management for unmanned aircraft, 2017, \url{https://doi.org/10.2514/6.2017-1293}.

\keyedreference{WL55}
Zdun, Uwe, Navarro, Elena, Leymann, Frank, Ensuring and assessing architecture conformance to microservice decomposition patterns, \textit{Lecture Notes in Computer Science (including subseries Lecture Notes in Artificial Intelligence and Lecture Notes in Bioinformatics)} 10601 LNCS (2017) 411 – 429, \url{https://doi.org/10.1007/978-3-319-69035-3_29}.

\keyedreference{WL56}
Bouras, Zine-Eddine, Maouche, Mourad, Software architectures evolution based merging, \textit{Informatica (Slovenia)} 41 (1) (2017) 111 – 120, \url{https://www.scopus.com/inward/record.uri?eid=2-s2.0-85018997238&partnerID=40&md5=f6f3db1e36ef2527a653fb9bc5baaa39}.

\keyedreference{WL57}
Haitzer, Thomas, Navarro, Elena, Zdun, Uwe, Reconciling software architecture and source code in support of software evolution, \textit{Journal of Systems and Software} 123 (2017) 119–144, \url{https://doi.org/10.1016/j.jss.2016.10.012}.

\keyedreference{WL58}
Fontana, Francesca Arcelli, Roveda, Riccardo, Zanoni, Marco, Raibulet, Claudia, Capilla, Rafael, An Experience Report on Detecting and Repairing Software Architecture Erosion, in: \textit{2016 13th Working IEEE/IFIP Conference on Software Architecture (WICSA)}, IEEE, 2016, pp. 21–30, \url{https://doi.org/10.1109/wicsa.2016.37}.

\keyedreference{WL59}
Miranda, Sergio, Rodrigues Jr, Elder, Valente, Marco Tulio, Terra, Ricardo, Architecture Conformance Checking in Dynamically Typed Languages, \textit{The Journal of Object Technology} 15 (3) (2016) 1:1, \url{https://doi.org/10.5381/jot.2016.15.3.a1}.

\keyedreference{WL60}
Castillo, José Carlos, Castro-González, Álvaro, Fernández-Caballero, Antonio, Latorre, José Miguel, Pastor, José Manuel, Fernández-Sotos, Alicia, Salichs, Miguel A., Software Architecture for Smart Emotion Recognition and Regulation of the Ageing Adult, \textit{Cognitive Computation} 8 (2) (2016) 357–367, \url{https://doi.org/10.1007/s12559-016-9383-y}.

\keyedreference{WL61}
De Silva, Mahesh, Perera, Indika, Preventing software architecture erosion through static architecture conformance checking, in: \textit{2015 IEEE 10th International Conference on Industrial and Information Systems (ICIIS)}, IEEE, 2015, pp. 43–48, \url{https://doi.org/10.1109/iciinfs.2015.7398983}.

\keyedreference{WL62}
Gerdes, Sebastian, Jasser, Stefanie, Riebisch, Matthias, Schr\"{o}der, Sandra, Soliman, Mohamed, Stehle, Tilmann, Towards the essentials of architecture documentation for avoiding architecture erosion, in: \textit{Proccedings of the 10th European Conference on Software Architecture Workshops}, ACM, 2016, pp. 1–4, \url{https://doi.org/10.1145/2993412.3004844}.

\keyedreference{WL63}
Greifenberg, Timo, M\"{u}ller, Klaus, Rumpe, Bernhard, Architectural Consistency Checking in Plugin-Based Software Systems, in: \textit{Proceedings of the 2015 European Conference on Software Architecture Workshops}, ACM, 2015, pp. 1–7, \url{https://doi.org/10.1145/2797433.2797493}.

\keyedreference{WL64}
Brunet, Joao, Murphy, Gail C., Serey, Dalton, Figueiredo, Jorge, Five Years of Software Architecture Checking: A Case Study of Eclipse, \textit{IEEE Software} 32 (5) (2015) 30–36, \url{https://doi.org/10.1109/ms.2014.106}.

\keyedreference{WL65}
Le, Duc Minh, Behnamghader, Pooyan, Garcia, Joshua, Link, Daniel, Shahbazian, Arman, Medvidovic, Nenad, An Empirical Study of Architectural Change in Open-Source Software Systems, in: \textit{2015 IEEE/ACM 12th Working Conference on Mining Software Repositories}, IEEE, 2015, pp. 235–245, \url{https://doi.org/10.1109/msr.2015.29}.

\keyedreference{WL66}
Rodriguez, Lina Maria Garces, Ampatzoglou, Apostolos, Avgeriou, Paris, Nakagawa, Elisa Yumi, A Comparative Analysis of Reference Architectures for Healthcare in the Ambient Assisted Living Domain, in: \textit{2015 IEEE 28th International Symposium on Computer-Based Medical Systems}, IEEE, 2015, pp. 270–275, \url{https://doi.org/10.1109/cbms.2015.29}.

\keyedreference{WL67}
Barros, Márcio de Oliveira, Farzat, Fábio de Almeida, Travassos, Guilherme Horta, Learning from optimization: A case study with Apache Ant, \textit{Information and Software Technology} 57 (2015) 684–704, \url{https://doi.org/10.1016/j.infsof.2014.07.015}.

\keyedreference{WL68}
Mokni, Abderrahman, Huchard, Marianne, Urtado, Christelle, Vauttier, Sylvain, Zhang, Huaxi Yulin, An evolution management model for multi-level component-based software architectures, in: \textit{Proceedings of the 27th International Conference on Software Engineering and Knowledge Engineering}, KSI Research Inc. and Knowledge Systems Institute Graduate School, 2015, pp. 674–679, \url{https://doi.org/10.18293/seke2015-172}.

\keyedreference{WL69}
Bandi, Ajay, Allen, Edward B., Williams, Byron J., Assessing code decay: A data-driven approach, 2015, pp. 95 – 102, \url{https://www.scopus.com/inward/record.uri?eid=2-s2.0-84964027504&partnerID=40&md5=4a95076c6cce0ec80867546a0919e9cd}.

\keyedreference{WL70}
Rathish Babu, T.K.S., Sankar Ram, N., A new framework for software architecture generation and fault rectification in software engineering, \textit{International Journal of Applied Engineering Research} 10 (16) (2015) 37813 – 37819, \url{https://www.scopus.com/inward/record.uri?eid=2-s2.0-84942033966&partnerID=40&md5=34357ec5fdb713b8d44743529c406333}.

\keyedreference{WL71}
Sanchez, Alejandro, Barbosa, Luis S., Madeira, Alexandre, Modelling and Verifying Smell-Free Architectures with the Archery Language, in: \textit{Software Engineering and Formal Methods}, Springer International Publishing, 2015, pp. 147–163.

\keyedreference{WL72}
Alencar, Felipe, Santos, Marcelo, Santana, Matheus, Fernandes, Stenio, How Software Aging affects SDN: A view on the controllers, in: \textit{2014 Global Information Infrastructure and Networking Symposium (GIIS)}, IEEE, 2014, pp. 1–6, \url{https://doi.org/10.1109/giis.2014.6934264}.

\keyedreference{WL73}
Guimaraes, Everton, Garcia, Alessandro, Cai, Yuanfang, Exploring Blueprints on the Prioritization of Architecturally Relevant Code Anomalies -- A Controlled Experiment, in: \textit{2014 IEEE 38th Annual Computer Software and Applications Conference}, IEEE, 2014, pp. 344–353, \url{https://doi.org/10.1109/compsac.2014.57}.

\keyedreference{WL74}
Herold, Sebastian, Mair, Matthias, Recommending Refactorings to Re-establish Architectural Consistency, in: \textit{Software Architecture}, Springer International Publishing, 2014, pp. 390–397.

\keyedreference{WL75}
Strasser, Arthur, Cool, Benjamin, Gernert, Christoph, Knieke, Christoph, K\"{o}rner, Marco, Niebuhr, Dirk, Peters, Henrik, Rausch, Andreas, Brox, Oliver, Jauns-Seyfried, Stefanie, Jelden, Hanno, Klie, Stefan, Kr\"{a}mer, Michael, Mastering Erosion of Software Architecture in Automotive Software Product Lines, in: \textit{SOFSEM 2014: Theory and Practice of Computer Science}, Springer International Publishing, 2014, pp. 491–502.

\keyedreference{WL76}
Stevanetic, Srdjan, Haitzer, Thomas, Zdun, Uwe, Supporting Software Evolution by Integrating DSL-based Architectural Abstraction and Understandability Related Metrics, in: \textit{Proceedings of the 2014 European Conference on Software Architecture Workshops}, ACM, 2014, pp. 1–8, \url{https://doi.org/10.1145/2642803.2642822}.

\keyedreference{WL77}
Mair, Matthias, Herold, Sebastian, Rausch, Andreas, Towards flexible automated software architecture erosion diagnosis and treatment, in: \textit{Proceedings of the WICSA 2014 Companion Volume}, ACM, 2014, pp. 1–6, \url{https://doi.org/10.1145/2578128.2578231}.

\keyedreference{WL78}
Nord, Robert L., Ozkaya, Ipek, Sangwan, Raghvinder S., Koontz, Ronald J., Architectural dependency analysis to understand rework costs for safety-critical systems, in: \textit{Companion Proceedings of the 36th International Conference on Software Engineering}, ACM, 2014, pp. 185–194, \url{https://doi.org/10.1145/2591062.2591185}.

\keyedreference{WL79}
Zhang, Xinhai, Persson, Magnus, Nyberg, Mattias, Mokhtari, Behrooz, Einarson, Anton, Linder, Henrik, Westman, Jonas, Chen, DeJiu, Torngren, Martin, Experience on applying software architecture recovery to automotive embedded systems, in: \textit{2014 Software Evolution Week - IEEE Conference on Software Maintenance, Reengineering, and Reverse Engineering (CSMR-WCRE)}, IEEE, 2014, pp. 379–382, \url{https://doi.org/10.1109/csmr-wcre.2014.6747199}.

\keyedreference{WL80}
Mirakhorli, Mehdi, Preventing erosion of architectural tactics through their strategic implementation, preservation, and visualization, in: \textit{2013 28th IEEE/ACM International Conference on Automated Software Engineering (ASE)}, IEEE, 2013, pp. 762–765, \url{https://doi.org/10.1109/ase.2013.6693152}.

\keyedreference{WL81}
Herold, Sebastian, Rausch, Andreas, Complementing model-driven development for the detection of software architecture erosion, in: \textit{2013 5th International Workshop on Modeling in Software Engineering (MiSE)}, IEEE, 2013, pp. 24–30, \url{https://doi.org/10.1109/mise.2013.6595292}.

\keyedreference{WL82}
Koziolek, Heiko, Domis, Dominik, Goldschmidt, Thomas, Vorst, Philipp, Measuring Architecture Sustainability, \textit{IEEE Software} 30 (6) (2013) 54–62, \url{https://doi.org/10.1109/ms.2013.101}.

\keyedreference{WL83}
Langhammer, Michael, Co-evolution of component-based architecture-model and object-oriented source code, in: \textit{Proceedings of the 18th international doctoral symposium on Components and architecture}, ACM, 2013, pp. 37–42, \url{https://doi.org/10.1145/2465498.2465504}.

\keyedreference{WL84}
Dimech, Claire, Balasubramaniam, Dharini, Maintaining Architectural Conformance during Software Development: A Practical Approach, in: \textit{Software Architecture}, Springer Berlin Heidelberg, 2013, pp. 208–223.

\keyedreference{WL85}
Mair, Matthias, Herold, Sebastian, Towards Extensive Software Architecture Erosion Repairs, in: \textit{Software Architecture}, Springer Berlin Heidelberg, 2013, pp. 299–306.

\keyedreference{WL86}
Baek, Seungjae, Choi, Jongmoo, Lee, Donghee, Noh, Sam H., Energy-efficient and high-performance software architecture for storage class memory, \textit{ACM Transactions on Embedded Computing Systems} 12 (3) (2013) 1–22, \url{https://doi.org/10.1145/2442116.2442131}.

\keyedreference{WL87}
Dragomir, Ana, Lichter, Horst, Run-Time Monitoring and Real-Time Visualization of Software Architectures, in: \textit{2013 20th Asia-Pacific Software Engineering Conference (APSEC)}, IEEE, 2013, pp. 396–403, \url{https://doi.org/10.1109/apsec.2013.60}.

\keyedreference{WL88}
Stal, Michael, Refactoring Software Architectures, in: \textit{Agile Software Architecture}, Elsevier, 2014, pp. 63–82.

\keyedreference{WL89}
Arcoverde, Roberta, Guimaraes, Everton, Macia, Isela, Garcia, Alessandro, Cai, Yuanfang, Prioritization of Code Anomalies Based on Architecture Sensitiveness, in: \textit{2013 27th Brazilian Symposium on Software Engineering}, IEEE, 2013, pp. 69–78, \url{https://doi.org/10.1109/sbes.2013.14}.

\keyedreference{WL90}
Potilde Janari, Robal Tarmo, Kalja Ahto, On Proving the Concept of an Ontology Aided Software Refactoring Tool, in: \textit{Databases and Information Systems VII}, IOS Press, 2013.

\keyedreference{WL91}
Koziolek, Heiko, Domis, Dominik, Goldschmidt, Thomas, Vorst, Philipp, Weiss, Roland J., MORPHOSIS: A Lightweight Method Facilitating Sustainable Software Architectures, in: \textit{2012 Joint Working IEEE/IFIP Conference on Software Architecture and European Conference on Software Architecture}, IEEE, 2012, pp. 253–257, \url{https://doi.org/10.1109/wicsa-ecsa.212.40}.

\keyedreference{WL92}
Haber, Arne, Rendel, Holger, Rumpe, Bernhard, Schaefer, Ina, Evolving Delta-Oriented Software Product Line Architectures, in: \textit{Large-Scale Complex IT Systems. Development, Operation and Management}, Springer Berlin Heidelberg, 2012, pp. 183–208.

\keyedreference{WL93}
Arcoverde, Roberta, Macia, Isela, Garcia, Alessandro, von Staa, Arndt, Automatically detecting architecturally-relevant code anomalies, in: \textit{2012 Third International Workshop on Recommendation Systems for Software Engineering (RSSE)}, IEEE, 2012, pp. 90–91, \url{https://doi.org/10.1109/rsse.2012.6233419}.

\keyedreference{WL94}
Larab, Ali, Conchon, Emmanuel, Bastide, Remi, Singer, Nicolas, A sustainable software architecture for home care monitoring applications, in: \textit{2012 6th IEEE International Conference on Digital Ecosystems and Technologies (DEST)}, IEEE, 2012, pp. 1–6, \url{https://doi.org/10.1109/dest.2012.6227928}.

\keyedreference{WL95}
Terra, Ricardo, Valente, Marco Tulio, Czarnecki, Krzysztof, Bigonha, Roberto S., Recommending Refactorings to Reverse Software Architecture Erosion, in: \textit{2012 16th European Conference on Software Maintenance and Reengineering}, IEEE, 2012, pp. 335–340, \url{https://doi.org/10.1109/csmr.2012.40}.

\keyedreference{WL96}
Dragomir, Ana, Lichter, Horst, Model-Based Software Architecture Evolution and Evaluation, in: \textit{2012 19th Asia-Pacific Software Engineering Conference}, IEEE, 2012, pp. 697–700, \url{https://doi.org/10.1109/apsec.2012.118}.

\keyedreference{WL97}
Schmidt, Frederik, MacDonell, Stephen G., Connor, Andrew M., An Automatic Architecture Reconstruction and Refactoring Framework, in: \textit{Software Engineering Research, Management and Applications 2011}, Springer Berlin Heidelberg, 2012, pp. 95–111.

\keyedreference{WL98}
Mirakhorli, Mehdi, Cleland-Huang, Jane, Using tactic traceability information models to reduce the risk of architectural degradation during system maintenance, in: \textit{2011 27th IEEE International Conference on Software Maintenance (ICSM)}, IEEE, 2011, pp. 123–132, \url{https://doi.org/10.1109/icsm.2011.6080779}.

\keyedreference{WL99}
Li, Zude, Long, Jun, A Case Study of Measuring Degeneration of Software Architectures from a Defect Perspective, in: \textit{2011 18th Asia-Pacific Software Engineering Conference}, IEEE, 2011, pp. 242–249, \url{https://doi.org/10.1109/apsec.2011.51}.

\keyedreference{WL100}
Zhang, Lei, Sun, Yanchun, Song, Hui, Chauvel, Franck, Mei, Hong, Detecting architecture erosion by design decision of architectural pattern, 2011, pp. 758 – 763, \url{https://www.scopus.com/inward/record.uri?eid=2-s2.0-84862922217&partnerID=40&md5=2f24ff4055589f2e906d87aa973fc4f8}.

\keyedreference{WL101}
Lavallée, Mathieu, Robillard, Pierre N., Causes of premature aging during software development: an observational study, in: \textit{Proceedings of the 12th International Workshop on Principles of Software Evolution and the 7th annual ERCIM Workshop on Software Evolution}, ACM, 2011, pp. 61–70, \url{https://doi.org/10.1145/2024445.2024458}.

\keyedreference{WL102}
Kavimandan, Amogh, Gokhale, Aniruddha, Karsai, Gabor, Gray, Jeff, Managing the quality of software product line architectures through reusable model transformations, in: \textit{Proceedings of the joint ACM SIGSOFT conference -- QoSA and ACM SIGSOFT symposium -- ISARCS on Quality of software architectures -- QoSA and architecting critical systems -- ISARCS}, ACM, 2011, pp. 13–22, \url{https://doi.org/10.1145/2000259.2000264}.

\keyedreference{WL103}
Mirakhorli, Mehdi, Cleland-Huang, Jane, Tracing architectural concerns in high assurance systems (NIER track), in: \textit{Proceedings of the 33rd International Conference on Software Engineering}, ACM, 2011, pp. 908–911, \url{https://doi.org/10.1145/1985793.1985942}.

\keyedreference{WL104}
Lee, Dongcheul, Rhee, Byung Ho, Efficient Abnormal Traffic Detection Software Architecture for a Seamless Network, \textit{KSII Transactions on Internet and Information Systems} (2011) 313–329, \url{https://doi.org/10.3837/tiis.2011.02.004}.

\keyedreference{WL105}
Masuta, Hiroyuki, Hiwada, Eriko, Kubota, Naoyuki, Control Architecture for Human Friendly Robots Based on Interacting with Human, in: \textit{Intelligent Robotics and Applications}, Springer Berlin Heidelberg, 2011, pp. 210–219.

\keyedreference{WL106}
Zhang, Huaxi, Urtado, Christelle, Vauttier, Sylvain, Architecture-centric development and evolution processes for component-based software, 2010, pp. 680 – 685, \url{https://www.scopus.com/inward/record.uri?eid=2-s2.0-79952422427&partnerID=40&md5=604ccba0975b6fd884feb979392bd45e}.

\keyedreference{WL107}
Kuusik, A., Reilent, E., Lõõbas, I., Luberg, A., Data acquisition software architecture for patient monitoring devices, \textit{Elektronika ir Elektrotechnika} (9) (2010) 97 – 100, \url{https://www.scopus.com/inward/record.uri?eid=2-s2.0-79961146686&partnerID=40&md5=a237b2fa67389e7887059d786642f373}.

\keyedreference{WL108}
Haimei Zhang, Kerong Ben, Architectural design of the Health Watch system with an integrated aspect-oriented modeling approach, in: \textit{2010 International Conference On Computer Design and Applications}, IEEE, 2010, pp. V1--624--V1--628, \url{https://doi.org/10.1109/iccda.2010.5540893}.

\keyedreference{WL109}
Cortellessa, Vittorio, Trubiani, Catia, Mostarda, Leonardo, Dulay, Naranker, An Architectural Framework for Analyzing Tradeoffs between Software Security and Performance, in: \textit{Architecting Critical Systems}, Springer Berlin Heidelberg, 2010, pp. 1–18.

\keyedreference{WL110}
Merkle, Bernhard, Stop the software architecture erosion: building better software systems, in: \textit{Proceedings of the ACM international conference companion on Object oriented programming systems languages and applications companion}, ACM, 2010, pp. 129–138, \url{https://doi.org/10.1145/1869542.1869563}.

\keyedreference{WL111}
Riaz, Mehwish, Sulayman, Muhammad, Naqvi, Husnain, Architectural Decay during Continuous Software Evolution and Impact of ‘Design for Change’ on Software Architecture, in: \textit{Advances in Software Engineering}, Springer Berlin Heidelberg, 2009, pp. 119–126.

\keyedreference{WL112}
Rui, Wang, Xiao-Guang, Mao, Zi-Ying, Dai, Yan-Ni, Wang, Extending UML for Aspect-Oriented Architecture Modeling, in: \textit{2009 Second International Workshop on Computer Science and Engineering}, IEEE, 2009, pp. 362–366, \url{https://doi.org/10.1109/wcse.2009.688}.

\keyedreference{WL113}
Anquetil, Nicolas, Royer, Jean-Claude, Andre, Pascal, Ardourel, Gilles, Hnetynka, Petr, Poch, Tomas, Petrascu, Dragos, Petrascu, Vladiela, JavaCompExt: Extracting Architectural Elements from Java Source Code, in: \textit{2009 16th Working Conference on Reverse Engineering}, IEEE, 2009, pp. 317–318, \url{https://doi.org/10.1109/wcre.2009.53}.

\keyedreference{WL114}
Capiluppi, Andrea, Beecher, Karl, Structural Complexity and Decay in FLOSS Systems: An Inter-repository Study, in: \textit{2009 13th European Conference on Software Maintenance and Reengineering}, IEEE, 2009, pp. 169–178, \url{https://doi.org/10.1109/csmr.2009.37}.

\keyedreference{WL115}
Tejedor, Jorge Aurelio, Fernández, Luis, Arroyo, Fernando, Bravo, Ginés, An architecture for attacking the communication bottleneck in P systems, \textit{Artificial Life and Robotics} 12 (1–2) (2008) 236–240, \url{https://doi.org/10.1007/s10015-007-0474-4}.

\keyedreference{WL116}
Sarela, Antti, Bidargaddi, Niranjan, Karunanithi, Mohan, A software architecture and data model for community-based healthcare environments, in: \textit{2008 Second International Conference on Pervasive Computing Technologies for Healthcare}, IEEE, 2008, pp. 161–164, \url{https://doi.org/10.1109/pcthealth.2008.4571059}.

\keyedreference{WL117}
Nguyen, Tien N., Managing Software Architectural Evolution at Multiple Levels of Abstraction, \textit{Journal of Software} 3 (3) (2008), \url{https://doi.org/10.4304/jsw.3.3.60-70}.

\keyedreference{WL118}
Williams, Byron J., Carver, Jeffrey C., Characterizing software architecture changes: An initial study, 2007, pp. 410 – 419, \url{https://doi.org/10.1109/ESEM.2007.28}.

\keyedreference{WL119}
Garcia, Alessandro, Chavez, Christina, Batista, Thais, Sant’anna, Claudio, Kulesza, Uirá, Rashid, Awais, Lucena, Carlos, On the Modular Representation of Architectural Aspects, in: \textit{Software Architecture}, Springer Berlin Heidelberg, 2006, pp. 82–97.

\keyedreference{WL120}
Anders, B., Fellmann, J., Lindvall, M., Rus, I., Experimenting with Software Architecture Flexibility Using an Implementation of the Tactical Separation Assisted Flight Environment, in: \textit{29th Annual IEEE/NASA Software Engineering Workshop}, IEEE, 2005, pp. 275–284, \url{https://doi.org/10.1109/sew.2005.24}.

\keyedreference{WL121}
Everitt, T., Tvedt, R.T., Tvedt, J.D., Validating and improving an existing software architectural evaluation process, in: \textit{20th IEEE International Conference on Software Maintenance, 2004. Proceedings.}, IEEE, 2004, pp. 417–421, \url{https://doi.org/10.1109/icsm.2004.1357827}.

\keyedreference{WL122}
Lindvall, Mikael, Tvedt, Roseanne Tesoriero, Costa, Patricia, An Empirically-Based Process for Software Architecture Evaluation, \textit{Empirical Software Engineering} 8 (1) (2003) 83–108, \url{https://doi.org/10.1023/a:1021772917036}.

\keyedreference{WL123}
Shelton, Charles, Koopman, Philip, Using Architectural Properties to Model and Measure Graceful Degradation, in: \textit{Architecting Dependable Systems}, Springer Berlin Heidelberg, 2003, pp. 267–289.

\keyedreference{WL124}
Tvedt, R.T., Lindvall, M., Costa, P., A process for software architecture evaluation using metrics, in: \textit{27th Annual NASA Goddard/IEEE Software Engineering Workshop, 2002. Proceedings.}, IEEE, 2002, pp. 191–196, \url{https://doi.org/10.1109/sew.2002.1199475}.

\keyedreference{WL125}
Lindvall, M., Tesoriero, R., Costa, P., Avoiding architectural degeneration: an evaluation process for software architecture, in: \textit{Proceedings Eighth IEEE Symposium on Software Metrics}, IEEE, 2002, pp. 77–86, \url{https://doi.org/10.1109/metric.2002.1011327}.

\keyedreference{WL126}
Tvedt, Roseanne Tesoriero, Costa, Patricia, Lindvall, Mikael, Does the code match the design? A process for architecture evaluation, 2002, pp. 393 – 401, \url{https://www.scopus.com/inward/record.uri?eid=2-s2.0-0036441910&partnerID=40&md5=993628d331716d34c29c4198b9cd836c}.

\keyedreference{WL127}
von Mayrhauser, A., Wang, J., Ohlsson, M.C., Wohlin, C., Deriving a fault architecture from defect history, 1999, pp. 295 – 303, \url{https://www.scopus.com/inward/record.uri?eid=2-s2.0-0033344933&partnerID=40&md5=646c8585e70de439d9d1d7b6f2e13d47}.

\keyedreference{WL128}
Hulse, Christine, Edgerton, Scott, Ubnoske, Michael, Vazquez, Louis, Reducing maintenance costs through the application of modern software architecture principles, 1999, pp. 101 – 110, \url{https://doi.org/10.1145/319294.319310}.

\keyedreference{WL129}
Le, Duc Minh, Carrillo, Carlos, Capilla, Rafael, Medvidovic, Nenad, Relating Architectural Decay and Sustainability of Software Systems, in: \textit{2016 13th Working IEEE/IFIP Conference on Software Architecture (WICSA)}, IEEE, 2016, pp. 178–181, \url{https://doi.org/10.1109/wicsa.2016.15}.

\keyedreference{WL130}
Zhao, Yong, Kerkhoff, Hans G., Unit-Based Functional IDDT Testing for Aging Degradation Monitoring in a VLIW Processor, in: \textit{2015 Euromicro Conference on Digital System Design}, IEEE, 2015, pp. 353–358, \url{https://doi.org/10.1109/dsd.2015.113}.

\keyedreference{WL131}
Le, Duc Minh, Link, Daniel, Shahbazian, Arman, Medvidovic, Nenad, An Empirical Study of Architectural Decay in Open-Source Software, in: \textit{2018 IEEE International Conference on Software Architecture (ICSA)}, IEEE, 2018, pp. 176–17609, \url{https://doi.org/10.1109/icsa.2018.00027}.

\keyedreference{WL132}
Hassaine, Salima, Guéhéneuc, Yann-Gaël, Hamel, Sylvie, Antoniol, Giuliano, ADvISE: Architectural Decay in Software Evolution, in: \textit{2012 16th European Conference on Software Maintenance and Reengineering}, IEEE, 2012, pp. 267–276, \url{https://doi.org/10.1109/csmr.2012.34}.

\keyedreference{WL133}
Guimaraes, Everton, Garcia, Alessandro, Figueiredo, Eduardo, Cai, Yuanfang, Prioritizing software anomalies with software metrics and architecture blueprints, in: \textit{2013 5th International Workshop on Modeling in Software Engineering (MiSE)}, IEEE, 2013, pp. 82–88, \url{https://doi.org/10.1109/mise.2013.6595301}.

\keyedreference{WL134}
O’Reilly, C., Morrow, P., Bustard, D., Lightweight prevention of architectural erosion, in: \textit{Sixth International Workshop on Principles of Software Evolution, 2003. Proceedings.}, IEEE, \url{https://doi.org/10.1109/iwpse.2003.1231211}.

\keyedreference{WL135}
Le, Duc, Architectural-Based Speculative Analysis to Predict Bugs in a Software System, in: \textit{2016 IEEE/ACM 38th International Conference on Software Engineering Companion (ICSE-C)}, 2016, pp. 807--810.

\keyedreference{WL136}
Li, Ruiyin, Soliman, Mohamed, Liang, Peng, Avgeriou, Paris, Symptoms of Architecture Erosion in Code Reviews: A Study of Two OpenStack Projects, in: \textit{2022 IEEE 19th International Conference on Software Architecture (ICSA)}, IEEE, 2022, pp. 24–35, \url{https://doi.org/10.1109/icsa53651.2022.00011}.

\keyedreference{WL137}
Castoldi, Andrea, Guazzoni, Chiara, Signorelli, Davide, Multiple-event sensitive DAQ for high-resolution position-sensing Silicon Drift Detectors: Hardware, firmware and software architecture, in: \textit{2007 IEEE Nuclear Science Symposium Conference Record}, IEEE, 2007, pp. 515–520, \url{https://doi.org/10.1109/nssmic.2007.4436382}.

\keyedreference{WL139}
Paing, Aye Myat Myat, Analysis of Availability Model Based on Software Aging in SDN Controllers with Rejuvenation, in: \textit{2020 IEEE Conference on Computer Applications(ICCA)}, IEEE, 2020, pp. 1–7, \url{https://doi.org/10.1109/icca49400.2020.9022818}.

\keyedreference{WL140}
Li, Ruiyin, Liang, Peng, Soliman, Mohamed, Avgeriou, Paris, Understanding Architecture Erosion: The Practitioners’ Perceptive, in: \textit{2021 IEEE/ACM 29th International Conference on Program Comprehension (ICPC)}, IEEE, 2021, pp. 311–322, \url{https://doi.org/10.1109/icpc52881.2021.00037}.

\keyedreference{WL141}
Mo, Ran, Garcia, Joshua, Cai, Yuanfang, Medvidovic, Nenad, Mapping architectural decay instances to dependency models, in: \textit{2013 4th International Workshop on Managing Technical Debt (MTD)}, IEEE, 2013, pp. 39–46, \url{https://doi.org/10.1109/mtd.2013.6608677}.

\keyedreference{WL142}
Jin, Wuxia, Zhang, Yuyun, Shang, Jiaowei, Hou, Yi, Fan, Ming, Liu, Ting, Identifying Code Changes for Architecture Decay via a Metric Forest Structure, in: \textit{2023 ACM/IEEE International Conference on Technical Debt (TechDebt)}, IEEE, 2023, pp. 62–71, \url{https://doi.org/10.1109/techdebt59074.2023.00014}.

\keyedreference{WL143}
Garcia, Joshua, Kouroshfar, Ehsan, Ghorbani, Negar, Malek, Sam, Forecasting Architectural Decay From Evolutionary History, \textit{IEEE Transactions on Software Engineering} 48 (7) (2022) 2439–2454, \url{https://doi.org/10.1109/tse.2021.3060068}.

\keyedreference{WL144}
Li, Bixin, Liao, Li, Si, Jingwen, A technique to evaluate software evolution based on architecture metric, in: \textit{2016 IEEE 14th International Conference on Software Engineering Research, Management and Applications (SERA)}, IEEE, 2016, pp. 1–8, \url{https://doi.org/10.1109/sera.2016.7516156}.

\keyedreference{WL145}
Yue, Jing, Wu, Xiaojun, Xue, Yunqing, Microservice Aging and Rejuvenation, in: \textit{2020 World Conference on Computing and Communication Technologies (WCCCT)}, IEEE, 2020, pp. 1–5, \url{https://doi.org/10.1109/wccct49810.2020.9170005}.

\keyedreference{WL146}
Luke, J., Bittorie, J.W., Cannon, W.J., Haldeman, D.G., A commercial off-the-shelf based replacement strategy for aging avionics computers, in: \textit{Proceedings of the IEEE 1998 National Aerospace and Electronics Conference. NAECON 1998. Celebrating 50 Years (Cat. No.98CH36185)}, IEEE, pp. 177–181, \url{https://doi.org/10.1109/naecon.1998.710113}.

\keyedreference{WL147}
Sababha, Belal H., Rawashdeh, Osamah A., Sa'deh, Waseem A., A real-time gracefully degrading avionics system for unmanned aerial vehicles, in: \textit{2012 IEEE National Aerospace and Electronics Conference (NAECON)}, 2012, pp. 171--177, \url{https://doi.org/10.1109/NAECON.2012.6531050}.

\keyedreference{WL148}
Andrade, Ermeson, Machida, Fumio, Analysis of Software Aging Impacts on Plant Anomaly Detection with Edge Computing, in: \textit{2019 IEEE International Symposium on Software Reliability Engineering Workshops (ISSREW)}, IEEE, 2019, pp. 204–210, \url{https://doi.org/10.1109/issrew.2019.00073}.

\keyedreference{WL149}
Le, Duc Minh, Karthik, Suhrid, Laser, Marcelo Schmitt, Medvidovic, Nenad, Architectural Decay as Predictor of Issue- and Change-Proneness, in: \textit{2021 IEEE 18th International Conference on Software Architecture (ICSA)}, IEEE, 2021, pp. 92–103, \url{https://doi.org/10.1109/icsa51549.2021.00017}.

\keyedreference{WL150}
Siozios, Kostas, Soudris, Dimitrios, A Methodology for Alleviating the Performance Degradation of TMR Solutions, \textit{IEEE Embedded Systems Letters} 2 (4) (2010) 111–114, \url{https://doi.org/10.1109/les.2010.2083632}.

\keyedreference{WL151}
Feng, Qiong, Cai, Yuanfang, Kazman, Rick, Cui, Di, Liu, Ting, Fang, Hongzhou, Active Hotspot: An Issue-Oriented Model to Monitor Software Evolution and Degradation, in: \textit{2019 34th IEEE/ACM International Conference on Automated Software Engineering (ASE)}, IEEE, 2019, pp. 986–997, \url{https://doi.org/10.1109/ase.2019.00095}.

\keyedreference{WL152}
Hanif, Muhammad Abdullah, Shafique, Muhammad, DNN-Life: An Energy-Efficient Aging Mitigation Framework for Improving the Lifetime of On-Chip Weight Memories in Deep Neural Network Hardware Architectures, in: \textit{2021 Design, Automation \& Test in Europe Conference \& Exhibition (DATE)}, IEEE, 2021, pp. 729–734, \url{https://doi.org/10.23919/date51398.2021.9473943}.

\keyedreference{WL153}
Khan, O., Kundu, S., A self-adaptive system architecture to address transistor aging, in: \textit{2009 Design, Automation \& Test in Europe Conference \& Exhibition}, IEEE, 2009, pp. 81–86, \url{https://doi.org/10.1109/date.2009.5090637}.

\keyedreference{WL154}
Oliveira, Felipe, Araujo, Jean, Matos, Rubens, Lins, Luan, Rodrigues, Andre, Maciel, Paulo, Experimental Evaluation of Software Aging Effects in a Container-Based Virtualization Platform, in: \textit{2020 IEEE International Conference on Systems, Man, and Cybernetics (SMC)}, IEEE, 2020, pp. 414–419, \url{https://doi.org/10.1109/smc42975.2020.9283358}.

\keyedreference{WL155}
Parri, Jacopo, Sampietro, Samuele, Scommegna, Leonardo, Vicario, Enrico, Evaluation of software aging in component-based Web Applications subject to soft errors over time, in: \textit{2021 IEEE International Symposium on Software Reliability Engineering Workshops (ISSREW)}, IEEE, 2021, pp. 25–32, \url{https://doi.org/10.1109/issrew53611.2021.00040}.

\keyedreference{WL156}
Corici, Marius, Buda, Teodora Sandra, Shrestha, Ranjan, Cau, Eleonora, Metin, Taner, Assem, Havtham, Practical Performance Degradation Mitigation Solution Using Anomaly Detection for Carrier-Grade Software Networks, in: \textit{2018 IEEE Conference on Standards for Communications and Networking (CSCN)}, IEEE, 2018, pp. 1–6, \url{https://doi.org/10.1109/cscn.2018.8581818}.

\keyedreference{WL157}
Rahimi, Abbas, Benini, Luca, Gupta, Rajesh K., Aging-aware compiler-directed VLIW assignment for GPGPU architectures, in: \textit{Proceedings of the 50th Annual Design Automation Conference}, ACM, 2013, pp. 1–6, \url{https://doi.org/10.1145/2463209.2488754}.

\keyedreference{WL158}
Shelton, C.P., Koopman, P., Nace, W., A framework for scalable analysis and design of system-wide graceful degradation in distributed embedded systems, in: \textit{Proceedings of the Eighth International Workshop on Object-Oriented Real-Time Dependable Systems, 2003. (WORDS 2003).}, IEEE, pp. 156–163, \url{https://doi.org/10.1109/words.2003.1218078}.

\keyedreference{WL159}
Bai, Jing, Chang, Xiaolin, Machida, Fumio, Trivedi, Kishor S., Understanding Container-Based Services Under Software Aging: Dependability and Performance Views, \textit{IEEE Transactions on Sustainable Computing} 10 (3) (2025) 562–575, \url{https://doi.org/10.1109/tsusc.2024.3506213}.

\keyedreference{WL160}
Bai, Jing, Chang, Xiaolin, Machida, Fumio, Jiang, Lili, Han, Zhen, Trivedi, Kishor S., Impact of Service Function Aging on the Dependability for MEC Service Function Chain, \textit{IEEE Transactions on Dependable and Secure Computing} 20 (4) (2023) 2811–2824, \url{https://doi.org/10.1109/tdsc.2022.3150782}.

\keyedreference{WL161}
Beierlieb, Lukas, Iffl\"{a}nder, Lukas, Milenkoski, Aleksandar, Gon\c{c}alves, Charles F., Antunes, Nuno, Kounev, Samuel, Towards Testing the Software Aging Behavior of Hypervisor Hypercall Interfaces, in: \textit{2019 IEEE International Symposium on Software Reliability Engineering Workshops (ISSREW)}, IEEE, 2019, pp. 218–224, \url{https://doi.org/10.1109/issrew.2019.00075}.

\keyedreference{WL162}
Karray, Hassen, Paulitsch, Michael, Koppenhoefer, Bernd, Geiger, Dietmar, Design and implementation of a degraded vision landing aid application on a multicore processor architecture for safety-critical application, in: \textit{16th IEEE International Symposium on Object/component/service-oriented Real-time distributed Computing (ISORC 2013)}, IEEE, 2013, pp. 1–8, \url{https://doi.org/10.1109/isorc.2013.6913229}.

\keyedreference{WL163}
Luis Silva, Henrique Madeira, Software Aging and Rejuvenation in a SOAP-based Server, in: \textit{Fifth IEEE International Symposium on Network Computing and Applications (NCA’06)}, IEEE, pp. 56–65, \url{https://doi.org/10.1109/nca.2006.51}.

\keyedreference{WL164}
Hua, Xiayu, Guo, Chunhui, Wu, Hao, Lautner, Douglas, Ren, Shangping, Schedulability Analysis for Real-Time Task Set on Resource with Performance Degradation and Dual-Level Periodic Rejuvenations, \textit{IEEE Transactions on Computers} 66 (3) (2017) 553–559, \url{https://doi.org/10.1109/tc.2016.2602833}.

\keyedreference{WL165}
Weiss, Philipp, Weichslgartner, Andreas, Reimann, Felix, Steinhorst, Sebastian, Fail-Operational Automotive Software Design Using Agent-Based Graceful Degradation, in: \textit{2020 Design, Automation \& Test in Europe Conference \& Exhibition (DATE)}, IEEE, 2020, pp. 1169–1174, \url{https://doi.org/10.23919/date48585.2020.9116322}.

\keyedreference{WL166}
Herold, Sebastian, Mair, Matthias, Rausch, Andreas, Schindler, Ingrid, Checking Conformance with Reference Architectures: A Case Study, in: \textit{2013 17th IEEE International Enterprise Distributed Object Computing Conference}, IEEE, 2013, pp. 71–80, \url{https://doi.org/10.1109/edoc.2013.17}.

\keyedreference{WL167}
Deissenboeck, Florian, Heinemann, Lars, Hummel, Benjamin, Juergens, Elmar, Flexible architecture conformance assessment with ConQAT, in: \textit{Proceedings of the 32nd ACM/IEEE International Conference on Software Engineering - Volume 2}, ACM, 2010, pp. 247–250, \url{https://doi.org/10.1145/1810295.1810343}.

\keyedreference{WL168}
Tan, Xueyong, Liu, Jing, GRCEM: Generating Optimal Software Rejuvenation Strategies for Cloud-Edge Collaborative Systems Based on MADRL, in: \textit{2025 IEEE 49th Annual Computers, Software, and Applications Conference (COMPSAC)}, IEEE, 2025, pp. 1360–1369, \url{https://doi.org/10.1109/compsac65507.2025.00170}.

\keyedreference{WL169}
De Stefano, Manuel, Pecorelli, Fabiano, Di Nucci, Dario, De Lucia, Andrea, A preliminary evaluation on the relationship among architectural and test smells, in: \textit{2022 IEEE 22nd International Working Conference on Source Code Analysis and Manipulation (SCAM)}, IEEE, 2022, pp. 66–70, \url{https://doi.org/10.1109/scam55253.2022.00013}.

\keyedreference{WL170}
Zhao, Ming, Jianfeng Yang, A DCA-based method for software prognostics and health management, in: \textit{Proceedings of the IEEE 2012 Prognostics and System Health Management Conference (PHM-2012 Beijing)}, IEEE, 2012, pp. 1–5, \url{https://doi.org/10.1109/phm.2012.6228830}.

\keyedreference{WL171}
Shahbazian, Arman, Nam, Daye, Medvidovic, Nenad, Toward Predicting Architectural Significance of Implementation Issues, in: \textit{2018 IEEE/ACM 15th International Conference on Mining Software Repositories (MSR)}, 2018, pp. 215--219.

\keyedreference{WL172}
Xinyi Dong, Godfrey, M.W., Identifying Architectural Change Patterns in Object-Oriented Systems, in: \textit{2008 16th IEEE International Conference on Program Comprehension}, IEEE, 2008, pp. 33–42, \url{https://doi.org/10.1109/icpc.2008.19}.

\keyedreference{WL173}
Rachow, Paula, Riebisch, Matthias, An architecture smell knowledge base for managing architecture technical debt, in: \textit{Proceedings of the International Conference on Technical Debt}, ACM, 2022, pp. 1–10, \url{https://doi.org/10.1145/3524843.3528092}.

\keyedreference{WL174}
Macia, I., Garcia, A., Chavez, C., von Staa, A., Enhancing the Detection of Code Anomalies with Architecture-Sensitive Strategies, in: \textit{2013 17th European Conference on Software Maintenance and Reengineering}, IEEE, 2013, pp. 177–186, \url{https://doi.org/10.1109/csmr.2013.27}.

\keyedreference{WL175}
Assunção, Wesley K. G., Software Modernization: from industry needs to developers’ perception, in: \textit{2024 IEEE/ACM Workshop on Software Engineering Challenges in Financial Firms (FinanSE)}, 2024, pp. 21--22.

\keyedreference{WL176}
Peercy, M., Banerjee, P., Design and analysis of software reconfiguration strategies for hypercube multicomputers under multiple faults, in: \textit{[1992] Digest of Papers. FTCS-22: The Twenty-Second International Symposium on Fault-Tolerant Computing}, IEEE, pp. 448–455, \url{https://doi.org/10.1109/ftcs.1992.243590}.

\keyedreference{WL177}
Mendoza, Camilo, Garcés, Kelly, Casallas, Rubby, Bocanegra, José, Detecting Architectural Issues During the Continuous Integration Pipeline, in: \textit{2019 ACM/IEEE 22nd International Conference on Model Driven Engineering Languages and Systems Companion (MODELS-C)}, IEEE, 2019, pp. 589–597, \url{https://doi.org/10.1109/models-c.2019.00090}.

\keyedreference{WL178}
Xinyuan, Fan, Guozhi, Xu, Rendong, Ying, Hao, Zhang, Modeling and analysis of software rejuvenation on dispatcher-based web server farms, \textit{Journal of Systems Engineering and Electronics} 16 (2) (2005) 460--465.

\keyedreference{WL179}
Pigazzini, Ilaria, Di Nucci, Dario, Fontana, Francesca Arcelli, Belotti, Marco, Exploiting dynamic analysis for architectural smell detection: a preliminary study, in: \textit{2022 48th Euromicro Conference on Software Engineering and Advanced Applications (SEAA)}, IEEE, 2022, pp. 282–289, \url{https://doi.org/10.1109/seaa56994.2022.00051}.

\keyedreference{WL180}
Hull, Andrew, Rearchitecting legacy LRUs to support distributed IMA a strategy to integrate HMD systems into legacy aircraft, in: \textit{2011 IEEE/AIAA 30th Digital Avionics Systems Conference}, IEEE, 2011, pp. 7A4--1--7A4--10, \url{https://doi.org/10.1109/dasc.2011.6096115}.

\keyedreference{WL181}
Sarkar, Santonu, Rama, Girish Maskeri, R., Shubha, A Method for Detecting and Measuring Architectural Layering Violations in Source Code, in: \textit{2006 13th Asia Pacific Software Engineering Conference (APSEC’06)}, IEEE, 2006, pp. 165–172, \url{https://doi.org/10.1109/apsec.2006.7}.

\keyedreference{WL182}
Martini, Antonio, Bosch, Jan, Architectural Technical Debt in Embedded Systems, 2019, \url{https://doi.org/10.1002/9781119513957.ch4}.

\keyedreference{WL183}
Gnoyke, Philipp, Schulze, Sandro, Kr\"{u}ger, Jacob, On Developing and Improving Tools for Architecture-Smell Tracking in Java Systems, in: \textit{2023 IEEE 23rd International Working Conference on Source Code Analysis and Manipulation (SCAM)}, IEEE, 2023, pp. 248–253, \url{https://doi.org/10.1109/scam59687.2023.00034}.

\keyedreference{WL184}
Tan, Yuanlong, Chen, Shuoshuo, Emmerson, Steve, Zhang, Yizhe, Veeraraghavan, Malathi, Advances in Reliable File-Stream Multicasting over Multi-Domain Software Defined Networks (SDN), in: \textit{2019 28th International Conference on Computer Communication and Networks (ICCCN)}, IEEE, 2019, pp. 1–11, \url{https://doi.org/10.1109/icccn.2019.8847110}.

\keyedreference{WL185}
Diaz-Pace, Jorge Andres, Tommasel, Antonela, Godoy, Daniela, [Research Paper] Towards Anticipation of Architectural Smells Using Link Prediction Techniques, in: \textit{2018 IEEE 18th International Working Conference on Source Code Analysis and Manipulation (SCAM)}, IEEE, 2018, pp. 62–71, \url{https://doi.org/10.1109/scam.2018.00015}.

\keyedreference{WL186}
Lee, Ganghee, Chang, Kyungwook, Choi, Kiyoung, Automatic mapping of control-intensive kernels onto coarse-grained reconfigurable array architecture with speculative execution, in: \textit{2010 IEEE International Symposium on Parallel \& Distributed Processing, Workshops and Phd Forum (IPDPSW)}, IEEE, 2010, pp. 1–4, \url{https://doi.org/10.1109/ipdpsw.2010.5470746}.

\keyedreference{WL187}
Jayaseelan, Ramkumar, Mitra, Tulika, Dynamic thermal management via architectural adaptation, in: \textit{2009 46th ACM/IEEE Design Automation Conference}, 2009, pp. 484--489.

\keyedreference{WL188}
Knodel, Jens, Muthig, Dirk, Rost, Dominik, Constructive architecture compliance checking — an experiment on support by live feedback, in: \textit{2008 IEEE International Conference on Software Maintenance}, IEEE, 2008, pp. 287–296, \url{https://doi.org/10.1109/icsm.2008.4658077}.

\keyedreference{WL189}
Tanaphalan, K., Autonomous control system for satellite formation flying, in: \textit{IEE Seminar on Autonomous Agents in Control}, IEE, 2005, pp. 85–91, \url{https://doi.org/10.1049/ic:20050185}.

\keyedreference{WL190}
Ozols, J., Pedersen, D.A., Fault tolerant virtual machine architecture for advanced avionics applications, in: \textit{[1993 Proceedings] AIAA/IEEE Digital Avionics Systems Conference}, IEEE, pp. 231–236, \url{https://doi.org/10.1109/dasc.1993.283542}.

\keyedreference{WL191}
Kyungwook Chang, Choi, Kiyoung, Mapping control intensive kernels onto coarse-grained reconfigurable array architecture, in: \textit{2008 International SoC Design Conference}, IEEE, 2008, pp. I--362--I–365, \url{https://doi.org/10.1109/socdc.2008.4815647}.

\keyedreference{WL192}
Dantas, Leandro Poloni, de Azevedo, Rodolfo J., Gimenez, Salvador Pinillos, A Novel Processor Architecture With a Hardware Microkernel to Improve the Performance of Task-Based Systems, \textit{IEEE Embedded Systems Letters} 11 (2) (2019) 46–49, \url{https://doi.org/10.1109/les.2018.2864094}.

\keyedreference{WL193}
Caracciolo, Andrea, Lungu, Mircea Filip, Nierstrasz, Oscar, A Unified Approach to Architecture Conformance Checking, in: \textit{2015 12th Working IEEE/IFIP Conference on Software Architecture}, IEEE, 2015, pp. 41–50, \url{https://doi.org/10.1109/wicsa.2015.11}.

\keyedreference{WL194}
De Simone, L., Di Mauro, M., Natella, R., Postiglione, F., Performability Evaluation of Softwarized Multi-Tenant 5G Architectures with Rejuvenation, in: \textit{2024 14th International Workshop on Resilient Networks Design and Modeling (RNDM)}, IEEE, 2024, pp. 1–6, \url{https://doi.org/10.1109/rndm64105.2024.10820438}.

\keyedreference{WL195}
Chatterjee, Debapriya, Koyfman, Anatoly, Morad, Ronny, Ziv, Avi, Bertacco, Valeria, Checking architectural outputs instruction-by-instruction on acceleration platforms, in: \textit{Proceedings of the 49th Annual Design Automation Conference}, ACM, 2012, pp. 955–961, \url{https://doi.org/10.1145/2228360.2228531}.

\keyedreference{WL196}
Basu, Kashinath, Hamdullah, Aws, Ball, Frank, Architecture of a Cloud-based Fault-Tolerant Control Platform for improving the QoS of Social Multimedia Applications on SD-WAN, in: \textit{2020 13th International Conference on Communications (COMM)}, IEEE, 2020, pp. 495–500, \url{https://doi.org/10.1109/comm48946.2020.9142038}.

\keyedreference{WL197}
Liu, Zhuanzhuan, Liu, Yiming, Tan, Xueyong, Liu, Jing, Determine When and How to Perform Edge Rejuvenation Effectively for Cloud-Edge Collaborative System, in: \textit{2024 IEEE 48th Annual Computers, Software, and Applications Conference (COMPSAC)}, IEEE, 2024, pp. 1382–1387, \url{https://doi.org/10.1109/compsac61105.2024.00182}.

\keyedreference{WL198}
Krikhaar, R., Postma, A., Sellink, A., Stroucken, M., Verhoef, C., A two-phase process for software architecture improvement, in: \textit{Proceedings IEEE International Conference on Software Maintenance - 1999 (ICSM’99). “Software Maintenance for Business Change” (Cat. No.99CB36360)}, IEEE, 1999, pp. 371–380, \url{https://doi.org/10.1109/icsm.1999.792635}.

\keyedreference{WL199}
Bertran, Isela Macia, Detecting architecturally-relevant code smells in evolving software systems, in: \textit{Proceedings of the 33rd International Conference on Software Engineering}, ACM, 2011, pp. 1090–1093, \url{https://doi.org/10.1145/1985793.1986003}.

\keyedreference{WL200}
Hochstein, L., Lindvall, M., Diagnosing architectural degeneration, in: \textit{28th Annual NASA Goddard Software Engineering Workshop, 2003. Proceedings.}, IEEE, 2003, pp. 137–142, \url{https://doi.org/10.1109/sew.2003.1270736}.

\keyedreference{WL201}
Sousa, Leonardo, Oizumi, Willian, Garcia, Alessandro, Oliveira, Anderson, Cedrim, Diego, Lucena, Carlos, When Are Smells Indicators of Architectural Refactoring Opportunities: A Study of 50 Software Projects, in: \textit{Proceedings of the 28th International Conference on Program Comprehension}, ACM, 2020, pp. 354–365, \url{https://doi.org/10.1145/3387904.3389276}.

\keyedreference{WL202}
Tran, J.B., Godfrey, M.W., Lee, E.H.S., Holt, R.C., Architectural repair of open source software, in: \textit{Proceedings IWPC 2000. 8th International Workshop on Program Comprehension}, IEEE Comput. Soc, pp. 48–59, \url{https://doi.org/10.1109/wpc.2000.852479}.

\keyedreference{WL203}
Torres, Romina, Bencomo, Nelly, Astudillo, Hernan, Addressing the QoS drift in specification models of self-adaptive service-based systems, in: \textit{2013 2nd International Workshop on Realizing Artificial Intelligence Synergies in Software Engineering (RAISE)}, IEEE, 2013, pp. 28–34, \url{https://doi.org/10.1109/raise.2013.6615201}.

\keyedreference{WL204}
Sierra, Giancarlo, Tahmid, Ahmad, Shihab, Emad, Tsantalis, Nikolaos, Is Self-Admitted Technical Debt a Good Indicator of Architectural Divergences?, in: \textit{2019 IEEE 26th International Conference on Software Analysis, Evolution and Reengineering (SANER)}, IEEE, 2019, pp. 534–543, \url{https://doi.org/10.1109/saner.2019.8667999}.

\keyedreference{WL206}
Verdecchia, Roberto, Malavolta, Ivano, Lago, Patricia, Architectural technical debt identification: the research landscape, Association for Computing Machinery, New York, NY, USA, 2018, pp. 11–20, \url{https://doi.org/10.1145/3194164.3194176}.

\keyedreference{WL207}
Witte, Thomas, Tichy, Matthias, Checking consistency of robot software architectures in ROS, in: \textit{Proceedings of the 1st International Workshop on Robotics Software Engineering}, Association for Computing Machinery, New York, NY, USA, 2018, pp. 1–8, \url{https://doi.org/10.1145/3196558.3196559}.

\keyedreference{WL208}
Liu, Jingwen, Jin, Wuxia, Zhou, Junhui, Feng, Qiong, Fan, Ming, Wang, Haijun, Liu, Ting, 3Erefactor: Effective, Efficient and Executable Refactoring Recommendation for Software Architectural Consistency, \textit{IEEE Transactions on Software Engineering} 50 (10) (2024) 2633–2655, \url{https://doi.org/10.1109/tse.2024.3449564}.

\keyedreference{WL209}
Hoch, Ralph, Kaindl, Hermann, Popp, Roman, Zeidler, Christian, Aligning Architectures of Business and Software: Software Driven by Business Process Models and Its User Interface, in: \textit{2016 49th Hawaii International Conference on System Sciences (HICSS)}, IEEE, 2016, pp. 4556–4565, \url{https://doi.org/10.1109/hicss.2016.566}.

\keyedreference{WL210}
Kramer, J., Magee, J., Analysing dynamic change in software architectures: a case study, in: \textit{Proceedings. Fourth International Conference on Configurable Distributed Systems (Cat. No.98EX159)}, IEEE Comput. Soc, pp. 91–100, \url{https://doi.org/10.1109/cds.1998.675762}.

\keyedreference{WL211}
Moonzoo Kim, Jaejoon Lee, Youngjin Hong, Seokwon Bang, Kyo Chul Kang, Re-engineering software architecture of home service robots: a case study, in: \textit{Proceedings. 27th International Conference on Software Engineering, 2005. ICSE 2005.}, IEEe, pp. 505–513, \url{https://doi.org/10.1109/icse.2005.1553596}.

\keyedreference{WL212}
Julier, S., King, R., Colbert, B., Durbin, J., Rosenblum, L., The software architecture of a real-time battlefield visualization virtual environment, in: \textit{Proceedings IEEE Virtual Reality (Cat. No. 99CB36316)}, IEEE Comput. Soc, pp. 29–36, \url{https://doi.org/10.1109/vr.1999.756920}.

\keyedreference{WL213}
Rahimi, Razieh, Khosravi, Ramtin, Architecture conformance checking of multi-language applications, in: \textit{ACS/IEEE International Conference on Computer Systems and Applications - AICCSA 2010}, IEEE, 2010, pp. 1–8, \url{https://doi.org/10.1109/aiccsa.2010.5587025}.

\keyedreference{WL214}
Wei-Kuo Liao, Pang-Hsin Shih, Architecture of proxy partial caching using HTTP for supporting interactive video and cache consistency, in: \textit{Proceedings. Eleventh International Conference on Computer Communications and Networks}, IEEE, pp. 216–221, \url{https://doi.org/10.1109/icccn.2002.1043069}.

\keyedreference{WL215}
Jahic, Jasmin, Kumar, Varun, Antonino, Pablo Oliveira, Wirrer, Gerhard, Testing the Implementation of Concurrent AUTOSAR Drivers Against Architecture Decisions, in: \textit{2019 IEEE International Conference on Software Architecture (ICSA)}, IEEE, 2019, pp. 171–180, \url{https://doi.org/10.1109/icsa.2019.00026}.

\keyedreference{WL216}
Chen, Xinghua, Chen, Jinchang, Yu, Hao, Ning, Yuanyuan, Huang, Lixian, Digit-Module Decoupling Platform Architecture for Safety Automation Equipment, in: \textit{2023 8th Asia Conference on Power and Electrical Engineering (ACPEE)}, IEEE, 2023, pp. 97–103, \url{https://doi.org/10.1109/acpee56931.2023.10135693}.

\keyedreference{WL217}
Hanai, Ryo, Saito, Hajime, Nakabo, Yoshihiro, Fujiwara, Kiyoshi, Ogure, Takuya, Mizuguchi, Daichi, Homma, Keiko, Ohba, Kohtaro, Proposal of architecture and implementation process for IEC61508 compliant, dependable robot systems, in: \textit{2012 IEEE International Conference on Robotics and Biomimetics (ROBIO)}, IEEE, 2012, pp. 1218–1223, \url{https://doi.org/10.1109/robio.2012.6491136}.

\keyedreference{WL219}
Zhou, Yu, A runtime architecture-based approach for the dynamic evolution of distributed component-based systems, in: \textit{Companion Proceedings of the 30th International Conference on Software Engineering}, Association for Computing Machinery, New York, NY, USA, 2008, pp. 979–982, \url{https://doi.org/10.1145/1370175.1370217}.

\keyedreference{WL220}
Scutari, Marco, Bayesian Network Constraint-Based Structure Learning Algorithms: Parallel and Optimized Implementations in the bnlearn R Package, \textit{Journal of Statistical Software} 77 (2) (2017), \url{https://doi.org/10.18637/jss.v077.i02}.

\keyedreference{WL221}
Griswold, W.G., Notkin, D., Architectural tradeoffs for a meaning-preserving program restructuring tool, \textit{IEEE Transactions on Software Engineering} 21 (4) (1995) 275–287, \url{https://doi.org/10.1109/32.385967}.

\keyedreference{WL222}
Silva, Cícero A., Aquino, Gibeon S., Melo, Sávio R. M., Egídio, Dannylo J. B., A Fog Computing-Based Architecture for Medical Records Management, \textit{Wireless Communications and Mobile Computing} 2019 (2019) 1–16, \url{https://doi.org/10.1155/2019/1968960}.

\keyedreference{WL223}
Mazkatli, Manar, Monschein, David, Armbruster, Martin, Heinrich, Robert, Koziolek, Anne, Continuous integration of architectural performance models with parametric dependencies – the CIPM approach, \textit{Automated Software Engineering} 32 (2) (2025), \url{https://doi.org/10.1007/s10515-025-00521-9}.

\keyedreference{WL224}
Vanya, Adam, Klusener, Steven, Premraj, Rahul, van Vliet, Hans, Supporting software architects to improve their software system’s decomposition – lessons learned, \textit{Journal of Software: Evolution and Process} 25 (3) (2011) 219–232, \url{https://doi.org/10.1002/smr.574}.

\keyedreference{WL225}
Jaktman, Catherine, Leaney, John, Liu, Ming, Structural Analysis of the Software Architecture — A Maintenance Assessment Case Study, 1999, pp. 455--470, \url{https://doi.org/10.1007/978-0-387-35563-4_26}.

\keyedreference{WL226}
Zhang, Yuanpeng, Hu, Xing, Chen, Xi, Yuan, Zhihang, Li, Cong, Zhu, Jingchen, Wang, Zhao, Zhang, Chenguang, Si, Xin, Gao, Wei, Wu, Qiang, Wang, Runsheng, Sun, Guangyu, AIM: Software and Hardware Co-design for Architecture-level IR-drop Mitigation in High-performance PIM, in: \textit{Proceedings of the 52nd Annual International Symposium on Computer Architecture}, Association for Computing Machinery, New York, NY, USA, 2025, pp. 849–866, \url{https://doi.org/10.1145/3695053.3730987}.

\keyedreference{WL227}
Olsson, Tobias, Toll, Daniel, Ericsson, Morgan, Wingkvist, Anna, Evaluation of an architectural conformance checking software service, in: \textit{Proccedings of the 10th European Conference on Software Architecture Workshops}, Association for Computing Machinery, New York, NY, USA, 2016, \url{https://doi.org/10.1145/2993412.3003391}.

\keyedreference{WL228}
Nicolodi, Luciane Baldo, Colanzi, Thelma Elita, Assun\c{c}\~{a}o, Wesley K. G., Architectural Feature Re-Modularization for Software Product Line Evolution, in: \textit{Proceedings of the 14th Brazilian Symposium on Software Components, Architectures, and Reuse}, Association for Computing Machinery, New York, NY, USA, 2020, pp. 31–40, \url{https://doi.org/10.1145/3425269.3425271}.

\keyedreference{WL229}
Fittkau, Florian, Stelzer, Phil, Hasselbring, Wilhelm, Live Visualization of Large Software Landscapes for Ensuring Architecture Conformance, in: \textit{Proceedings of the 2014 European Conference on Software Architecture Workshops}, Association for Computing Machinery, New York, NY, USA, 2014, \url{https://doi.org/10.1145/2642803.2642831}.

\keyedreference{WL230}
Goldstein, Maayan, Segall, Itai, Automatic and Continuous Software Architecture Validation, in: \textit{2015 IEEE/ACM 37th IEEE International Conference on Software Engineering}, 2015, pp. 59--68, \url{https://doi.org/10.1109/ICSE.2015.135}.

\keyedreference{WL231}
Cool, Benjamin, Knieke, Christoph, Rausch, Andreas, Schindler, Mirco, Strasser, Arthur, Vogel, Martin, Brox, Oliver, Jauns-Seyfried, Stefanie, From product architectures to a managed automotive software product line architecture, in: \textit{Proceedings of the 31st Annual ACM Symposium on Applied Computing}, Association for Computing Machinery, New York, NY, USA, 2016, pp. 1350–1353, \url{https://doi.org/10.1145/2851613.2851964}.

\keyedreference{WL232}
Olsson, Tobias, Ericsson, Morgan, Wingkvist, Anna, The relationship of code churn and architectural violations in the open source software JabRef, in: \textit{Proceedings of the 11th European Conference on Software Architecture: Companion Proceedings}, Association for Computing Machinery, New York, NY, USA, 2017, pp. 152–158, \url{https://doi.org/10.1145/3129790.3129810}.

\keyedreference{WL233}
Medvidovic, Nenad, Adapting our view of software adaptation: an architectural perspective (keynote), in: \textit{Proceedings of the 9th International Symposium on Software Engineering for Adaptive and Self-Managing Systems}, Association for Computing Machinery, New York, NY, USA, 2014, pp. 5–6, \url{https://doi.org/10.1145/2593929.2593949}.

\keyedreference{WL234}
C\'{a}mara, Javier, Correia, Pedro, de Lemos, Rog\'{e}rio, Vieira, Marco, Empirical resilience evaluation of an architecture-based self-adaptive software system, in: \textit{Proceedings of the 10th International ACM Sigsoft Conference on Quality of Software Architectures}, Association for Computing Machinery, New York, NY, USA, 2014, pp. 63–72, \url{https://doi.org/10.1145/2602576.2602577}.

\keyedreference{WL235}
Schmitt Laser, Marcelo, Medvidovic, Nenad, Le, Duc Minh, Garcia, Joshua, ARCADE: an extensible workbench for architecture recovery, change, and decay evaluation, in: \textit{Proceedings of the 28th ACM Joint Meeting on European Software Engineering Conference and Symposium on the Foundations of Software Engineering}, Association for Computing Machinery, New York, NY, USA, 2020, pp. 1546–1550, \url{https://doi.org/10.1145/3368089.3417941}.

\keyedreference{WL236}
Maribondo, Paulo D S, Fernandes, Natalia C, Avoiding Voice Traffic Degradation in IP Enterprise Networks Using CAoS, in: \textit{Proceedings of the 2016 Workshop on Fostering Latin-American Research in Data Communication Networks}, Association for Computing Machinery, New York, NY, USA, 2016, pp. 34–36, \url{https://doi.org/10.1145/2940116.2940129}.

\keyedreference{WL237}
Neto, Valdemar V. Graciano, Manzano, Wallace, Garc\'{e}s, Lina, Guessi, Milena, Oliveira, Brauner, Volpato, Tiago, Nakagawa, Elisa Yumi, Back-SoS: towards a model-based approach to address architectural drift in systems-of-systems, in: \textit{Proceedings of the 33rd Annual ACM Symposium on Applied Computing}, Association for Computing Machinery, New York, NY, USA, 2018, pp. 1461–1463, \url{https://doi.org/10.1145/3167132.3167422}.

\keyedreference{WL238}
Pinto, Felipe, Kulesza, Uir\'{a}, Silva, Leo, Guerra, Eduardo, A Metadata-based Framework for Quality Attribute Degradation Analysis in Web Systems, in: \textit{Proceedings of the 20th Brazilian Symposium on Multimedia and the Web}, Association for Computing Machinery, New York, NY, USA, 2014, pp. 171–178, \url{https://doi.org/10.1145/2664551.2664570}.

\keyedreference{WL239}
Perez, Boris, Correal, Dario, Astudillo, Hernan, A Proposed Model-Driven Approach to Manage Architectural Technical Debt Life Cycle, in: \textit{2019 IEEE/ACM International Conference on Technical Debt (TechDebt)}, 2019, pp. 73--77, \url{https://doi.org/10.1109/TechDebt.2019.00025}.

\keyedreference{WL240}
de Freitas Apolin\'{a}rio, Daniel Rodrigo, de Fran\c{c}a, Breno Bernard Nicolau, Towards a method for monitoring the coupling evolution of microservice-based architectures, in: \textit{Proceedings of the 14th Brazilian Symposium on Software Components, Architectures, and Reuse}, Association for Computing Machinery, New York, NY, USA, 2020, pp. 71–80, \url{https://doi.org/10.1145/3425269.3425273}.

\keyedreference{WL241}
Pigazzini, Ilaria, Automatic detection of architectural bad smells through semantic representation of code, in: \textit{Proceedings of the 13th European Conference on Software Architecture - Volume 2}, Association for Computing Machinery, New York, NY, USA, 2019, pp. 59–62, \url{https://doi.org/10.1145/3344948.3344951}.

\keyedreference{WL242}
Haitzer, Thomas, Navarro, Elena, Zdun, Uwe, Architecting for decision making about code evolution, in: \textit{Proceedings of the 2015 European Conference on Software Architecture Workshops}, Association for Computing Machinery, New York, NY, USA, 2015, \url{https://doi.org/10.1145/2797433.2797487}.

\keyedreference{WL243}
Guimar\~{a}es, Everton, Garcia, Alessandro, Cai, Yuanfang, Architecture-sensitive heuristics for prioritizing critical code anomalies, in: \textit{Proceedings of the 14th International Conference on Modularity}, Association for Computing Machinery, New York, NY, USA, 2015, pp. 68–80, \url{https://doi.org/10.1145/2724525.2724567}.

\keyedreference{WL244}
Rybalkin, Vladimir, Bukhari, Syed Saqib, Ghaffar, Muhammad Mohsin, Ghafoor, Aqib, Wehn, Norbert, Dengel, Andreas, iDocChip: A Configurable Hardware Architecture for Historical Document Image Processing: Percentile Based Binarization, in: \textit{Proceedings of the ACM Symposium on Document Engineering 2018}, Association for Computing Machinery, New York, NY, USA, 2018, \url{https://doi.org/10.1145/3209280.3209538}.

\keyedreference{WL245}
Nagamine, Motoi, Nakajima, Tsuyoshi, Kuno, Noriyoshi, A case study of applying software product line engineering to the air conditioner domain, in: \textit{Proceedings of the 20th International Systems and Software Product Line Conference}, Association for Computing Machinery, New York, NY, USA, 2016, pp. 220–226, \url{https://doi.org/10.1145/2934466.2934489}.

\keyedreference{WL246}
Fontana, Francesca Arcelli, Roveda, Riccardo, Zanoni, Marco, Tool support for evaluating architectural debt of an existing system: an experience report, in: \textit{Proceedings of the 31st Annual ACM Symposium on Applied Computing}, Association for Computing Machinery, New York, NY, USA, 2016, pp. 1347–1349, \url{https://doi.org/10.1145/2851613.2851963}.

\keyedreference{WL247}
Ferreira, Manuele, Barbosa, Eiji, Macia, Isela, Arcoverde, Roberta, Garcia, Alessandro, Detecting architecturally-relevant code anomalies: a case study of effectiveness and effort, in: \textit{Proceedings of the 29th Annual ACM Symposium on Applied Computing}, Association for Computing Machinery, New York, NY, USA, 2014, pp. 1158–1163, \url{https://doi.org/10.1145/2554850.2555036}.

\keyedreference{WL248}
Heinrich, Robert, Architectural Run-time Models for Performance and Privacy Analysis in Dynamic Cloud Applications, \textit{SIGMETRICS Perform. Eval. Rev.} 43 (4) (2016) 13–22, \url{https://doi.org/10.1145/2897356.2897359}.

\keyedreference{WL249}
Yonglin Lei, Ning Zhu, Jian Yao, Zhi Zhu, Sarjoughian, Hessam S., Model-architecture oriented combat system effectiveness simulation, in: \textit{2015 Winter Simulation Conference (WSC)}, 2015, pp. 3190--3191, \url{https://doi.org/10.1109/WSC.2015.7408464}.

\keyedreference{WL250}
Larsson, Stig, Wall, Anders, Wallin, Peter, Assessing the influence on processes when evolving the software architecture, in: \textit{Ninth International Workshop on Principles of Software Evolution: In Conjunction with the 6th ESEC/FSE Joint Meeting}, Association for Computing Machinery, New York, NY, USA, 2007, pp. 59–66, \url{https://doi.org/10.1145/1294948.1294963}.

\keyedreference{WL251}
Mishra, Prabhat, Shrivastava, Aviral, Dutt, Nikil, Architecture description language (ADL)-driven software toolkit generation for architectural exploration of programmable SOCs, \textit{ACM Trans. Des. Autom. Electron. Syst.} 11 (3) (2004) 626–658, \url{https://doi.org/10.1145/1142980.1142985}.

\keyedreference{WL252}
Rosik, Jacek, Le Gear, Andrew, Buckley, Jim, Ali Babar, Muhammad, An industrial case study of architecture conformance, in: \textit{Proceedings of the Second ACM-IEEE International Symposium on Empirical Software Engineering and Measurement}, Association for Computing Machinery, New York, NY, USA, 2008, pp. 80–89, \url{https://doi.org/10.1145/1414004.1414019}.

\keyedreference{WL253}
Ghafari, Mohammad, Jamshidi, Pooyan, Shahbazi, Saeed, Haghighi, Hassan, An architectural approach to ensure globally consistent dynamic reconfiguration of component-based systems, in: \textit{Proceedings of the 15th ACM SIGSOFT Symposium on Component Based Software Engineering}, Association for Computing Machinery, New York, NY, USA, 2012, pp. 177–182, \url{https://doi.org/10.1145/2304736.2304765}.

\keyedreference{WL254}
Schanz, Travis, Izurieta, Clemente, Object oriented design pattern decay: a taxonomy, in: \textit{Proceedings of the 2010 ACM-IEEE International Symposium on Empirical Software Engineering and Measurement}, Association for Computing Machinery, New York, NY, USA, 2010, \url{https://doi.org/10.1145/1852786.1852796}.

\keyedreference{WL255}
Dabrowski, Christopher, Mills, Kevin, Elder, Jesse, Understanding consistency maintenance in service discovery architectures during communication failure, in: \textit{Proceedings of the 3rd International Workshop on Software and Performance}, Association for Computing Machinery, New York, NY, USA, 2002, pp. 168–178, \url{https://doi.org/10.1145/584369.584396}.

\keyedreference{WL256}
Macia, Isela, Garcia, Joshua, Popescu, Daniel, Garcia, Alessandro, Medvidovic, Nenad, von Staa, Arndt, Are automatically-detected code anomalies relevant to architectural modularity? an exploratory analysis of evolving systems, in: \textit{Proceedings of the 11th Annual International Conference on Aspect-Oriented Software Development}, Association for Computing Machinery, New York, NY, USA, 2012, pp. 167–178, \url{https://doi.org/10.1145/2162049.2162069}.

\keyedreference{WL257}
Vogel, Thomas, Neumann, Stefan, Hildebrandt, Stephan, Giese, Holger, Becker, Basil, Model-driven architectural monitoring and adaptation for autonomic systems, in: \textit{Proceedings of the 6th International Conference on Autonomic Computing}, Association for Computing Machinery, New York, NY, USA, 2009, pp. 67–68, \url{https://doi.org/10.1145/1555228.1555249}.

\keyedreference{WL258}
Stopford, Benjamin, Counsell, Steve, A framework for the simulation of structural software evolution, \textit{ACM Trans. Model. Comput. Simul.} 18 (4) (2008), \url{https://doi.org/10.1145/1391978.1391983}.

\keyedreference{WL259}
Mercadal, Julien, Enard, Quentin, Consel, Charles, Loriant, Nicolas, A domain-specific approach to architecturing error handling in pervasive computing, \textit{SIGPLAN Not.} 45 (10) (2010) 47–61, \url{https://doi.org/10.1145/1932682.1869465}.

\keyedreference{WL260}
Roshandel, Roshanak, Hoek, Andr\'{e} Van Der, Mikic-Rakic, Marija, Medvidovic, Nenad, Mae---a system model and environment for managing architectural evolution, \textit{ACM Trans. Softw. Eng. Methodol.} 13 (2) (2004) 240–276, \url{https://doi.org/10.1145/1018210.1018213}.

\keyedreference{WL261}
Osterloh, Christoph, Litza, Marek, Maehle, Erik, Hard- and Software Architecture of a Small Autonomous Underwater Vehicle for Environmental Monitoring Tasks, in: \textit{Advances in Robotics Research: Theory, Implementation, Application}, Springer Berlin Heidelberg, Berlin, Heidelberg, 2009, pp. 347--356.

\keyedreference{WL262}
Ksenzov, M. V., Architectural refactoring of corporate program systems, \textit{Programming and Computer Software} 32 (1) (2006) 31–43, \url{https://doi.org/10.1134/s036176880601004x}.

\keyedreference{WL263}
Mao, Song, Kanungo, Tapas, Software architecture of PSET: a page segmentation evaluation toolkit, \textit{International Journal on Document Analysis and Recognition} 4 (3) (2002) 205–217, \url{https://doi.org/10.1007/s100320200070}.

\keyedreference{WL264}
Garcia, Alessandro F., Rubira, Cecìlia M. F., An Architectural-Based Reflective Approach to Incorporating Exception Handling into Dependable Software, in: \textit{Advances in Exception Handling Techniques}, Springer Berlin Heidelberg, 2001, pp. 189–206.

\keyedreference{WL265}
Everton Cavalcante, Thais Batista, Using Software Architecture Descriptions to Detect Architectural Smells at Design Time, in: \textit{Proceedings of the Brazilian Symposium on Software Engineering (SBES)}, IEEE, Brazil, 2015, \url{https://doi.org/10.5753/cibse.2023.24697}.

\keyedreference{WL266}
Zhou, Yu, Chen, Taolue, Software Adaptation in an Open Environment: A Software Architecture Perspective, 2017, \url{http://dx.doi.org/10.1201/9781315181721}.

\keyedreference{GL267}
Ruichen Hu, \textit{An Automated Approach to Check Software Architecture Erosion}, Master's thesis, Eindhoven University of Technology, Eindhoven, The Netherlands, 2023, \url{https://research.tue.nl/files/319375239/Hu_R.pdf}.

\keyedreference{GL268}
Ajay Bandi, \textit{Assessing Code Decay by Detecting Software Architecture Violations}, Ph.D. thesis, Mississippi State University, Mississippi State, MS, USA, 2014, \url{https://search.proquest.com/openview/557e28562aa9966bfe834739c9e79361/1?pq-origsite=gscholar&cbl=18750}.

\keyedreference{GL269}
Camilo Andrés Mendoza Arrieta, \textit{Architectural Issues Detection and Visualization in the Continuous Integration Pipeline}, Master's thesis, University of Manitoba, Winnipeg, MB, Canada, 2020, \url{https://repositorio.uniandes.edu.co/items/fef7183d-2e1d-439d-a401-87d92f733359}.

\keyedreference{GL270}
Zude Li, \textit{Characterizing and Diagnosing Architectural Degeneration of Software Systems from Defect Perspective}, Ph.D. thesis, The University of Western Ontario, London, ON, Canada, 2010, \url{https://search.proquest.com/openview/82c81771a4a3e82189ed46116c0f87c7/1?pq-origsite=gscholar&cbl=18750&diss=y}.

\keyedreference{GL271}
Ruiyin Li, \textit{Understanding, Analysis, and Handling of Software Architecture Erosion}, Ph.D. thesis, University of Groningen, Groningen, The Netherlands, 2019, \url{https://research.rug.nl/en/publications/understanding-analysis-and-handling-of-software-architecture-eros}.

\keyedreference{GL272}
Christian Rehn, Software Architectural Tactics and Patterns for Safety and Security, TU Kaiserslautern, Technical Report, Kaiserslautern, Germany, 2010, \url{http://www.christian-rehn.de/wp-content/uploads/downloads/2010/06/seminar_safe_sec.pdf}.

\keyedreference{GL273}
Anastassiya Samuilova, Samal Nursapa, \textit{Using a Transformer to Evaluate the Impact of Architectural Smells on Software Performance}, Master's thesis, M{\"a}lardalen University, Sweden, 2025, \url{https://www.diva-portal.org/smash/record.jsf?pid=diva2:1977604}.

\keyedreference{GL274}
Roy T. Fielding, \textit{Architectural Styles and the Design of Network-based Software Architectures}, Ph.D. thesis, University of California, Irvine, Irvine, California, USA, 2000, \url{https://ics.uci.edu/~fielding/pubs/dissertation/fielding_dissertation.pdf}.

\keyedreference{GL275}
David Benjaminsson, Edvin Bengtsson, \textit{Observing Software Architectural Gaps in Industry: A Case Study}, Master's thesis, Chalmers University of Technology and University of Gothenburg, Gothenburg, Sweden, 2021, \url{https://odr.chalmers.se/bitstreams/78d3ec09-e448-456e-baea-db59d16d3989/download}.

\keyedreference{GL276}
Salama, Maria, Bahsoon, Rami, Buyya, Rajkumar, Architectural Stability Reasoning using Self-Awareness Principles: Case of Self-Adaptive Cloud Architectures, 2019, \url{https://arxiv.org/abs/1912.06469}.

\keyedreference{GL277}
Hugo Ari Rodrigues Drumond, \textit{Software Architecture by Component Selection}, Ph.D. thesis, Universidade do Porto, Porto, Portugal, 2019, \url{https://search.proquest.com/openview/4ec34fb9a24cf047142951cbd13ba478/1?pq-origsite=gscholar&cbl=2026366&diss=y}.

\keyedreference{GL278}
Duarte Jorge Guedes Oliveira Dias Figueiredo, \textit{Recovery of Software Architecture Models From Infrastructure-As-Code Specifications}, Ph.D. thesis, Universidade do Porto, Porto, Portugal, 2024, \url{https://search.proquest.com/openview/3c3c157dd3aa583694318c3d2000038f/1?pq-origsite=gscholar&cbl=2026366&diss=y}.

\keyedreference{GL279}
Evertson Croes, \textit{Software Architectural Styles in the Internet of Things}, Master's thesis, Radboud University Nijmegen, Nijmegen, The Netherlands, 2015, \url{https://www.cs.ru.nl/masters-theses/2015/E_Croes___Software_architectural_styles_in_the_internet_of_things.pdf}.

\keyedreference{GL280}
Grady Andersen, MoldStud Research Team, Architectural Erosion: Preventing Degradation of Software Systems over Time, 2024, \url{https://moldstud.com/articles/p-architectural-erosion-preventing-degradation-of-software-systems-over-time}.

\keyedreference{GL281}
Mario Bittencourt, Graceful Degradation --- Why It Is Important and How To Achieve It, 2023, \url{https://medium.com/@mbneto/graceful-degradation-why-it-is-important-and-how-to-achieve-it-07f6d4a5f7d5}.

\keyedreference{GL282}
{Quora user}, Will the Code Degrade Gracefully?, \url{https://www.quora.com/Will-the-code-degrade-gracefully}.

\keyedreference{GL283}
Terry Bollinger, The First Nobel Prize for Insidious Software Degradation, 2024, \url{https://www.linkedin.com/posts/terry-bollinger-8a976_the-first-nobel-prize-for-insidious-software-activity-7258662235151953920-e4gf}.

\keyedreference{GL284}
1ijk, Can Software Architects Predict of How Microservice Architectures Will Mature over Time?, 2018, \url{https://stackoverflow.com/questions/50455134}.

\end{description}

\section{Appendix}

\subsection{Search and Selection Round Inter-Rater Agreement}

% \begin{figure}[tb]
%     \centering
%     \includegraphics[width=0.5\linewidth]{Figures/interagreement/title_abstract_cohen'sK.png}
%     \caption{Search and Selection}
%     \label{fig:search-selection-gwet-ac}
% \end{figure}

% \begin{figure}[tb]
%     \centering
%     \includegraphics[width=0.5\linewidth]{Figures/interagreement/Gwet_AC_title&Abstract.png}
%     \caption{Search and Selection}
%     \label{fig:search-selection-cohen-k}
% \end{figure}

\begin{figure}[b]
    \centering

    \begin{subfigure}[t]{0.48\textwidth}
        \centering
        \includegraphics[width=\linewidth]{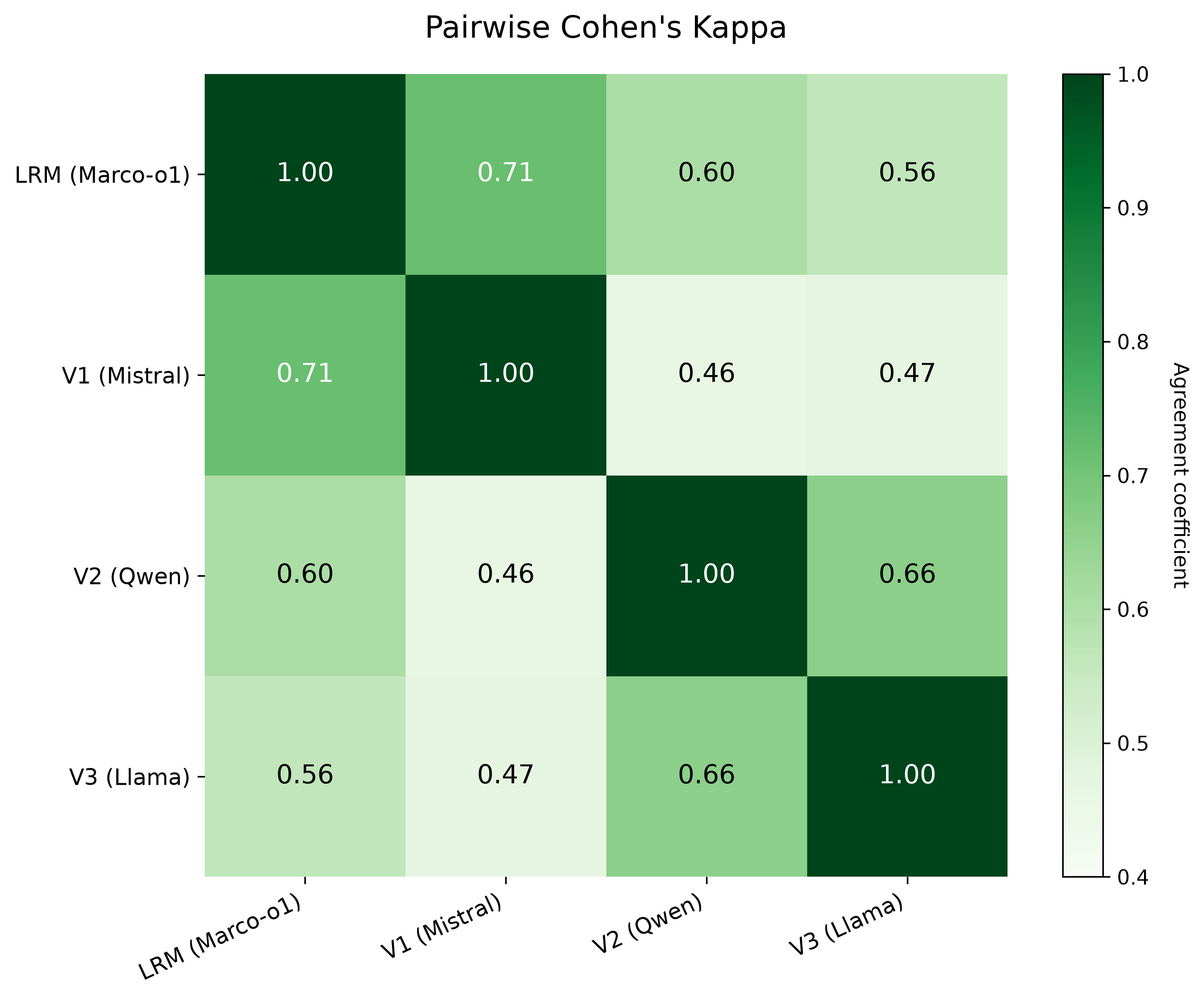}
        \caption{Cohen's Kappa}
        \label{fig:search-selection-cohen-k}
    \end{subfigure}
    \hfill
    \begin{subfigure}[t]{0.48\textwidth}
        \centering
        \includegraphics[width=\linewidth]{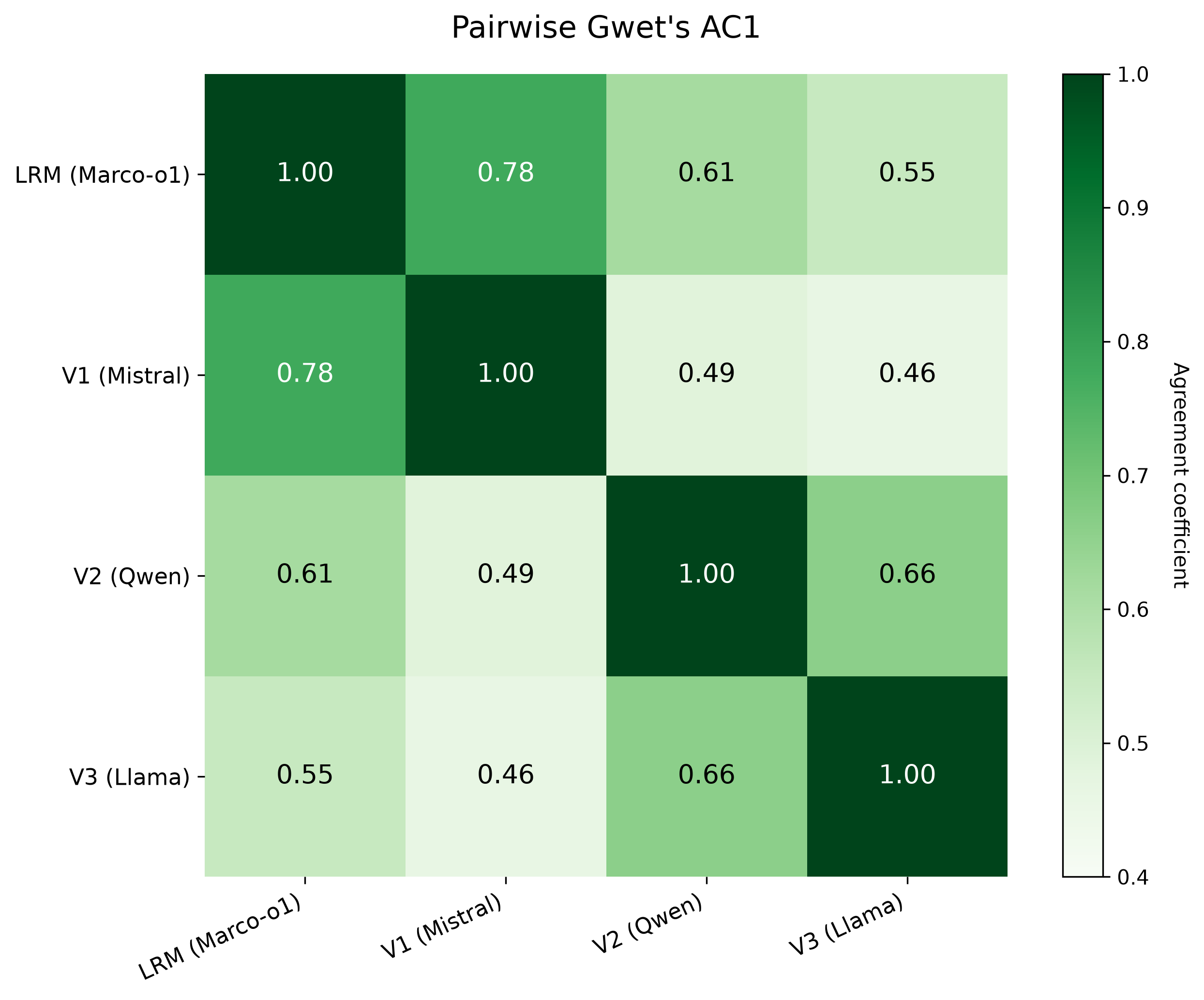}
        \caption{Gwet's AC}
        \label{fig:search-selection-gwet-ac}
    \end{subfigure}

    \caption{Inter-rater agreement for the title and abstract screening phase.}
    \label{fig:search-selection-interagreement}
\end{figure}

Figure~\ref{fig:search-selection-interagreement} shows the pairwise inter-rater agreement among the four raters during the title and abstract screening phase, measured using Cohen’s Kappa (K) and Gwet’s AC1. The results showed moderate to substantial agreement across the raters. For Cohen’s Kappa, the pairwise agreement values range from 0.46 to 0.71. The highest agreement is observed between the LRM (Marco-o1) and V1 (Mistral) raters (K = 0.71), followed by V2 (Qwen) and V3 (Llama) (K = 0.66). In contrast, the lowest agreement is found between V1 (Mistral) and V2 (Qwen) (K = 0.46).
A similar pattern is observed for Gwet’s AC1, with agreement coefficients ranging from 0.46 to 0.78. The strongest agreement is again between LRM (Marco-o1) and V1 (Mistral), reaching an AC1 value of 0.78, while V2 (Qwen) and V3 (Llama) show an agreement of 0.66. The lowest AC1 value (0.46) is observed between V1 (Mistral) and V3 (Llama). These results show some variation across the raters, with certain pairs showing substantially stronger agreement than others.

\subsection{Data Extraction}
\subsubsection{(RQ$_1$) Approaches to measure degradation}
% \begin{figure}
%     \centering
%     \includegraphics[width=0.5\linewidth]{Figures//interagreement/cohens_kappa_heatmap_approaches_full.png}
%     \caption{RQ1 Approaches to measure architectural degradation}
%     \label{fig:approachesfullreadcohenk}
% \end{figure}
% \begin{figure}
%     \centering
%     \includegraphics[width=0.5\linewidth]{Figures//interagreement/gwets_ac1_heatmap_approaches_full.png}
%     \caption{RQ1 Approaches to measure architectural degradation}
%     \label{fig:approachesfullreadgwetac}
% \end{figure} 

\begin{figure}[tb]
    \centering
 \begin{subfigure}[t]{0.48\textwidth}
        \centering
        \includegraphics[width=\linewidth]{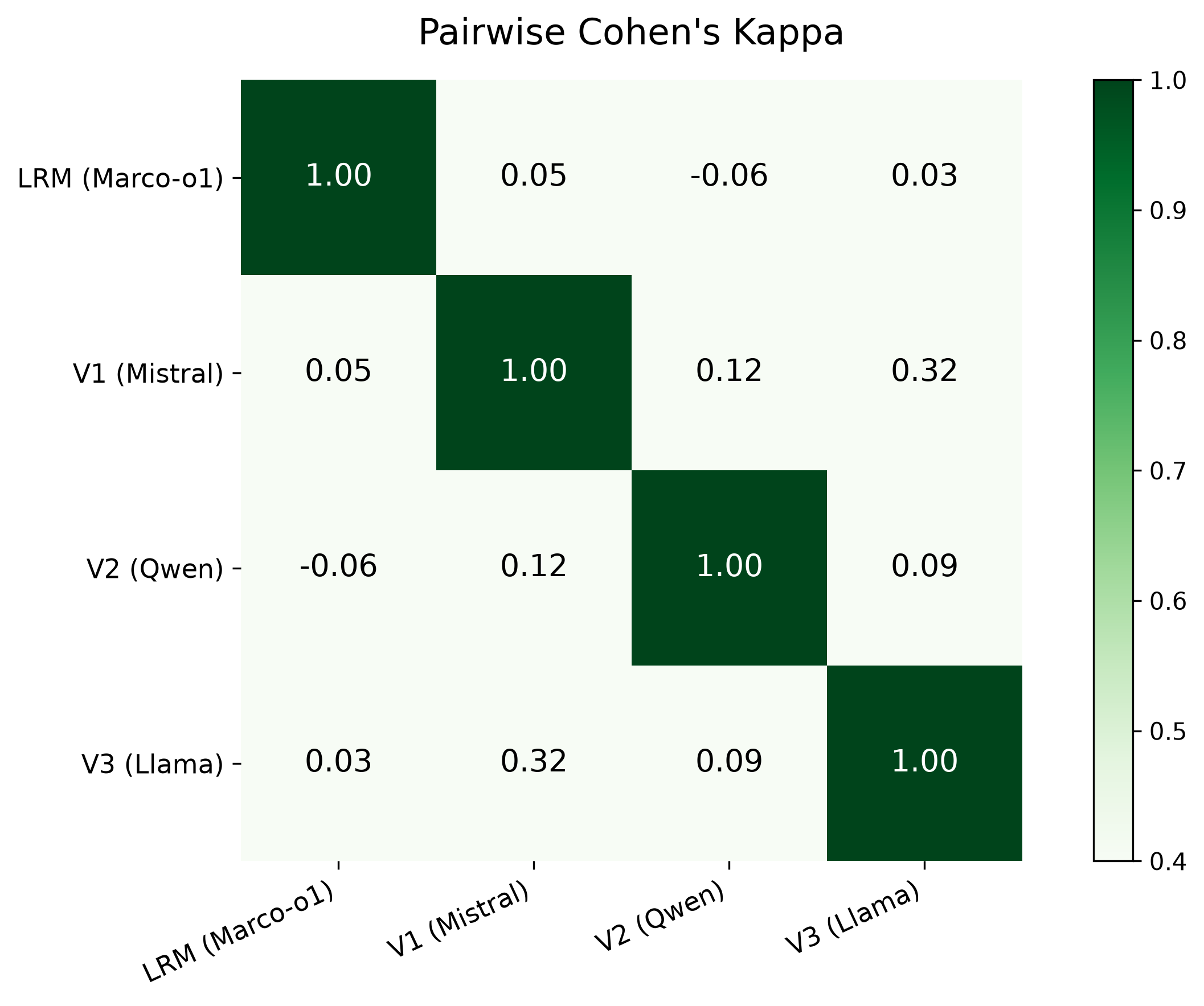}
        \caption{Cohen's Kappa}
        \label{fig:approachesfullreadcohenk}
    \end{subfigure}
    \hfill
    \begin{subfigure}[t]{0.48\textwidth}
        \centering
        \includegraphics[width=\linewidth]{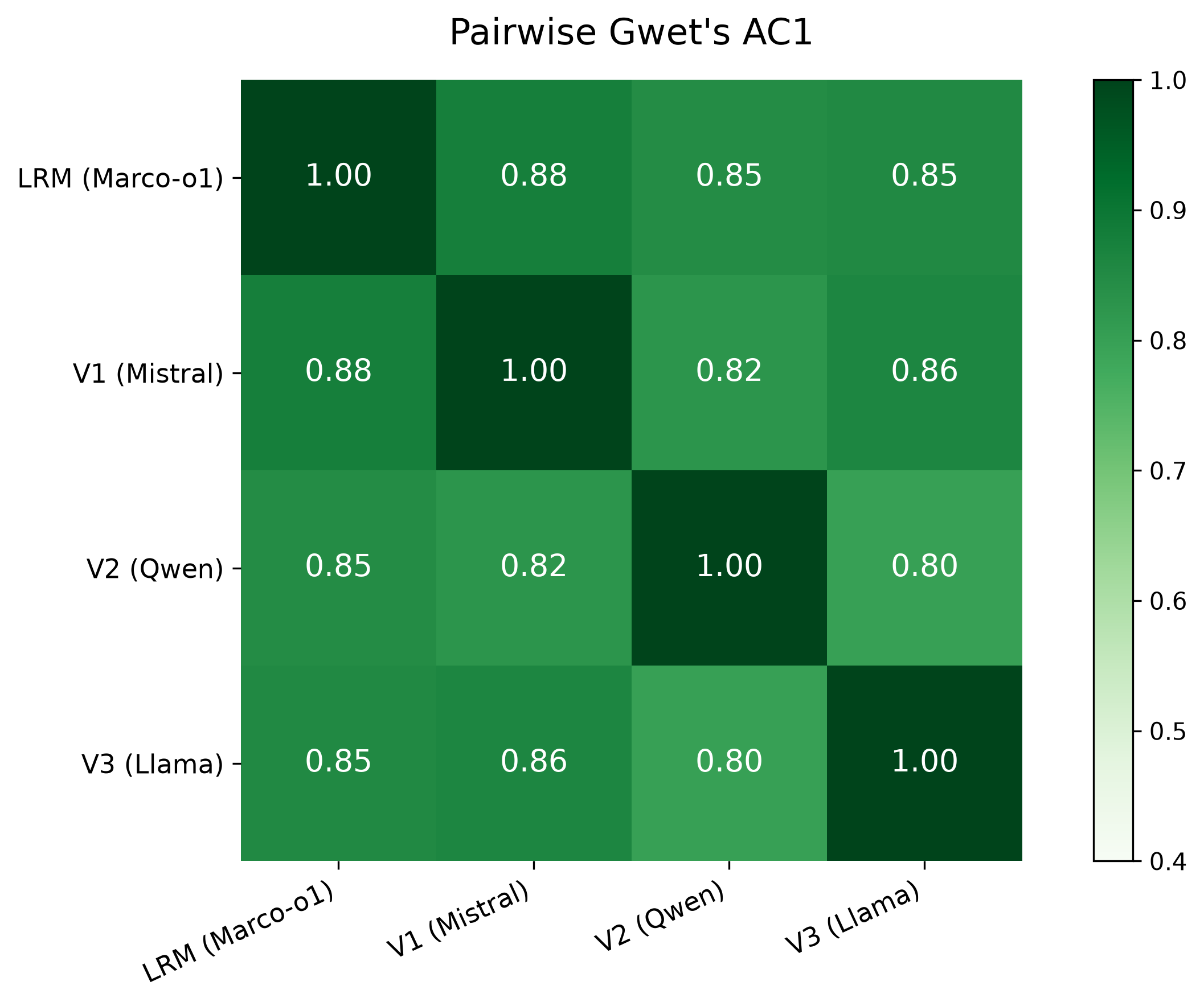}
        \caption{Gwet's AC}
        \label{fig:approachesfullreadgwetac}
    \end{subfigure}

    \caption{Approaches to measure architectural degradation (RQ$_1$).}
    \label{fig:approachesfullread}
\end{figure}

Figure~\ref{fig:approachesfullread} presents the pairwise inter-rater agreement for the RQ$_1$ data-extraction round, which focused on identifying approaches used to measure architectural degradation. The results show a considerable difference between Cohen's Kappa (K) and Gwet's AC1. Cohen's Kappa values are generally low, ranging from $-0.06$ to $0.32$. The highest pairwise agreement is observed between V1 (Mistral) and V3 (Llama) (K = $0.32$), while the lowest value occurs between the LRM (Marco-o1) and V2 (Qwen) (K = $-0.06$). The remaining pairwise K values range from $0.03$ to $0.12$, indicating limited agreement according to this measure.
In contrast, Gwet’s AC1 indicates consistently high agreement across all rater pairs, with coefficients ranging from 0.80 to 0.88. The highest agreement is observed between the LRM (Marco-o1) and V1 (Mistral) (AC1 = 0.88), followed by V1 (Mistral) and V3 (Llama) (AC1 = 0.86). The lowest agreement, between V2 (Qwen) and V3 (Llama), remains high (AC1 = 0.80).

\subsubsection{(RQ$_2$) Metrics for measuring architectural degradation}
% \begin{figure}
%     \centering
%     \includegraphics[width=0.5\linewidth]{Figures//interagreement/cohens_kappa_heatmap(metricsfullread).png}
%     \caption{RQ2 Metrics for measuring architectural degradation}
%     \label{fig:metricsfullreadcohenk}
% \end{figure}
% \begin{figure}
%     \centering
%     \includegraphics[width=0.5\linewidth]{Figures//interagreement/gwets_ac1_heatmap_fullreadmetrics.png}
%     \caption{RQ2 Metrics for measuring architectural degradation}
%     \label{fig:metricsfullreadgwetac}
% \end{figure}

\begin{figure}[tb]
    \centering

    \begin{subfigure}[t]{0.48\textwidth}
        \centering
        \includegraphics[width=\linewidth]{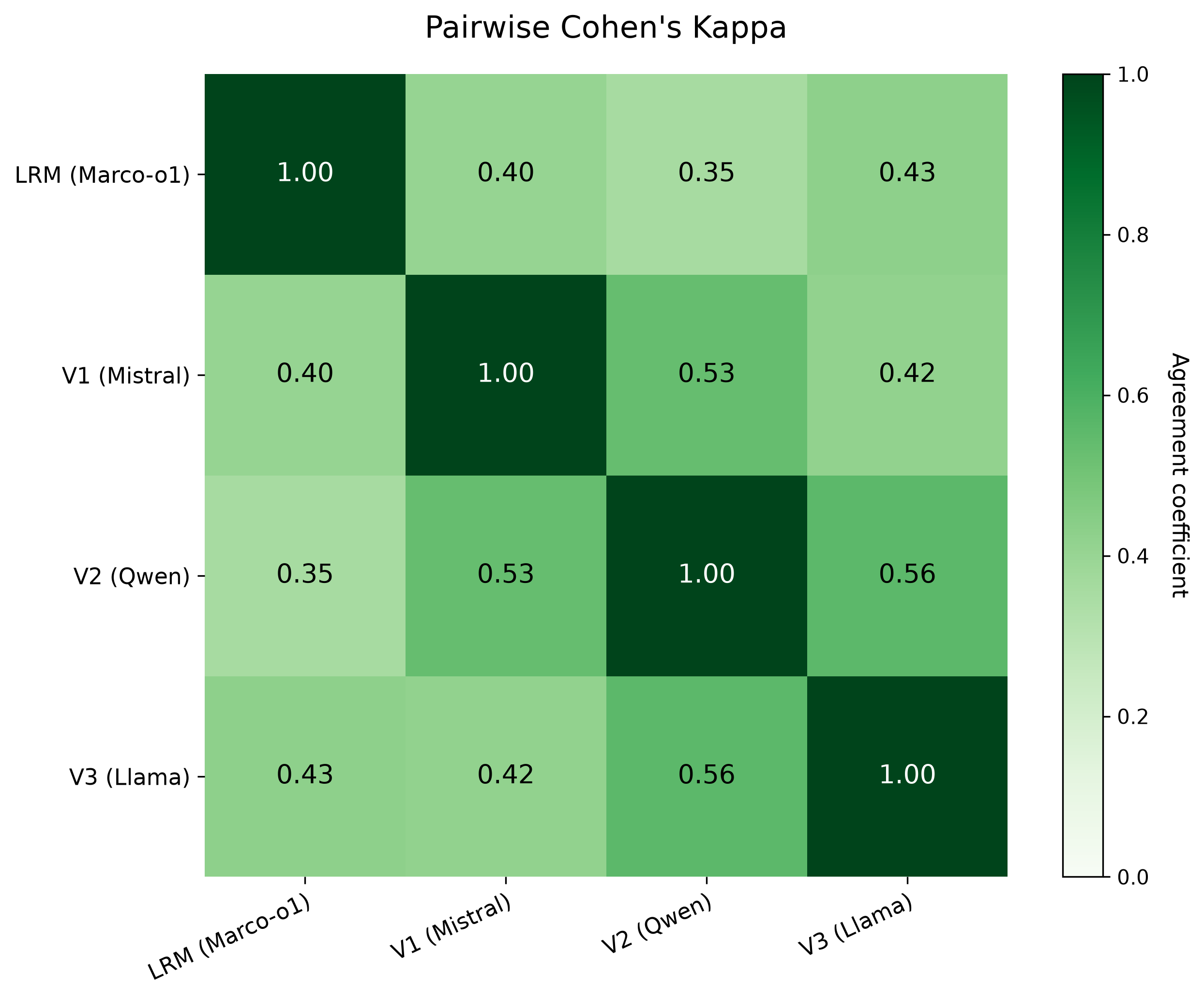}
        \caption{Cohen's Kappa}
        \label{fig:metricsfullreadcohenk}
    \end{subfigure}
    \hfill
    \begin{subfigure}[t]{0.48\textwidth}
        \centering
        \includegraphics[width=\linewidth]{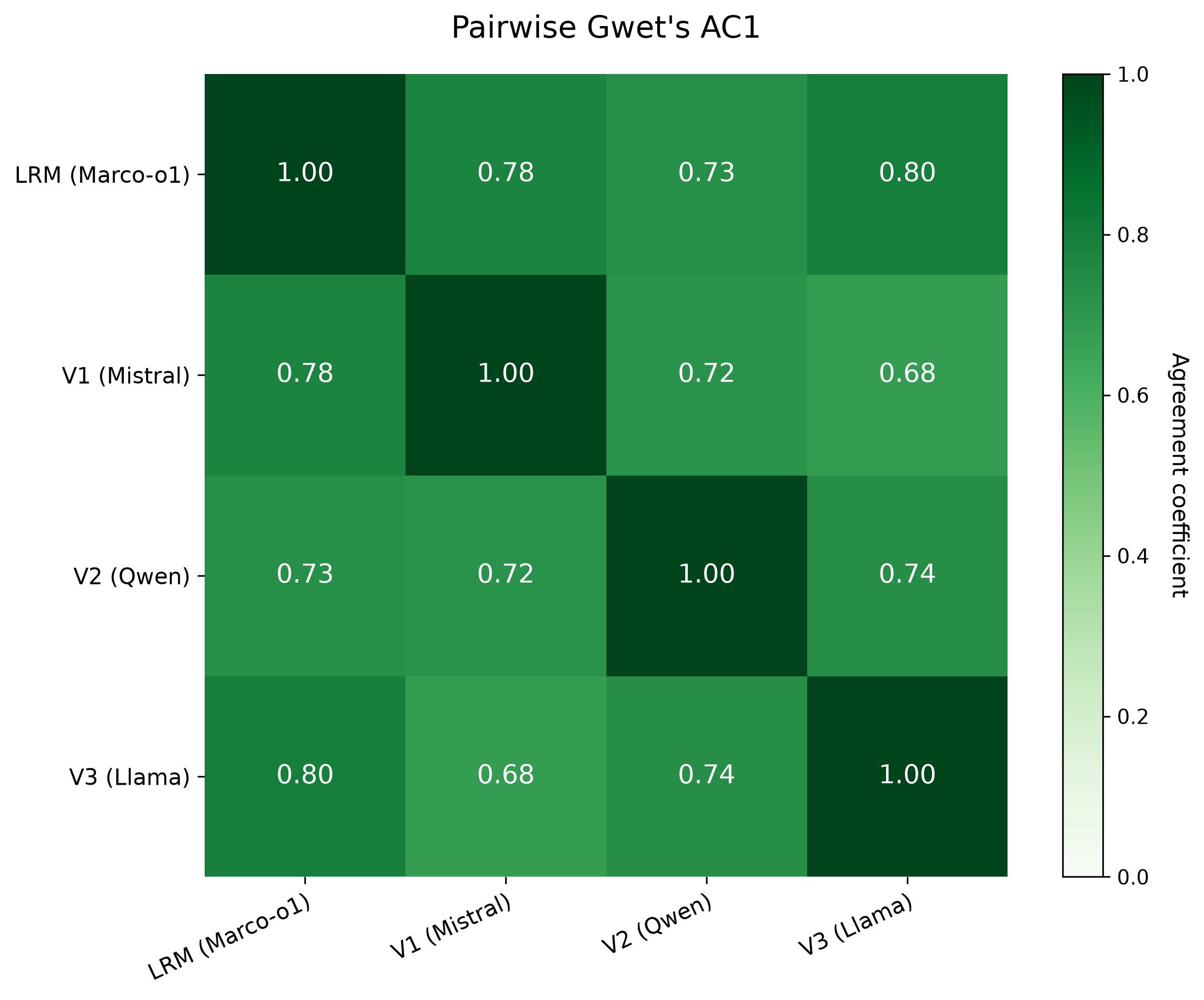}
        \caption{Gwet's AC}
        \label{fig:metricsfullreadgwetac}
    \end{subfigure}

    \caption{Metrics for measuring architectural degradation (RQ$_2$).}
    \label{fig:metricsfullread}
\end{figure}

Figure~\ref{fig:metricsfullread} shows the pairwise inter-rater agreement for the RQ$_2$ data extraction round, which concentrated on identifying metrics used to measure architectural degradation. Cohen’s Kappa (K) values range from 0.35 to 0.56, showing varying levels of agreement among the raters. The highest agreement is observed between V2 (Qwen) and V3 (Llama) (K = 0.56), followed by V1 (Mistral) and V2 (Qwen) (K = 0.53). The lowest agreement occurs between the LRM (Marco-o1) and V2 (Qwen) (K = 0.35), while the remaining pairwise values range from 0.40 to 0.43.
Gwet’s AC1 shows consistently higher agreement, with pairwise coefficients ranging from 0.68 to 0.80. The highest agreement is observed between the LRM (Marco-o1) and V3 (Llama) (AC1 = 0.80), followed by the LRM and V1 (Mistral) (AC1 = 0.78). The lowest agreement is between V1 (Mistral) and V3 (Llama) (AC1 = 0.68). So, it shows a moderate level of agreement according to Cohen’s Kappa and a stronger level of agreement according to Gwet’s AC1.

\subsubsection{(RQ$_3$) Tools to support architectural degradation analysis}
% \begin{figure}
%     \centering
%     \includegraphics[width=0.5\linewidth]{Figures//interagreement/cohens_kappa_heatmaptoolfullread.png}
%     \caption{RQ3 Tools to support architectural degradation analysis}
%     \label{fig:toolsfullreadcohenk}
% \end{figure}
% \begin{figure}
%     \centering
%     \includegraphics[width=0.5\linewidth]{Figures//interagreement/gwets_ac1_heatmaptoolsfullread.png}
%     \caption{RQ3 Tools to support architectural degradation analysis}
%     \label{fig:toolsfullreadgwetac}
% \end{figure}

\begin{figure}[tb]
    \centering

    \begin{subfigure}[t]{0.48\textwidth}
        \centering
        \includegraphics[width=\linewidth]{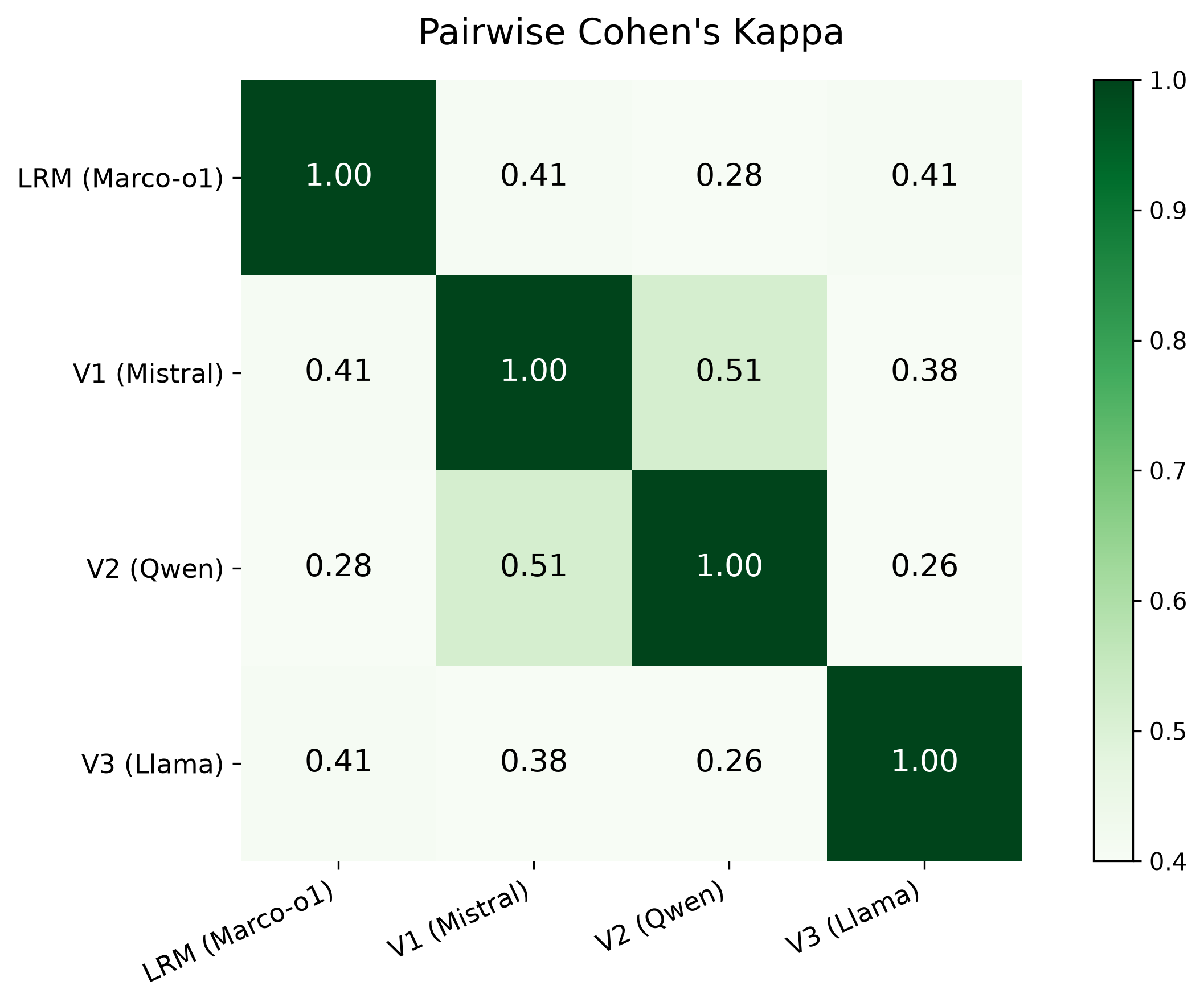}
        \caption{Cohen's Kappa}
        \label{fig:toolsfullreadcohenk}
    \end{subfigure}
    \hfill
    \begin{subfigure}[t]{0.48\textwidth}
        \centering
        \includegraphics[width=\linewidth]{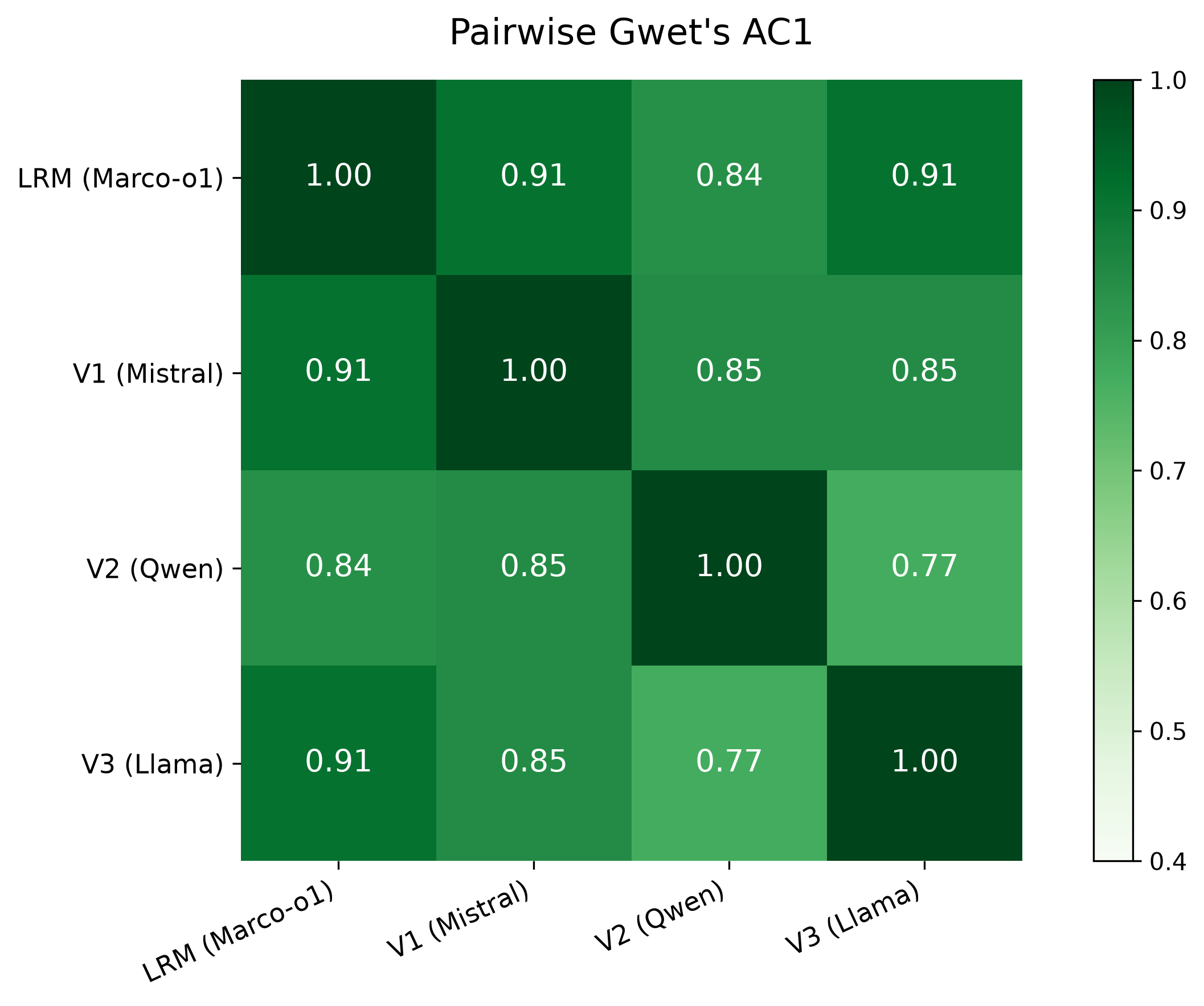}
        \caption{Gwet's AC}
        \label{fig:toolsfullreadgwetac}
    \end{subfigure}

    \caption{Tools to support architectural degradation analysis (RQ$_3$).}
    \label{fig:toolsfullread}
\end{figure}

Figure~\ref{fig:toolsfullread} provides the pairwise inter-rater agreement for the RQ$_3$ data extraction round, that is focused on identifying tools used to support architectural degradation analysis. Cohen’s Kappa (K) values range from 0.26 to 0.51, indicating varying levels of agreement among the raters. The highest agreement is observed between V1 (Mistral) and V2 (Qwen) (K = 0.51). The LRM (Marco-o1) shows the same level of agreement with V1 (Mistral) and V3 (Llama) (K = 0.41), while the lowest agreement is observed between V2 (Qwen) and V3 (Llama) (K = 0.26).
However, Gwet’s AC1 indicates consistently high agreement across the rater pairs, with coefficients ranging from 0.77 to 0.91. The highest agreement is observed between the LRM (Marco-o1) and both V1 (Mistral) and V3 (Llama) (AC1 = 0.91). The lowest agreement is between V2 (Qwen) and V3 (Llama) (AC1 = 0.77), which nevertheless indicates a relatively high level of agreement.

\subsubsection{(RQ$_4$) Remediation Approaches}
% \begin{figure}
%     \centering
%     \includegraphics[width=0.5\linewidth]{Figures//interagreement/cohens_kappa_heatmap_fullread_remediation.png}
%     \caption{RQ4 Remediation Approaches}
%     \label{fig:Remediationfullreadcohenk}
% \end{figure}
% \begin{figure}
%     \centering
%     \includegraphics[width=0.5\linewidth]{Figures//interagreement/gwets_ac1_heatmap_fullread_remediation.png}
%     \caption{RQ4 Remediation Approaches}
%     \label{fig:Remediaitonfullreadgwetac}
% \end{figure}

\begin{figure}[tb]
    \centering

    \begin{subfigure}[t]{0.48\textwidth}
        \centering
        \includegraphics[width=\linewidth]{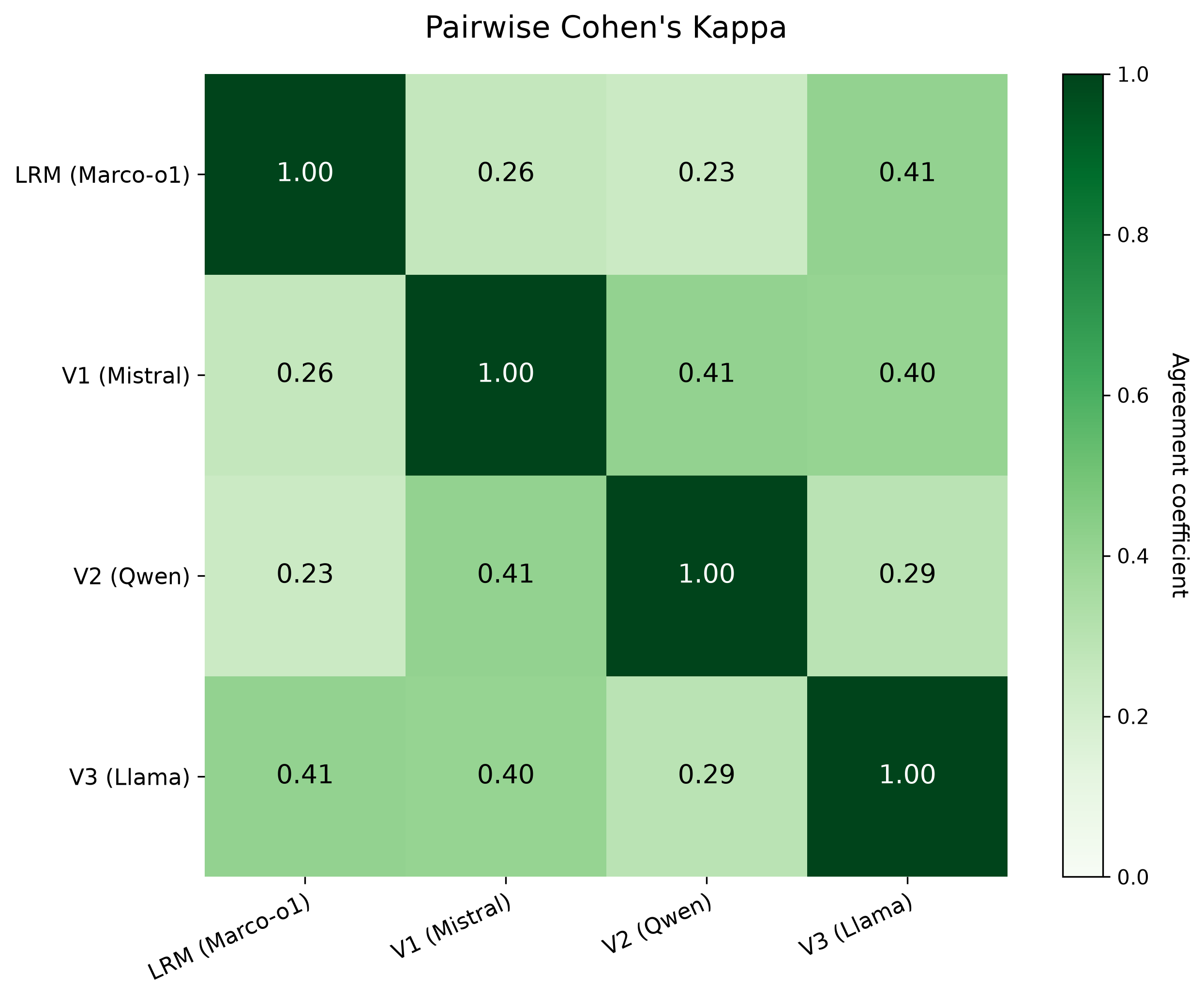}
        \caption{Cohen's Kappa}
        \label{fig:remediationfullreadcohenk}
    \end{subfigure}
    \hfill
    \begin{subfigure}[t]{0.48\textwidth}
        \centering
        \includegraphics[width=\linewidth]{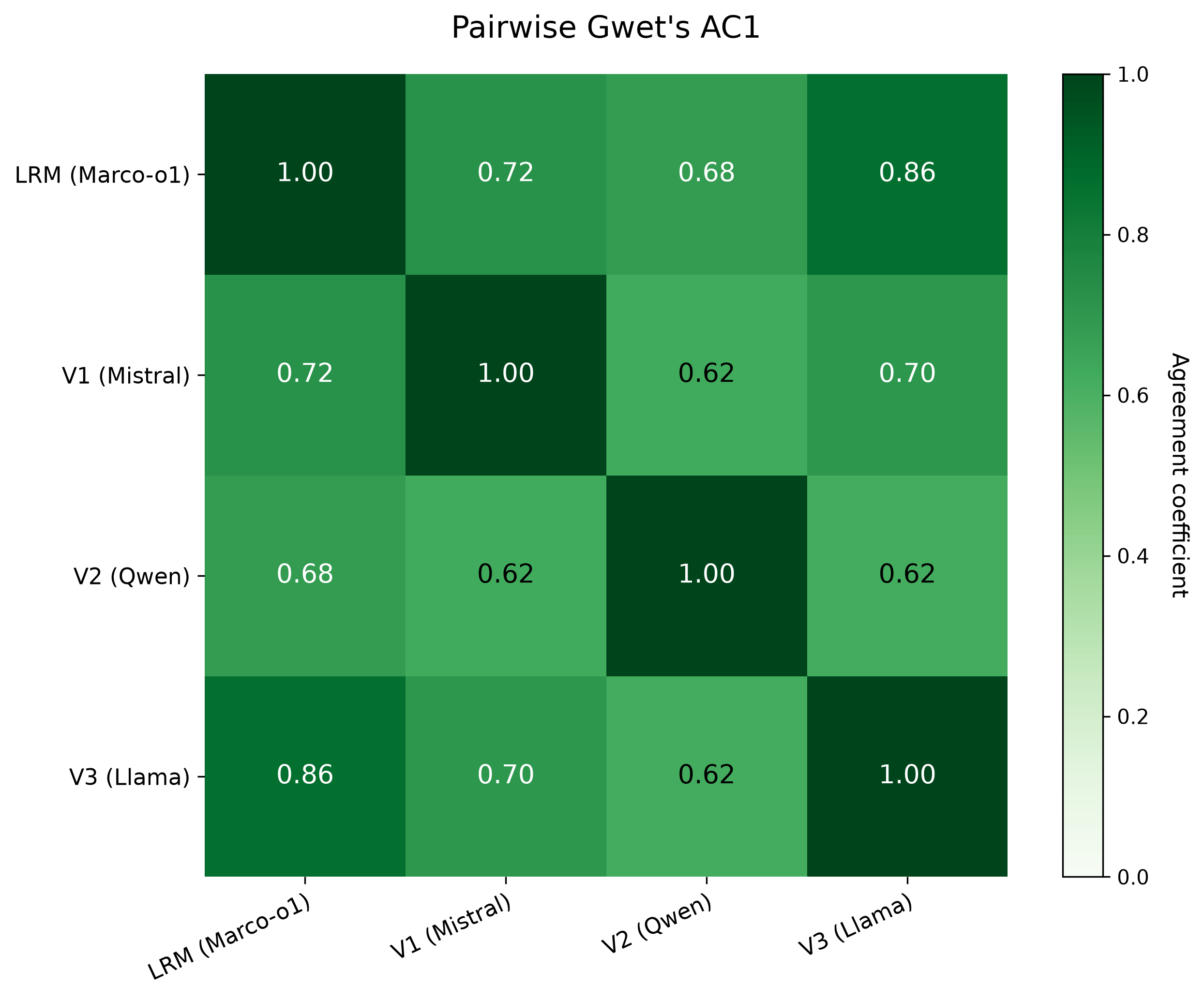}
        \caption{Gwet's AC}
        \label{fig:remediationfullreadgwetac}
    \end{subfigure}

    \caption{Remediation approaches (RQ$_4$).}
    \label{fig:remediationfullread}
\end{figure}

Figure~\ref{fig:remediationfullread} presents the pairwise inter-rater agreement for the RQ$_4$ data extraction round, which aimed to find remediation approaches for architectural degradation. Cohen’s Kappa (K) values range from 0.23 to 0.41, indicating varying levels of agreement among the raters. The highest agreement is observed between the LRM (Marco-o1) and V3 (Llama), as well as between V1 (Mistral) and V2 (Qwen) (K = 0.41). The lowest agreement occurs between the LRM (Marco-o1) and V2 (Qwen) (K = 0.23), while the remaining pairwise values range from 0.26 to 0.40.
Gwet’s AC1 produces consistently higher agreement coefficients, ranging from 0.62 to 0.86. The highest agreement is observed between the LRM (Marco-o1) and V3 (Llama) (AC1 = 0.86), followed by the LRM and V1 (Mistral) (AC1 = 0.72). The lowest agreement is observed for both V1 (Mistral)–V2 (Qwen) and V2 (Qwen)–V3 (Llama) (AC1 = 0.62).  

\subsection{Data Classification}
\subsubsection{(RQ$_1$) Approaches to measure architectural degradation}
% \begin{figure}
%     \centering
%     \includegraphics[width=0.5\linewidth]{Figures//interagreement/cohens_kappa_heatmap_final_approachestomeasure.png}
%     \caption{Approaches to measure architectural degradation (RQ$_1$)}
%     \label{fig:appraochesfinalcohenk}
% \end{figure}

% \begin{figure}
%     \centering
%     \includegraphics[width=0.5\linewidth]{Figures//interagreement/gwets_ac1_heatmap_approachesmeasure_final.png}
%     \caption{Approaches to measure architectural degradation (RQ$_1$) }
%     \label{fig:appraochesfinalgwetAc}
% \end{figure}

\begin{figure}[tb]
    \centering

    \begin{subfigure}[t]{0.48\textwidth}
        \centering
        \includegraphics[width=\linewidth]{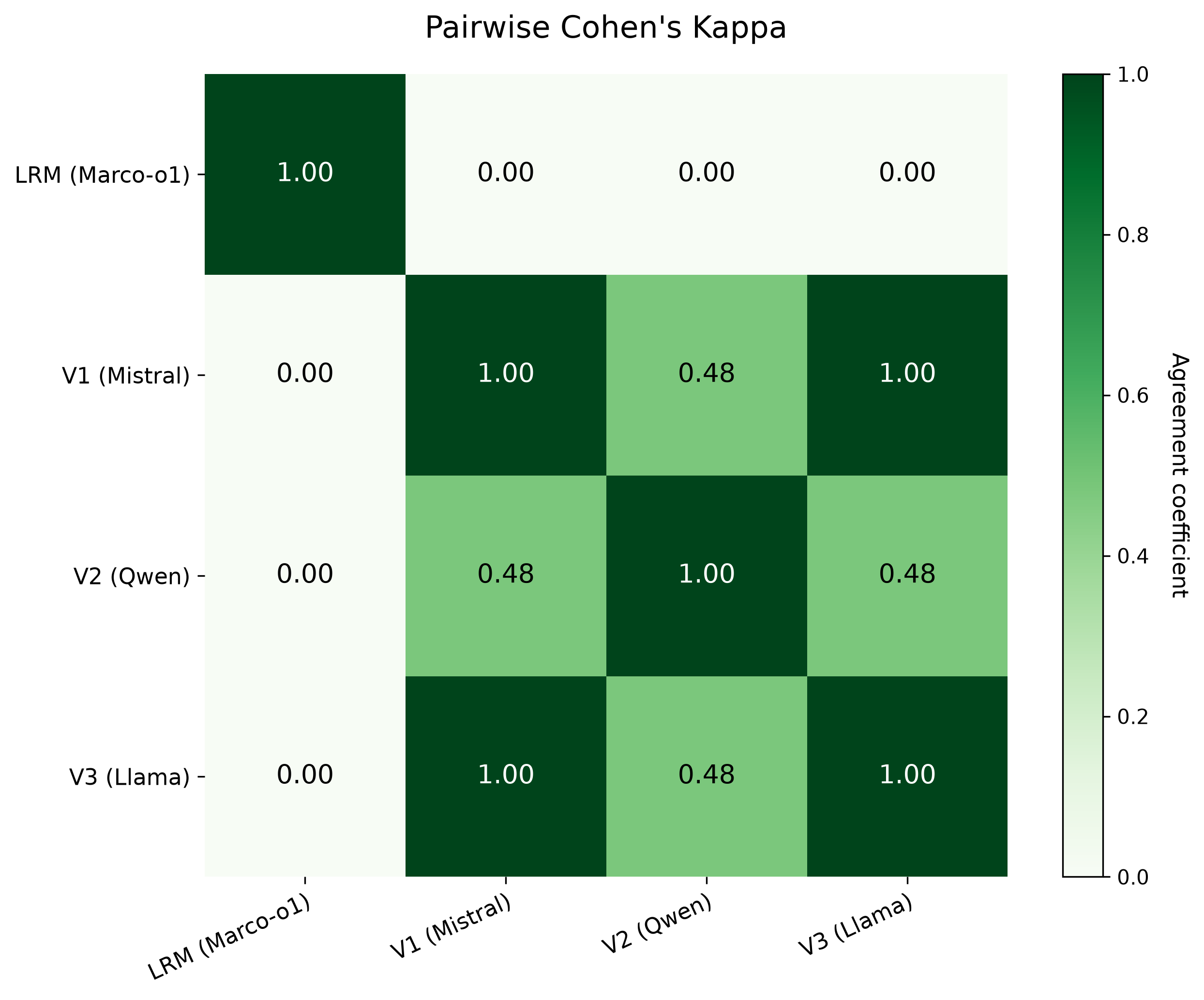}
        \caption{Cohen's Kappa}
        \label{fig:approachesfinalcohenk}
    \end{subfigure}
    \hfill
    \begin{subfigure}[t]{0.48\textwidth}
        \centering
        \includegraphics[width=\linewidth]{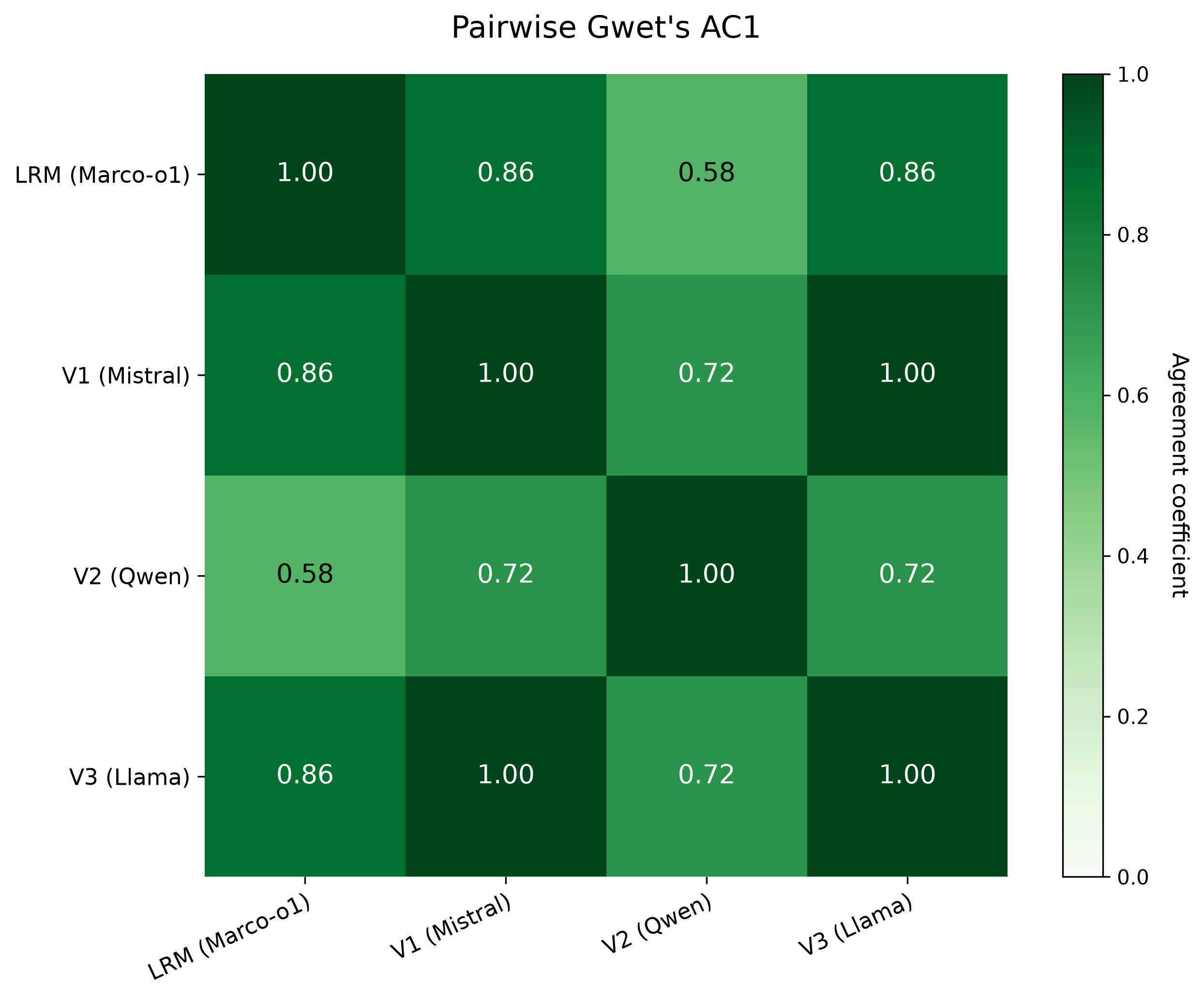}
        \caption{Gwet's AC}
        \label{fig:approachesfinalgwetac}
    \end{subfigure}

    \caption{Approaches to measure architectural degradation (RQ$_1$).}
    \label{fig:approachesfinal}
\end{figure}
Figure~\ref{fig:approachesfinal} presents the pairwise inter-rater agreement the data classification phase of RQ$_1$, which focused on classifying the identified approaches for measuring architectural degradation. Cohen’s Kappa (K) shows notable variation across the rater pairs, with values ranging from 0.00 to 1.00. Perfect agreement is observed between V1 (Mistral) and V3 (Llama) (K = 1.00), while V1–V2 and V2–V3 achieve K = 0.48. In contrast, the LRM (Marco-o1) obtains K = 0.00 with all three other raters.
However, Gwet’s AC1 shows substantially stronger agreement overall, with pairwise values ranging from 0.58 to 1.00. V1 (Mistral) and V3 (Llama) again demonstrate perfect agreement (AC1 = 1.00). The LRM shows high agreement with V1 and V3 (AC1 = 0.86), despite the corresponding Cohen’s Kappa values being zero, while its agreement with V2 is lower (AC1 = 0.58). V2 achieves an AC1 of 0.72 with both V1 and V3. 

\subsubsection{(RQ$_2$) Metrics for measuring architectural degradation}
% \begin{figure}
%     \centering
%     \includegraphics[width=0.5\linewidth]{Figures//interagreement/cohens_kappa_heatmap_metrics_final.png}
%     \caption{Metrics for measuring architectural degradation (RQ$_2$)}
%     \label{fig:metricsfinalcohenk}
% \end{figure}
% \begin{figure}
%     \centering
%     \includegraphics[width=0.5\linewidth]{Figures//interagreement/gwets_ac1_heatmap_final_metrics.png}
%     \caption{Metrics for measuring architectural degradation (RQ$_2$)}
%     \label{fig:metricsfinalgwetac}
% \end{figure}

\begin{figure}[tb]
    \centering

    \begin{subfigure}[t]{0.48\textwidth}
        \centering
        \includegraphics[width=\linewidth]{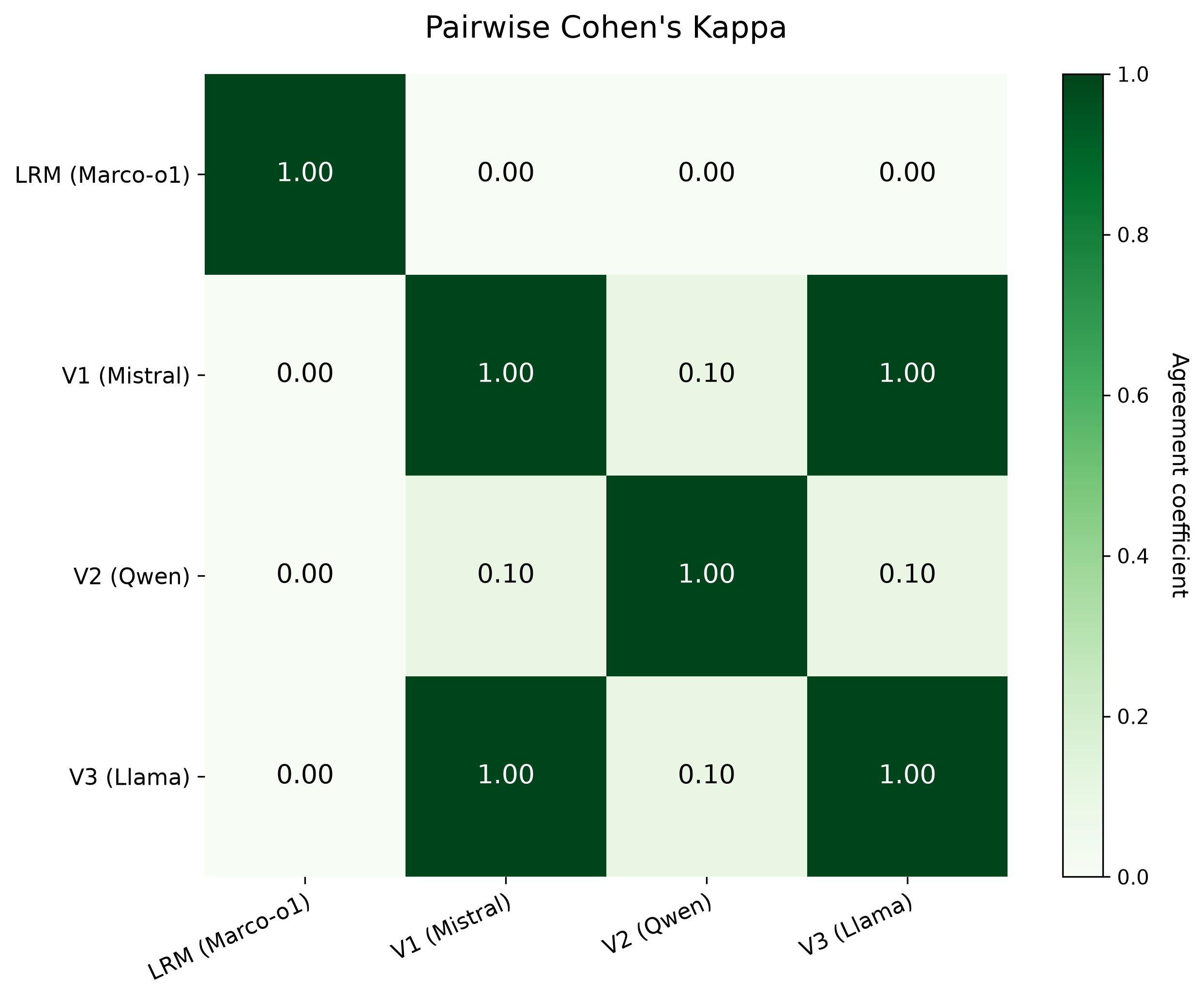}
        \caption{Cohen's Kappa}
        \label{fig:metricsfinalcohenk}
    \end{subfigure}
    \hfill
    \begin{subfigure}[t]{0.48\textwidth}
        \centering
        \includegraphics[width=\linewidth]{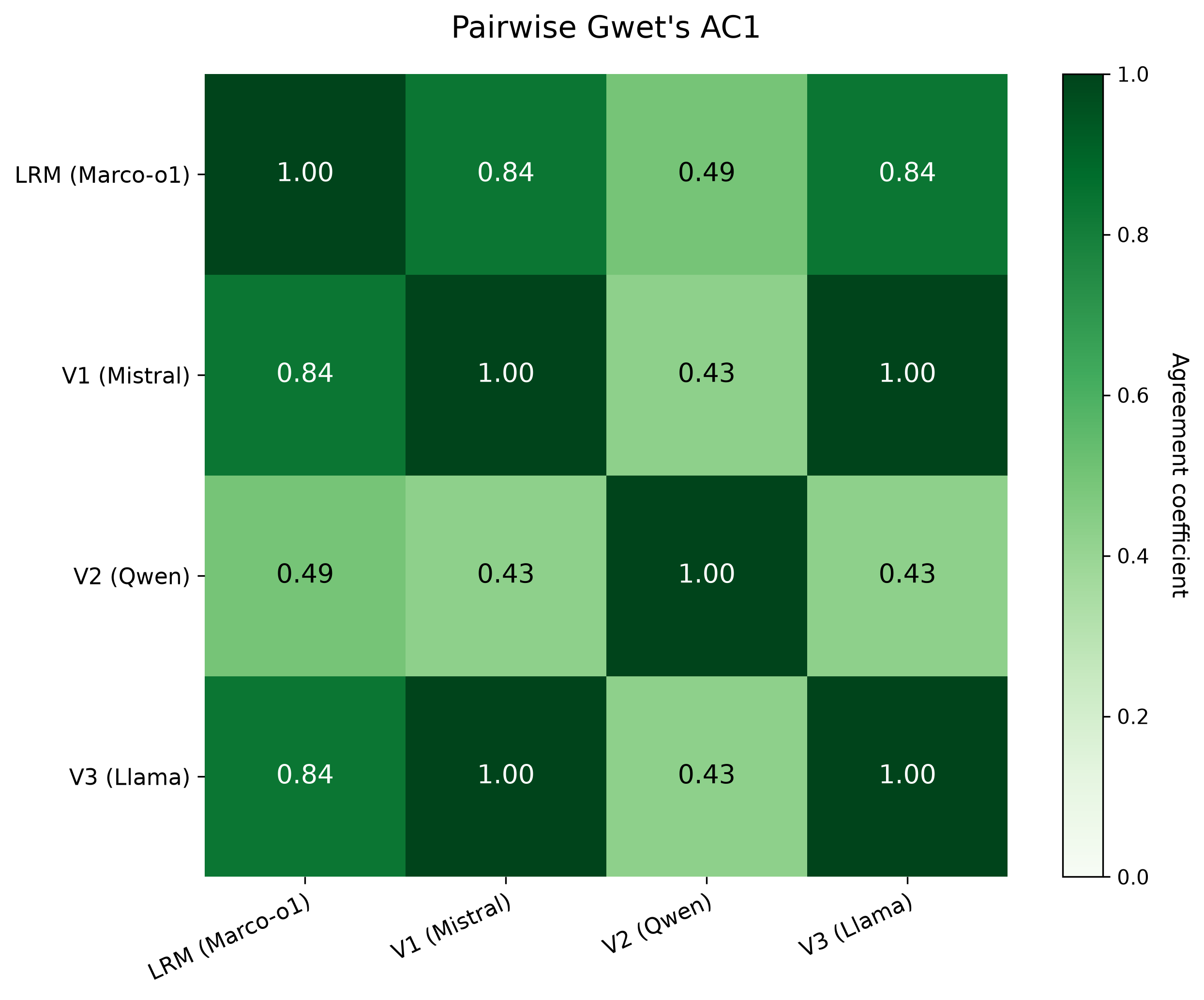}
        \caption{Gwet's AC}
        \label{fig:metricsfinalgwetac}
    \end{subfigure}

    \caption{Metrics for measuring architectural degradation (RQ$_2$).}
    \label{fig:metricsfinal}
\end{figure}

Figure~\ref{fig:metricsfinal} provides the pairwise inter-rater agreement for the RQ$_2$ data classification, which concentrated on the identified metrics for measuring architectural degradation. Cohen's Kappa (K) shows considerable variation across the rater pairs, with values ranging from $0.00$ to $1.00$. Perfect agreement is observed between V1 (Mistral) and V3 (Llama) (K = $1.00$). In contrast, V2 (Qwen) shows low agreement with both V1 and V3 (K = $0.10$), while the LRM (Marco-o1) obtains K = $0.00$ with all three other raters.
Gwet’s AC1 provides higher agreement estimates overall, with values ranging from 0.43 to 1.00. V1 (Mistral) and V3 (Llama) again achieve perfect agreement (AC1 = 1.00). The LRM showed high agreement with both V1 and V3 (AC1 = 0.84), despite the corresponding Cohen’s Kappa values being zero. In comparison, agreement involving V2 is notably lower, with AC1 values of 0.49 with the LRM and 0.43 with both V1 and V3.

\subsubsection{(RQ$_3$) Tools to support architectural degradation analysis}

% \begin{figure}
%     \centering
%     \includegraphics[width=0.5\linewidth]{Figures//interagreement/cohens_kappa_heatmap_tools_final.png}
%     \caption{Tools to support architectural degradation analysis (RQ$_3$)}
%     \label{fig:toolsfinalcohenk}
% \end{figure}

% \begin{figure}
%     \centering
%     \includegraphics[width=0.5\linewidth]{Figures//interagreement/gwets_ac1_heatmap_tools_final.png}
%     \caption{Tools to support architectural degradation analysis (RQ$_3$)}
%     \label{fig:toolsfinalgwetAc}
% \end{figure}

\begin{figure}[tb]
    \centering

    \begin{subfigure}[t]{0.48\textwidth}
        \centering
        \includegraphics[width=\linewidth]{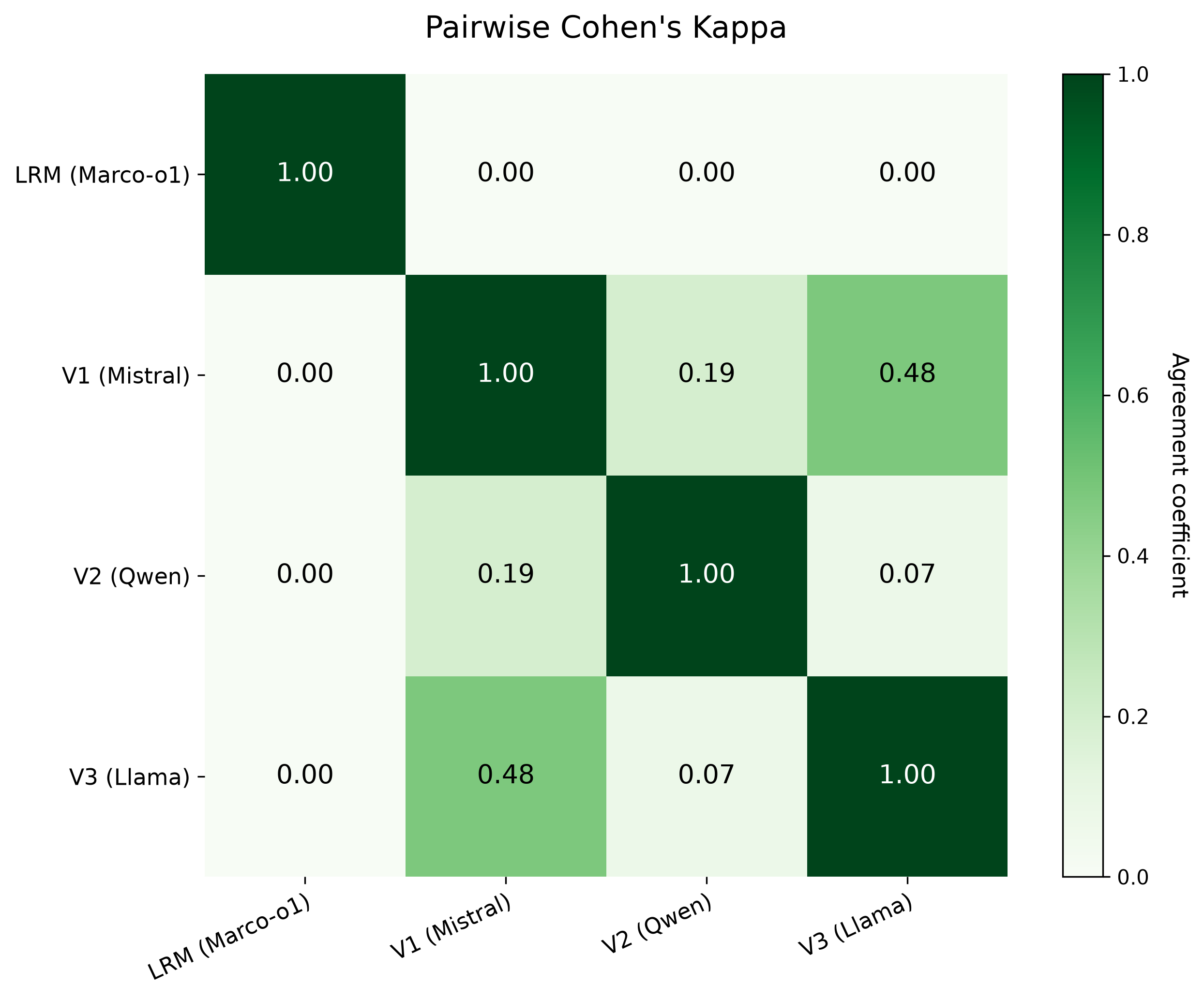}
        \caption{Cohen's Kappa}
        \label{fig:toolsfinalcohenk}
    \end{subfigure}
    \hfill
    \begin{subfigure}[t]{0.48\textwidth}
        \centering
        \includegraphics[width=\linewidth]{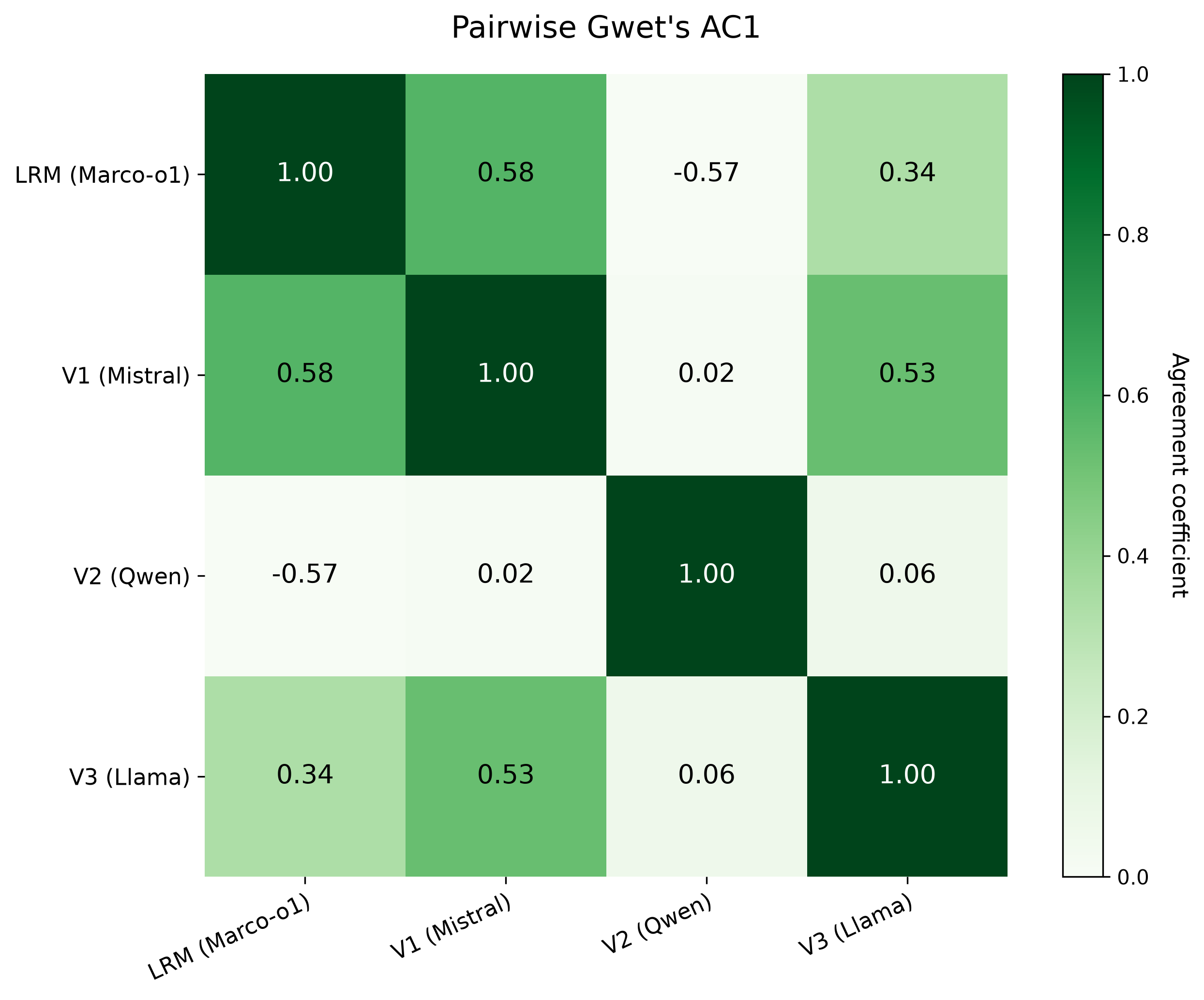}
        \caption{Gwet's AC}
        \label{fig:toolsfinalgwetac}
    \end{subfigure}

    \caption{Tools to support architectural degradation analysis (RQ$_3$).}
    \label{fig:toolsfinal}
\end{figure}
Figure~\ref{fig:toolsfinal} illustrates the pairwise inter-rater agreement for the RQ$_3$ data classification concerning the classification of tools supporting architectural degradation analysis. The Cohen’s Kappa (K) results indicate generally low agreement, with values ranging from 0.00 to 0.48. The strongest agreement is found between V1 (Mistral) and V3 (Llama) (K = 0.48), followed by V1 and V2 (Qwen) (K = 0.19). Agreement between V2 and V3 is considerably lower (K = 0.07), while the LRM (Marco-o1) records K = 0.00 with each of the other raters.
A more varied pattern emerges from Gwet’s AC1, where coefficients range from $-0.57$ to $0.58$. The highest agreement occurs between the LRM and V1 (AC1 = $0.58$), followed by V1 and V3 (AC1 = $0.53$). The LRM and V3 reach a lower agreement of $0.34$. In contrast, agreement involving V2 is particularly weak: V1--V2 and V2--V3 obtain AC1 values of $0.02$ and $0.06$, respectively. Most notably, the LRM--V2 comparison yields a negative AC1 value ($-0.57$), indicating disagreement beyond that expected by chance under this measure.
\subsubsection{(RQ$_4$) Remediation Approaches}

% \begin{figure}
%     \centering
%     \includegraphics[width=0.5\linewidth]{Figures//interagreement/cohens_kappa_heatmap_remediation_final.png}
%     \caption{Remediation Approaches (RQ$_4$)}
%     \label{fig:remediationapproachesfianlcohenk}
% \end{figure}

% \begin{figure}
%     \centering
%     \includegraphics[width=0.5\linewidth]{Figures//interagreement/gwets_ac1_heatmap_remediation_final.png}
%     \caption{Remediation Approaches (RQ$_4$)}
%     \label{fig:remediationapproachesfinalgwetac}
% \end{figure}

\begin{figure}[tb]
    \centering

    \begin{subfigure}[t]{0.48\textwidth}
        \centering
        \includegraphics[width=\linewidth]{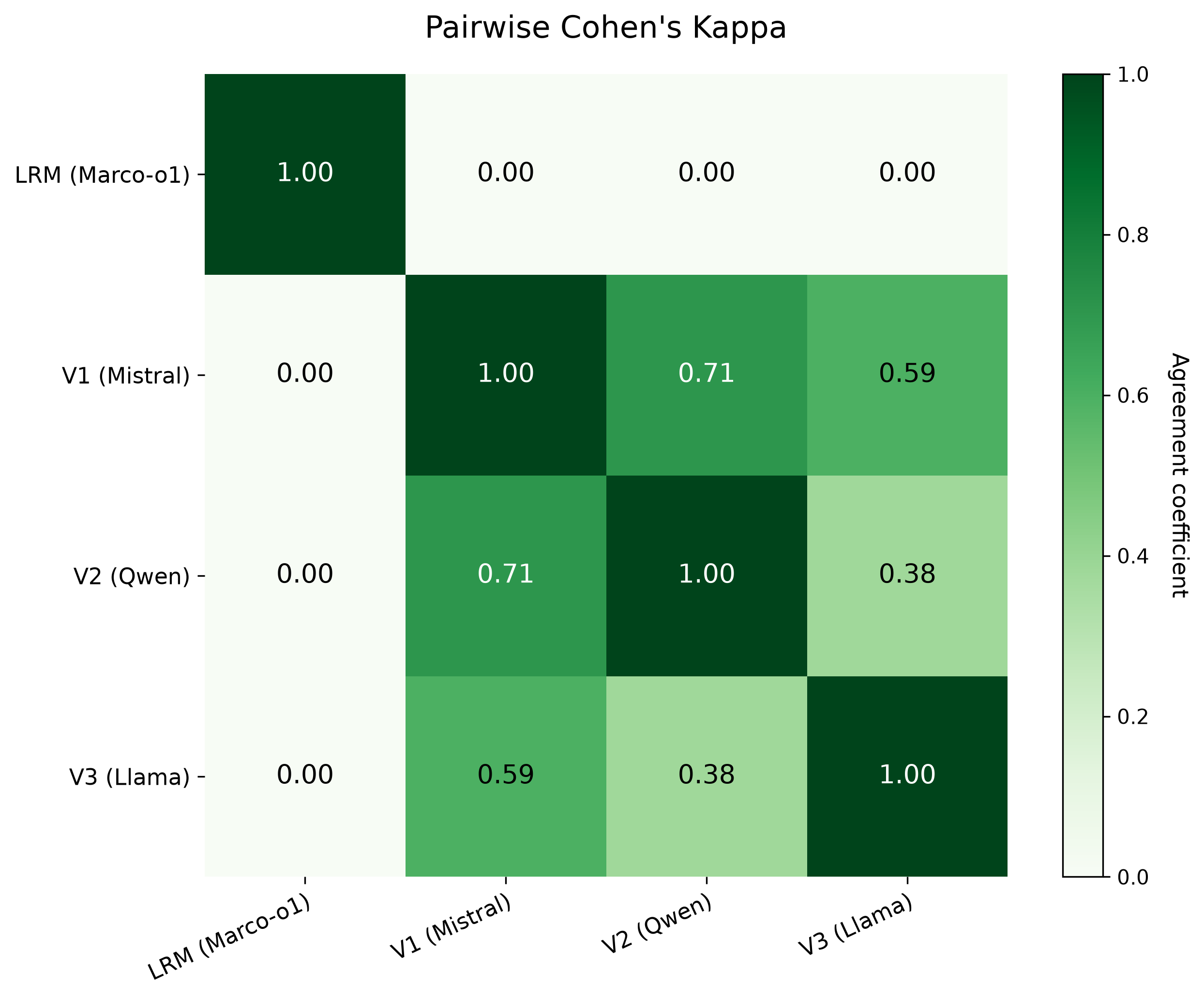}
        \caption{Cohen's Kappa}
        \label{fig:remediationapproachesfinalcohenk}
    \end{subfigure}
    \hfill
    \begin{subfigure}[t]{0.48\textwidth}
        \centering
        \includegraphics[width=\linewidth]{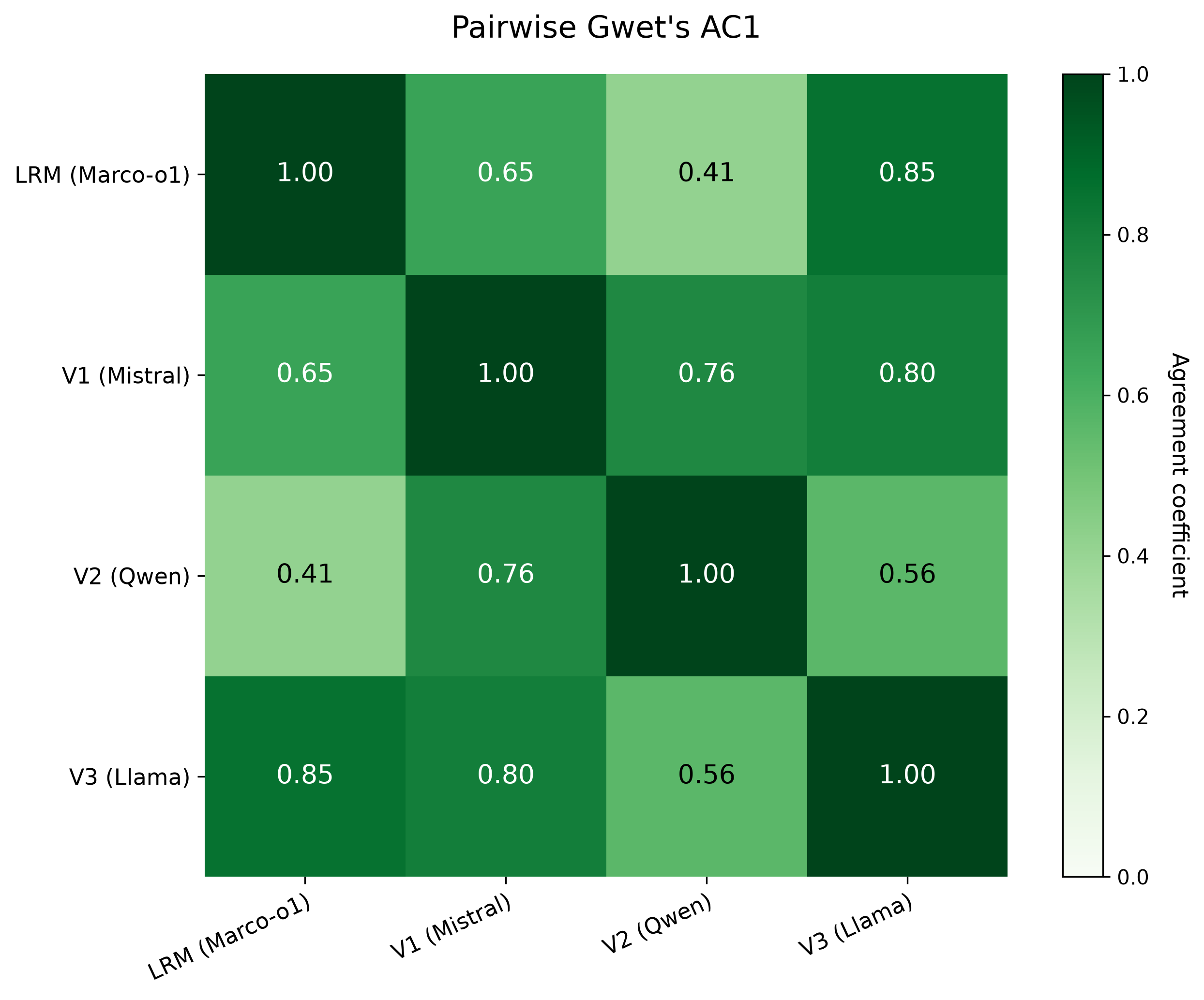}
        \caption{Gwet's AC}
        \label{fig:remediationapproachesfinalgwetac}
    \end{subfigure}

    \caption{Remediation approaches (RQ$_4$).}
    \label{fig:remediationapproachesfinal}
\end{figure}

Figure~\ref{fig:remediationapproachesfinal} summarizes the pairwise inter-rater agreement for the RQ$_4$ data-classification round, which concerned the classification of remediation approaches for architectural degradation. Cohen’s Kappa (K) reveals considerable variation among the rater pairs, with coefficients ranging from 0.00 to 0.71. The strongest agreement is achieved between V1 (Mistral) and V2 (Qwen) (K = 0.71), followed by V1 and V3 (Llama) (K = 0.59). Agreement between V2 and V3 is lower (K = 0.38). In contrast, the LRM (Marco-o1) obtains K = 0.00 with each of the other three raters.
Gwet’s AC1 yields higher agreement estimates, ranging from 0.41 to 0.85. The highest agreement is observed between the LRM and V3 (AC1 = 0.85), followed by V1 and V3 (AC1 = 0.80) and V1 and V2 (AC1 = 0.76). The lowest AC1 value occurs between the LRM and V2 (AC1 = 0.41), while V2 and V3 reach 0.56 and the LRM and V1 reach 0.65.

\end{document}